\documentclass[showpacs,preprintnumbers,amsmath,amssymb]{revtex4}
\usepackage{graphicx}
\usepackage{dcolumn}
\usepackage{bm}

\begin{document}

\title{FINE TUNING FREE PARADIGM OF TWO MEASURES  THEORY:
k-ESSENCE, ABSENCE OF  INITIAL SINGULARITY OF THE CURVATURE AND
INFLATION WITH GRACEFUL EXIT TO ZERO COSMOLOGICAL CONSTANT STATE}

\author
{E. I. Guendelman \thanks{guendel@bgu.ac.il} and A.  B. Kaganovich
\thanks{alexk@bgu.ac.il}}
\address{Physics Department, Ben Gurion University of the Negev, Beer
Sheva 84105, Israel}

\date{\today}

\begin{abstract}

The dilaton-gravity sector of the Two Measures Field Theory (TMT)
is explored in detail in the context of spatially flat FRW
cosmology. The model possesses scale invariance which is
spontaneously broken due to the intrinsic features of the TMT
dynamics. The dilaton $\phi$ dependence of the effective
Lagrangian appears only as a result of the spontaneous breakdown
of the scale invariance. If no fine tuning is made, the effective
$\phi$-Lagrangian $p(\phi,X)$ depends quadratically upon the
kinetic term $X$. Hence TMT represents an explicit example of {\it
the effective} $k$-{\it essence resulting from first principles
without any exotic term} in the underlying action intended for
obtaining this result. Depending of the choice of regions in the
parameter space (but without fine tuning), TMT exhibits different
possible outputs for cosmological dynamics: a) {\it Absence of
initial singularity of the curvature while its time derivative is
singular}. This is a sort of "sudden" singularities studied by
Barrow on purely kinematic grounds. b) Power law inflation in the
subsequent stage of evolution. Depending on the region in the
parameter space the inflation ends with a {\it graceful exit}
either into the state with zero cosmological constant (CC) or into
the state driven by both a small CC and the field $\phi$ with a
quintessence-like potential. c) Possibility of {\it resolution of
the old CC problem}. From the point of view of TMT, it becomes
clear why the old CC problem cannot be solved (without fine
tuning) in conventional field theories. d) TMT enables two ways
for achieving small CC without fine tuning of dimensionfull
parameters: either by a {\it seesaw} type mechanism or due to
 {\it a correspondence principle} between TMT and
conventional field theories (i.e theories with only the measure of
integration $\sqrt{-g}$  in the action).  e) There is a wide range
of the parameters such that in the late time universe: the
equation-of-state  $w=p/\rho <-1$; $w$ {\it asymptotically} (as
$t\rightarrow\infty$) {\it approaches} $-1$ {\it from below};
$\rho$ approaches a constant, the smallness of which does not
require fine tuning of dimensionfull parameters.
\end{abstract}

\maketitle

\section{Introduction}

The cosmological constant (CC) problem \cite{Weinberg1}-\cite{CC},
the accelerated expansion of the late time universe\cite{accel},
the cosmic coincidence \cite{coinc} are challenges for the
foundations of modern physics (see also reviews on dark
energy\cite{de-review}-\cite{Copeland}, dark matter
\cite{dm-review} and references therein). Numerous models have
been proposed with the aim to find answer to these puzzles, for
example: the quintessence\cite{quint},  coupled
quintessence\cite{Amendola},
$k$-essence\cite{k-essence},\cite{Caldwell-Steinhardt-Mukhanov},
variable mass particles\cite{vamp}, interacting
quintessence\cite{int-q}, Chaplygin gas\cite{Chapl}, phantom
field\cite{phantom}, tachyon matter cosmology\cite{tachyon},
abnormally weighting energy hypothesis\cite{Alimi},
 brane world scenarios\cite{brane}, etc.. These puzzles have also motivated an interest in
modifications and even radical revisions of the standard
gravitational theory (General Relativity
(GR))\cite{modified-gravity1},\cite{modified-gravity2}. Although
motivations for most of these models can be found in fundamental
theories like for example in brane world\cite{extra-dim},  the
questions concerning the Einstein GR limit and relation to the
regular particle physics, like standard model, still remain
unclear. One can add to the list of puzzles the problem of initial
singularity\cite{singular},\cite{HE-book}, including the
singularity theorems for scalar field-driven inflationary
cosmology\cite{Borde-Vilenkin-PRL}, resolution of which is perhaps
a crucial criteria for the true theory.

It is very hard to imagine that  it is possible  to propose ideas
 which are able to solve  the above mentioned problems  keeping
 at the same time unchanged the basis of fundamental physics, i.e.
 gravity and particle field theory. This paper may be regarded as
 {\it an attempt to find a way for satisfactory answers at least to a part of the
 fundamental questions on the basis of first principles, i.e. without
using semi-empirical models}. In this paper we explore a toy model
including gravity and a single scalar field $\phi$ in the
framework of the so called Two Measures Field Theory
(TMT)\cite{GK1}-\cite{hep-th/0603150}. In TMT, gravity (or more
exactly, geometry) and particle field theory intertwined in a very
non trivial manner, but the fifth force problem is resolved and
the Einstein's GR is restored when the local fermion matter energy
density (i.e in the space-time regions occupied by the fermions)
is much larger than the vacuum energy
density\cite{GK6},\cite{GK7}.

Here we {\it have no  purpose} of constructing a complete
realistic cosmological scenario. Instead, we concentrate our
attention on the possible role TMT may play in resolution of a
number of the above mentioned problems. All the novelty of the
field theory model we will study here results from the peculiar
structure of the TMT action while the Lagrangian densities do not
contain any exotic terms. The model is invariant under global
scale transformation of the metric accompanied  with an
appropriate shift of the dilaton field $\phi$. This scale symmetry
is spontaneously broken due to intrinsic features of the TMT
dynamics.  The obtained dynamics represents an explicit example of
$k$-essence resulting from first principles.

The organization of the paper is the following. In Sec.II we
present a review of the basic ideas of TMT. Sec.III starts from
formulation of a simple scale invariant model containing all the
terms respecting the symmetry of the model but without any exotic
terms. In Appendix A we present equations of motion in the
original frame. Using results of Appendix A, the complete set of
equations of motion in the Einstein frame is given in Sec.III as
well. It is shown there that if no fine tuning of the parameters
is made, the effective action of our model in the Einstein frame
turns out to be a k-essence type action quadratic in the kinetic
term. We start in Sec.IV from a simple case with fine tuned
parameters where the non-linear dependence of the kinetic term
disappears. Then three different shapes of the effective potential
are possible. For the spatially flat FRW universe we study some
features of the cosmological dynamics for each of the shapes of
the effective potential. For one of the shapes of the effective
potential, the zero vacuum energy is achieved without fine tuning
of dimensionfull parameters, integration constant and initial
conditions.

In Secs.V, VI and VII we study the cosmological dynamics of the
FRW universe without any fine tuning parameters. The structure of
the scalar field phase plane turns out to be very unexpected. When
continuing phase curves to the past, it is revealed that in a
finite cosmic time they arrive the repulsive line in the phase
plane. The energy density $\rho$, pressure $p$, the first two
derivatives of the scale factor $\dot{a}$ and $\ddot{a}$ remain
finite on this line but $\dot{p}$ and $\dddot{a}$ become singular.
 The subsequent stage of evolution is characterized by a power law inflation.
Depending on the region in the parameter space (but without fine
tuning) the inflation ends with a  graceful exit either into the
state with zero cosmological constant (CC), Sec.V, or into the
state driven by both a small CC and the field $\phi$ with a
quintessence-like potential, Sec.VI. The speed $c_s$ of
propagation of the cosmological perturbations varies and during
the power law inflation $c_s>1$.

In Sec.VII we show that there is a wide range of the parameters
such that: the equation-of-state in the late time universe
$w=p/\rho <-1$; $w$  asymptotically (as $t\rightarrow\infty$)
approaches $-1$  from below; $\rho$ approaches a constant, the
smallness of which does not require fine tuning of dimensionfull
parameters. It is shown that there is the possibility of a
superacceleration phase of the universe, and some details of the
dynamics are explored.

Sec.VIII is devoted to the  resolution of the CC problem. In
Sec.VIIIA we study two ways TMT enables  for resolution of the new
CC problem without fine tuning of dimensionfull parameters: either
by a  seesaw type mechanism or due to
  a correspondence principle between TMT and
conventional field theories (i.e theories with only the measure of
integration $\sqrt{-g}$  in the action). In Sec.VIIIB we analyze
in detail why the Weinberg's no-go cosmological constant
theorem\cite{Weinberg1} may be nonapplicable in our model. We
analyze also the difference between TMT and conventional field
theories (where only the measure of integration $\sqrt{-g}$ is
used in the fundamental action) which  allows to understand why
from the point of view of TMT the conventional field theories
failed to solve the old CC problem.

In Appendix B we shortly discuss what kind of model one would
obtain when choosing fine tuned couplings to the measures of
integration in the action. Some additional remarks concerning the
relation between the structure of TMT action and the CC problem
are given in Appendix C. In Appendix D we briefly discuss a
connection of a particular case of our model with the class of
models studied in Ref.\cite{Tsujikawa}

\section{Basis of Two Measures Field Theory}

TMT is a generally coordinate invariant theory where {\it all the
difference from the standard field theory in curved space-time
consists only of the following three additional assumptions}:

1. The main supposition  is that for describing the effective
action for 'gravity $+$ matter' at energies below the Planck
scale, the usual form of the action $S = \int L\sqrt{-g}d^{4}x$ is
not complete. Our positive hypothesis is that the effective action
has to be of the form\cite{GK3}-\cite{GK8}
\begin{equation}
    S = \int L_{1}\Phi d^{4}x +\int L_{2}\sqrt{-g}d^{4}x
\label{S}
\end{equation}
 including two Lagrangians $ L_{1}$ and $L_{2}$ and two
measures of integration $\sqrt{-g}$ and $\Phi$. One is the usual
measure of integration $\sqrt{-g}$ in the 4-dimensional space-time
manifold equipped with the metric
 $g_{\mu\nu}$.
Another  is the new measure of integration $\Phi$ in the same
4-dimensional space-time manifold. The measure  $\Phi$ being  a
scalar density and a total derivative\footnote{For applications of
the measure $\Phi$ in string and brane theories see
Ref.\cite{Mstring}.} may be defined
\begin{itemize}

\item
either by means of  four scalar fields $\varphi_{a}$
($a=1,2,3,4$), (compare with the approach by Wilczek\footnote{See:
 F. Wilczek, Phys.Rev.Lett. {\bf 80}, 4851
(1998). Wilczek's goal was to avoid the use of a fundamental
metric, and for this purpose he needs five scalar fields. In our
case we keep the standard role of the metric from the beginning,
but enrich the theory with a new metric independent density.}),
\begin{equation}
\Phi
=\varepsilon^{\mu\nu\alpha\beta}\varepsilon_{abcd}\partial_{\mu}\varphi_{a}
\partial_{\nu}\varphi_{b}\partial_{\alpha}\varphi_{c}
\partial_{\beta}\varphi_{d}.
\label{Phi}
\end{equation}

\item
or by means of a totally antisymmetric three index field
$A_{\alpha\beta\gamma}$
\begin{equation}
\Phi
=\varepsilon^{\mu\nu\alpha\beta}\partial_{\mu}A_{\nu\alpha\beta}.
\label{Aabg}
\end{equation}
\end{itemize}

To provide parity conservation in the case given by Eq.(\ref{Phi})
one can choose for example one of $\varphi_{a}$'s to be a
pseudoscalar; in the case given by Eq.(\ref{Aabg}) we must choose
$A_{\alpha\beta\gamma}$ to have negative parity. Some ideas
concerning the nature of the measure fields $\varphi_{a}$ are
discussed in Ref.\cite{GK8}. The idea of T.D. Lee on the
possibility of dynamical coordinates\cite{TDLee} may be related to
the measure fields $\varphi_{a}$; see also\cite{Reuter} and our
discussion in Sec.IX.C concerning the ideas of
Ref.\cite{Giddings}. A special case of the structure (\ref{S})
with definition (\ref{Aabg}) has been recently discussed in
Ref.\cite{hodge} in applications to supersymmetric theory and the
CC problem.

2. It is assumed that the Lagrangian densities $ L_{1}$ and
$L_{2}$ are functions of all matter fields, the dilaton field, the
metric, the connection
 but not of the
"measure fields" ($\varphi_{a}$ or $A_{\alpha\beta\gamma}$). In
such a case, i.e. when the measure fields  enter in the theory
only via the measure $\Phi$,
  the action (\ref{S}) possesses
an infinite dimensional symmetry. In the case given by
Eq.(\ref{Phi}) these symmetry transformations have the form
$\varphi_{a}\rightarrow\varphi_{a}+f_{a}(L_{1})$, where
$f_{a}(L_{1})$ are arbitrary functions of  $L_{1}$ (see details in
Ref.\cite{GK3}); in the case given by Eq.(\ref{Aabg}) they read
$A_{\alpha\beta\gamma}\rightarrow
A_{\alpha\beta\gamma}+\varepsilon_{\mu\alpha\beta\gamma}
f^{\mu}(L_{1})$ where $f^{\mu}(L_{1})$ are four arbitrary
functions of $L_{1}$ and $\varepsilon_{\mu\alpha\beta\gamma}$ is
numerically the same as $\varepsilon^{\mu\alpha\beta\gamma}$. One
can hope that this symmetry should prevent emergence of a measure
fields dependence in $ L_{1}$ and $L_{2}$ after quantum effects
are taken into account.

3. Important feature of TMT that is responsible for many
interesting and desirable results of the field theory models
studied so far\cite{GK3}-\cite{GK8}
 consists of the assumption that all fields, including
also metric, connection  and the {\it measure fields}
($\varphi_{a}$ or $A_{\alpha\beta\gamma}$) are independent
dynamical variables. All the relations between them are results of
equations of motion.  In particular, the independence of the
metric and the connection means that we proceed in the first order
formalism and the relation between connection and metric is not
necessarily according to Riemannian geometry.

We want to stress again that except for the listed three
assumptions we do not make any changes as compared with principles
of the standard field theory in curved space-time. In other words,
all the freedom in constructing different models in the framework
of TMT consists of the choice of the concrete matter content and
the Lagrangians $ L_{1}$ and $L_{2}$ that is quite similar to the
standard field theory.

Since $\Phi$ is a total derivative, a shift of $L_{1}$ by a
constant, $L_{1}\rightarrow L_{1}+const$, has no effect on the
equations of motion. Similar shift of $L_{2}$ would lead to the
change of the constant part of the Lagrangian coupled to the
volume element $\sqrt{-g}d^{4}x $. In the standard GR, this
constant term is the cosmological constant. However in TMT the
relation between the constant
 term of $L_{2}$ and the physical cosmological constant is very non
trivial (see \cite{GK3}-\cite{K},\cite{GK5}-\cite{GK7}).

In the case of the definition of $\Phi$ by means of
Eq.(\ref{Phi}), varying the measure fields $\varphi_{a}$, we get
\begin{equation}
B^{\mu}_{a}\partial_{\mu}L_{1}=0  \quad \text{where} \quad
B^{\mu}_{a}=\varepsilon^{\mu\nu\alpha\beta}\varepsilon_{abcd}
\partial_{\nu}\varphi_{b}\partial_{\alpha}\varphi_{c}
\partial_{\beta}\varphi_{d}.\label{varphiB}
\end{equation}
Since $Det (B^{\mu}_{a}) = \frac{4^{-4}}{4!}\Phi^{3}$ it follows
that if $\Phi\neq 0$,
\begin{equation}
 L_{1}=sM^{4} =const
\label{varphi}
\end{equation}
where $s=\pm 1$ and $M$ is a constant of integration with the
dimension of mass. In what follows we make the choice $s=1$.

In the case of the definition (\ref{Aabg}), variation of
$A_{\alpha\beta\gamma}$ yields
\begin{equation}
\varepsilon^{\mu\nu\alpha\beta}\partial_{\mu}L_{1}=0, \label{AdL1}
\end{equation}
that implies Eq.(\ref{varphi}) without the  condition $\Phi\neq 0$
needed in the model with four scalar fields $\varphi_{a}$.

 One should notice
 {\it the very important differences of
TMT from scalar-tensor theories with nonminimal coupling}: \\
 a) In general, the Lagrangian density $L_{1}$ (coupled to the measure
$\Phi$) may contain not only the scalar curvature term (or more
general gravity term) but also all possible matter fields terms.
This means that TMT modifies in general both the gravitational
sector  and the matter sector; b) If the field $\Phi$ were the
fundamental (non composite) one then instead of (\ref{varphi}),
the variation of $\Phi$ would result in the equation $L_{1}=0$ and
therefore the dimensionfull integration constant $M^4$ would not
appear in the theory.

Applying the Palatini formalism in TMT one can show (see for
example \cite{GK3} and Appendix A of the present paper)  that in
addition to the Christoffel's connection coefficients, the
resulting relation between connection and metric includes also the
gradient of the ratio of the two measures
\begin{equation}
\zeta \equiv\frac{\Phi}{\sqrt{-g}} \label{zeta}
\end{equation}
which is a scalar field. Consequently geometry of the space-time
with the metric $g_{\mu\nu}$ is non-Riemannian. The gravity and
matter field equations obtained by means of the first order
formalism contain both $\zeta$ and its gradient as well. It turns
out that at least at the classical level, the measure fields
affect the theory only through the scalar field $\zeta$.

The consistency condition of equations of motion has the form of a
constraint which determines $\zeta (x)$ as a function of matter
fields. The surprising feature of the theory is
 that neither Newton constant nor curvature appear in this constraint
which means that the {\it geometrical scalar field} $\zeta (x)$
{\it is determined by the matter fields configuration}  locally
and straightforward (that is without gravitational interaction).

By an appropriate change of the dynamical variables which includes
a  transformation of the metric, one can formulate the theory in a
Riemannian  space-time. The corresponding frame we call "the
Einstein frame". The big advantage of TMT is that in the very wide
class of models, {\it the gravity and all matter fields equations
of motion take canonical GR form in the Einstein frame}.
 All the novelty of TMT in the Einstein frame as compared
with the standard GR is revealed only
 in an unusual structure of the scalar fields
effective potential (produced in the Einstein frame), masses of
fermions  and their interactions with scalar fields as well as in
the unusual structure of fermion contributions to the
energy-momentum tensor: all these quantities appear to be $\zeta$
dependent\cite{GK5}-\cite{GK7}. This is why the scalar field
$\zeta (x)$ determined by the constraint as a function of matter
fields, has a key role in the dynamics of TMT models.

\section{Scale invariant model}
\subsection{Symmetries and Action}

The TMT models possessing a global scale
invariance\cite{G1}-\cite{GKatz},\cite{GK5}-\cite{GK7} are of
significant interest because they demonstrate the possibility of
spontaneous breakdown of the scale symmetry\footnote{The field
theory models with explicitly broken scale symmetry and their
application to the quintessential inflation type  cosmological
scenarios have been studied in Ref.\cite{K}. Inflation and
transition to slowly accelerated phase from higher curvature terms
was studied in Ref.\cite{GKatz}. }. In fact, if the action
(\ref{S}) is scale invariant then this classical field theory
effect results from Eq.(\ref{varphi}), namely  from solving the
equation of motion (\ref{varphiB}) or (\ref{AdL1}).
 One of the
interesting applications of the scale invariant TMT
models\cite{GK5} is a possibility to generate the exponential
potential for the scalar field $\phi$ by means of the mentioned
spontaneous symmetry breaking even  without introducing any
potentials for  $\phi$ in the Lagrangians  $ L_{1}$ and $L_{2}$ of
the action (\ref{S}). Some cosmological applications of this
effect have been studied in
 Ref.\cite{GK5} (see  also Appendix D of the present paper).

A dilaton field $\phi$ allows to realize a spontaneously broken
global scale invariance\cite{G1} and together with this it governs
the evolution of the universe on different stages: in the early
universe $\phi$ plays the role of inflaton and in the late time
universe it is transformed into a part of the dark energy.

According to the general prescriptions of TMT, we have to start
from studying the self-consistent system of gravity (metric
$g_{\mu\nu}$ and connection $\Gamma^{\mu}_{\alpha\beta}$), the
measure $\Phi$ degrees of freedom (i.e. measure fields $\varphi_a$
or $A_{\alpha\beta\gamma}$) and the dilaton field $\phi$
proceeding in the first order formalism.  We postulate that the
theory is invariant under the global scale transformations:
\begin{equation}
    g_{\mu\nu}\rightarrow e^{\theta }g_{\mu\nu}, \quad
\Gamma^{\mu}_{\alpha\beta}\rightarrow \Gamma^{\mu}_{\alpha\beta},
\quad \varphi_{a}\rightarrow \lambda_{ab}\varphi_{b}\quad
\text{where} \quad \det(\lambda_{ab})=e^{2\theta}, \quad
\phi\rightarrow \phi-\frac{M_{p}}{\alpha}\theta . \label{st}
\end{equation}
If the definition (\ref{Aabg}) is used for the measure $\Phi$ then
 the transformations of $\varphi_{a}$ in
(\ref{st}) should be changed by $A_{\alpha\beta\gamma}\rightarrow
e^{2\theta}A_{\alpha\beta\gamma}$. This global scale invariance
includes the shift symmetry of the dilaton $\phi$ and this is the
main factor why it is important for cosmological applications of
the theory\cite{G1}-\cite{GKatz},\cite{GK5}-\cite{GK7}.

We choose an action which, except for the modification of the
general structure caused by the basic assumptions of TMT,
 {\it does not contain
 any exotic terms and  fields} as like in the conventional formulation
 of the minimally coupled scalar-gravity system.
Keeping the general structure (\ref{S}), it is convenient to
represent the underlying action of our model in the following
form:
\begin{eqnarray}
&S=&\int d^{4}x e^{\alpha\phi /M_{p}}
\left[-\frac{1}{\kappa}R(\Gamma ,g)(\Phi +b_{g}\sqrt{-g})+(\Phi
+b_{\phi}\sqrt{-g})\frac{1}{2}g^{\mu\nu}\phi_{,\mu}\phi_{,\nu}-e^{\alpha\phi
/M_{p}}\left(\Phi V_{1} +\sqrt{-g}V_{2}\right)\right]
 \label{totaction}
\end{eqnarray}

In the action (\ref{totaction}) there are two types of the
gravitational terms and
 of the "kinetic-like terms"  which
respect the scale invariance : the terms of the one type coupled
to the
 measure $\Phi$ and those of the other type
coupled to the measure $\sqrt{-g}$. Using the freedom in
normalization of the measure fields ($\varphi_{a}$  in the case of
using Eq.(\ref{Phi}) or $A_{\alpha\beta\gamma}$ when using
Eq.(\ref{Aabg})), we set the coupling constant of the scalar
curvature to the measure $\Phi$ to be  $-\frac{1}{\kappa}$.
Normalizing all the fields such that their couplings to the
measure $\Phi$ have no additional factors, we are not able in
general to provide the same in the terms describing the
appropriate couplings to the measure $\sqrt{-g}$. This fact
explains the need to introduce the dimensionless real parameters
$b_g$ and $b_{\phi}$ and we will only assume\footnote{There is a
freedom to write down the underlying action with the alternative
choice of the parameters $b_g$ and $b_{\phi}$: one can normalize
all the fields such that their couplings to the measure
$\sqrt{-g}$ have no additional factors. Then generically one
should  introduce arbitrary dimensionless coupling constants to
the measure $\Phi$. It is clear that if for example the parameters
$b_g$ and $b_{\phi}$ in (\ref{totaction}) are large then  the new
parameters would be small and vice versa. Such formulation of the
underlying action results in the model equivalent to the model
(\ref{totaction}).} that {\it they are positive and have close
orders of magnitudes}. Note that in the case of the choice
$b_g=b_{\phi}$ we would proceed with {\it the fine tuned model}.
The real positive parameter $\alpha$ is assumed to be of the order
of unity; in all numerical solutions presented in this paper we
set $\alpha =0.2$. For the Newton constant we use $\kappa =16\pi G
$, \, $M_p=(8\pi G)^{-1/2}$.

One should  also point out the possibility of introducing two
different pre-potentials which are exponential functions of the
dilaton $\phi$ coupled to the measures $\Phi$ and $\sqrt{-g}$ with
factors $V_{1}$ and $V_{2}$. Such $\phi$-dependence provides the
scale symmetry (\ref{st}). However $V_{1}$ and $V_{2}$ might be
Higgs-dependent and then they play the role of the Higgs
pre-potentials\cite{hep-th/0603150}.

\subsection{Equations of motion in the Einstein
frame. }

In Appendix A  we present the equations of motion resulting from
the action (\ref{totaction}) when using the original set of
variables.  The common feature of all the equations in the
original frame is that the measure $\Phi$ degrees of freedom
appear only through dependence upon the scalar field $\zeta$,
Eq.(\ref{zeta}). In particular, all the equations of motion and
the solution for the connection coefficients include terms
proportional to $\partial_{\mu}\zeta$, that makes space-time non
Riemannian and all equations of motion - noncanonical.

It turns out that when working with the new metric ($\phi$
 remains the same)
\begin{equation}
\tilde{g}_{\mu\nu}=e^{\alpha\phi/M_{p}}(\zeta +b_{g})g_{\mu\nu},
\label{ct}
\end{equation}
which we call the Einstein frame,
 the connection  becomes Riemannian. Since
$\tilde{g}_{\mu\nu}$ is invariant under the scale transformations
(\ref{st}), spontaneous breaking of the scale symmetry (by means
of Eq.(\ref{varphi}) which for our model (\ref{totaction}) takes
the form (\ref{app1})) is reduced in the Einstein frame to the
{\it spontaneous breakdown of the shift symmetry}
\begin{equation}
 \phi\rightarrow\phi +const.
 \label{phiconst}
\end{equation}

Notice that the Goldstone theorem generically is not applicable in
this model\cite{G1}. The reason is the following. In fact, the
scale symmetry (\ref{st}) leads to a conserved dilatation current
$j^{\mu}$. However, for example in the spatially flat FRW universe
the spatial components of the current $j^{i}$ behave as
$j^{i}\propto M^4x^i$ as $|x^i|\rightarrow\infty$. Due to this
anomalous behavior at infinity, there is a flux of the current
leaking to infinity, which causes the non conservation of the
dilatation charge. The absence of the latter implies that one of
the conditions necessary for the Goldstone theorem is missing. The
non conservation of the dilatation charge is similar to the well
known effect of instantons in QCD where singular behavior in the
spatial infinity leads to the absence of the Goldstone boson
associated to the $U(1)$ symmetry.

 After the change of
variables  to the Einstein frame (\ref{ct}) and some simple
algebra, Eq.(\ref{app4}) takes the standard GR form
\begin{equation}
G_{\mu\nu}(\tilde{g}_{\alpha\beta})=\frac{\kappa}{2}T_{\mu\nu}^{eff}
 \label{gef}
\end{equation}
where  $G_{\mu\nu}(\tilde{g}_{\alpha\beta})$ is the Einstein
tensor in the Riemannian space-time with the metric
$\tilde{g}_{\mu\nu}$; the energy-momentum tensor
$T_{\mu\nu}^{eff}$ is now
\begin{eqnarray}
T_{\mu\nu}^{eff}&=&\frac{\zeta +b_{\phi}}{\zeta +b_{g}}
\left(\phi_{,\mu}\phi_{,\nu}-\frac{1}{2}
\tilde{g}_{\mu\nu}\tilde{g}^{\alpha\beta}\phi_{,\alpha}\phi_{,\beta}\right)
-\tilde{g}_{\mu\nu}\frac{b_{g}-b_{\phi}}{2(\zeta +b_{g})}
\tilde{g}^{\alpha\beta}\phi_{,\alpha}\phi_{,\beta}
+\tilde{g}_{\mu\nu}V_{eff}(\phi;\zeta,M)
 \label{Tmn}
\end{eqnarray}
where the function $V_{eff}(\phi;\zeta,M)$ is defined as
following:
\begin{equation}
V_{eff}(\phi;\zeta ,M)=
\frac{b_{g}\left[M^{4}e^{-2\alpha\phi/M_{p}}+V_{1}\right]
-V_{2}}{(\zeta +b_{g})^{2}}. \label{Veff1}
\end{equation}
 Putting $M$ in the
arguments of $V_{eff}$ we indicate explicitly that $V_{eff}$
incorporates our choice for the integration constant $M$ that
appears as a result of the spontaneous breakdown of the scale
symmetry. We will see in the next sections that $\zeta$-dependence
of $V_{eff}(\phi;\zeta ,M)$ in the form of {\it square} of $(\zeta
+b_{g})^{-1}$ has a key role in the resolution of the old CC
problem in TMT. The fact that only such $\zeta$-dependence emerges
in $V_{eff}(\phi;\zeta ,M)$, and a $\zeta$-dependence is absent
for example in the numerator of $V_{eff}(\phi;\zeta ,M)$, is a
direct result of our basic assumption that $L_1$ and $L_2$ are
independent of the measure fields (see item 2 in Sec.II).

The dilaton $\phi$ field equation (\ref{phi-orig}) in the Einstein
frame reads
\begin{eqnarray}
&&\frac{1}{\sqrt{-\tilde{g}}}\partial_{\mu}\left[\frac{\zeta
+b_{\phi}}{\zeta
+b_{g}}\sqrt{-\tilde{g}}\tilde{g}^{\mu\nu}\partial_{\nu}\phi\right]
\nonumber\\
 &-&\frac{\alpha}{M_{p}(\zeta +b_{g})^{2}} \left[(\zeta
+b_{g})M^{4}e^{-2\alpha\phi/M_{p}}-(\zeta -b_{g})V_{1}
-2V_{2}-\delta b_{g}(\zeta
+b_{g})\frac{1}{2}\tilde{g}^{\alpha\beta}\phi_{,\alpha}\phi_{,\beta}\right]
=0
 \label{phief}
\end{eqnarray}

The scalar field $\zeta$ in Eqs.(\ref{Tmn})-(\ref{phief}) is
determined by means of the constraint (\ref{app3}) which in the
Einstein frame (\ref{ct}) takes the form
\begin{eqnarray}
&&(b_{g}-\zeta)\left[M^{4}e^{-2\alpha\phi/M_{p}}+
V_{1}\right]-2V_{2}-\delta\cdot b_{g}(\zeta +b_{g})X
=0\label{constraint2}
\end{eqnarray}
where
\begin{equation}
X\equiv\frac{1}{2}\tilde{g}^{\alpha\beta}\phi_{,\alpha}\phi_{,\beta}
\qquad \text{and} \qquad \delta =\frac{b_{g}-b_{\phi}}{b_{g}}
\label{delta}
\end{equation}

Applying the constraint (\ref{constraint2}) to Eq.(\ref{phief})
one can reduce the latter to the form
\begin{equation}
\frac{1}{\sqrt{-\tilde{g}}}\partial_{\mu}\left[\frac{\zeta
+b_{\phi}}{\zeta
+b_{g}}\sqrt{-\tilde{g}}\tilde{g}^{\mu\nu}\partial_{\nu}\phi\right]-\frac{2\alpha\zeta}{(\zeta
+b_{g})^{2}M_{p}}M^{4}e^{-2\alpha\phi/M_{p}} =0,
\label{phi-after-con}
\end{equation}
where $\zeta$  is a solution of the constraint
(\ref{constraint2}).

The effective energy-momentum tensor (\ref{Tmn}) can be
represented in a form of that of  a perfect fluid
\begin{equation}
T_{\mu\nu}^{eff}=(\rho +p)u_{\mu}u_{\nu}-p\tilde{g}_{\mu\nu},
\qquad \text{where} \qquad
u_{\mu}=\frac{\phi_{,\mu}}{(2X)^{1/2}}\label{Tmnfluid}
\end{equation}
with the following energy and pressure densities resulting from
Eqs.(\ref{Tmn}) and (\ref{Veff1}) after inserting the solution
$\zeta =\zeta(\phi,X;M)$ of Eq.(\ref{constraint2}):
\begin{equation}
\rho(\phi,X;M) =X+ \frac{(M^{4}e^{-2\alpha\phi/M_{p}}+V_{1})^{2}-
2\delta b_{g}(M^{4}e^{-2\alpha\phi/M_{p}}+V_{1})X -3\delta^{2}
b_{g}^{2}X^2}{4[b_{g}(M^{4}e^{-2\alpha\phi/M_{p}}+V_{1})-V_{2}]},
\label{rho1}
\end{equation}
\begin{equation}
p(\phi,X;M) =X- \frac{\left(M^{4}e^{-2\alpha\phi/M_{p}}+V_{1}+
\delta b_{g}X\right)^2}
{4[b_{g}(M^{4}e^{-2\alpha\phi/M_{p}}+V_{1})-V_{2}]}. \label{p1}
\end{equation}

In a spatially flat FRW universe with the metric
$\tilde{g}_{\mu\nu}=diag(1,-a^2,-a^2,-a^2)$ filled with the
homogeneous scalar field $\phi(t)$, the $\phi$  field equation of
motion takes the form
\begin{equation}
Q_{1}\ddot{\phi}+ 3HQ_{2}\dot{\phi}- \frac{\alpha}{M_{p}}Q_{3}
M^{4}e^{-2\alpha\phi/M_{p}}=0 \label{phi1}
\end{equation}
 where $H$ is the Hubble parameter and we have used the following notations
\begin{equation}
\dot{\phi}\equiv \frac{d\phi}{dt} \label{phidot-vdot}
\end{equation}
\begin{equation}
Q_1=2[b_{g}(M^{4}e^{-2\alpha\phi/M_{p}}+V_{1})-V_{2}]\rho_{,X}
=(b_{g}+b_{\phi})(M^{4}e^{-2\alpha\phi/M_{p}}+V_{1})-
2V_{2}-3\delta^{2}b_{g}^{2}X \label{Q1}
\end{equation}
\begin{equation}
Q_2=2[b_{g}(M^{4}e^{-2\alpha\phi/M_{p}}+V_{1})-V_{2}]p_{,X}=
(b_{g}+b_{\phi})(M^{4}e^{-2\alpha\phi/M_{p}}+V_{1})-
2V_{2}-\delta^{2}b_{g}^{2}X\label{Q2}
\end{equation}
\begin{equation}
Q_{3}=\frac{1}{[b_{g}(M^{4}e^{-2\alpha\phi/M_{p}}+V_{1})-V_{2}]}
\left[(M^{4}e^{-2\alpha\phi/M_{p}}+V_{1})
[b_{g}(M^{4}e^{-2\alpha\phi/M_{p}}+V_{1})-2V_{2}] +2\delta
b_{g}V_{2}X+3\delta^{2}b_{g}^{3}X^{2}\right] \label{Q3}
\end{equation}
Note that
\begin{equation}
-\frac{\alpha}{M_{p}}Q_{3}
M^{4}e^{-2\alpha\phi/M_{p}}=2[b_{g}(M^{4}e^{-2\alpha\phi/M_{p}}+V_{1})-V_{2}]\rho_{,\phi}
\label{Q3-rho-phi}
\end{equation}

It is interesting that the non-linear $X$-dependence appears here
in the framework of the fundamental theory without exotic terms in
the Lagrangians $L_1$ and $L_2$, see Eqs.(\ref{S}) and
(\ref{totaction}).  This effect follows just from the fact that
there are no reasons to choose the parameters $b_{g}$ and
$b_{\phi}$ in the action (\ref{totaction}) to be equal in general;
on the contrary, the choice $b_{g}=b_{\phi}$ would be a fine
tuning. Besides one should stress that the $\phi$ dependence in
$\rho$, $p$ and in equations of motion emerges only in the form
$M^{4}e^{-2\alpha\phi/M_{p}}$ where $M$ is the integration
constant (see Eq.(A1)), i.e. due to the spontaneous breakdown of
the scale symmetry (\ref{st}) (or the shift symmetry
(\ref{phiconst}) in the Einstein frame). Thus the above equations
represent an {\it explicit example of $k$-essence}\cite{k-essence}
{\it resulting from first principles}. The system of equations
(\ref{gef}), (\ref{rho1})-(\ref{phi1}) accompanied with the
functions (\ref{Q1})-(\ref{Q3}) and written in the metric
$\tilde{g}_{\mu\nu}=diag(1,-a^2,-a^2,-a^2)$ can be obtained from
the k-essence type effective action
\begin{equation}
S_{eff}=\int\sqrt{-\tilde{g}}d^{4}x\left[-\frac{1}{\kappa}R(\tilde{g})
+p\left(\phi,X;M\right)\right] \label{k-eff},
\end{equation}
where $p(\phi,X;M)$ is given by Eq.(\ref{p1}). In contrast to the
simplified models studied in literature\cite{k-essence}, it is
impossible here to represent $p\left(\phi,X;M\right)$ in a
factorizable form like $\tilde{K}(\phi)\tilde{p}(X)$. The scalar
field effective Lagrangian, Eq.(\ref{p1}), can be represented in
the form
\begin{equation}
p\left(\phi,X;M\right)=K(\phi)X+
L(\phi)X^2-\frac{[V_{1}+M^{4}e^{-2\alpha\phi/M_{p}}]^{2}}
{4[b_{g}\left(V_{1}+M^{4}e^{-2\alpha\phi/M_{p}}\right)-V_{2}]}
\label{eff-L-ala-Mukhanov}
\end{equation}
where $K(\phi)$ and $L(\phi)$ depend on $\phi$ only via
$M^{4}e^{-2\alpha\phi/M_{p}}$. The obtained model belongs to a
more general class of models than those discussed recently in
Ref.\cite{Mukh-Vik-JCAP}. Note also that besides the presence of
the effective potential term, the Lagrangian
$p\left(\phi,X;M\right)$ differs from that of
Ref.\cite{k-inflation-Mukhanov} by the sign of $L(\phi)$: in our
case $L(\phi)<0$ provided the effective potential is non-negative.
{\it This result cannot be removed by a choice of the parameters}
of the underlying action (\ref{totaction}) while in
Ref.\cite{k-inflation-Mukhanov} the positivity of $L(\phi)$ was an
essential {\it assumption}. As we will see, this difference plays
a crucial role in a number of specific features of the scalar
field dynamics. In particular, {\it the absence of the initial
singularity of the curvature is directly related to the negative
sign of} $L(\phi)$.

It is interesting also to note that for the particular choice
$V_1=V_2=0$, $p\left(\phi,X;M\right)$ can be represented in the
form
$p\left(\phi,X;M\right)=Xg\left(Xe^{2\alpha\phi/M_{p}}\right)$.
Therefore with the particular choice of the parameters
$V_1=V_2=0$, the model (\ref{k-eff}) belongs to the class of
models developed in Ref.\cite{Tsujikawa}  to realize scaling
solutions in a general cosmological background (see also Appendix
D). We will see however that with a choice of non zero parameters
$V_1$ and $V_2$ one can realize  the equation of state $w<-1$ in
the late time universe.

\section{Cosmological Dynamics in  Fine Tuned $\delta =0$ Models
.}
\subsection{Equations of motion}

 The qualitative
analysis of equations is significantly simplified if $\delta =0$.
This is what we will assume in this section. Although it is a fine
tuning of the parameters (i.e. $b_{g}=b_{\phi}$), it allows us to
understand some of the general features of the model. The main
simplification in the case $\delta =0$ is that the effective
Lagrangian (\ref{p1}) takes the form of that of the scalar field
without higher powers of derivatives. Role of $\delta\neq 0$ in a
dynamical mechanism for avoidance of the initial singularity will
be studied in Secs.V and VI. A possibility to produce an effect of
a super-accelerated expansion of the late time universe (if
$\delta\neq 0$) will be studied in Sec.VII.

\begin{figure}[htb]
\includegraphics[width=18.0cm,height=12cm]{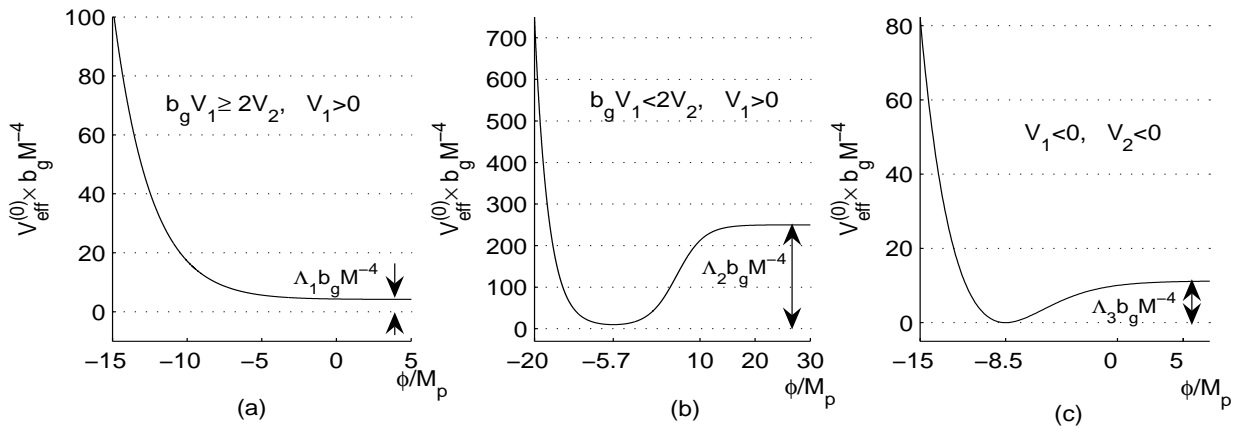}
\caption{Three possible shapes of the effective potential
$V_{eff}^{(0)}(\phi)$
 in the models with  $b_{g}V_{1}>V_{2}$:
 \quad Fig.(a) \, $b_{g}V_{1}\geq 2V_{2}$ (in the graph
 $V_{1}=10M^{4}$ and $V_{2}=4b_{g}M^{4}$); \quad Fig.(b) \, $b_{g}V_{1}<2V_{2}$
 (in the graph $V_{1}=10M^{4}$ and $V_{2}=9.9b_{g}M^{4}$).
 The value of $V_{eff}^{(0)}$ in the minimum $\phi_{min}=-5.7M_p$
 is larger than zero;
 \quad Fig.(c) \, $V_{1}<0$, $V_{2}<0$
 (in the graph $V_{1}=-30M^{4}$ and $V_{2}=-50b_{g}M^{4}$).
 $V_{eff}^{(0)}(\phi_{min})=0$ in the
 minimum $\phi_{min}=-8.5M_p$ .
In all the cases here as well as in all solutions presented in
this paper we choose $\alpha =0.2$.}\label{fig1}
\end{figure}

So let us study spatially flat FRW cosmological models governed by
the system of equations
\begin{equation}
\frac{\dot{a}^{2}}{a^{2}}=\frac{1}{3M_{p}^{2}}\rho
\label{cosm-phi}
\end{equation}
and (\ref{rho1})-(\ref{phi1}) where one should  set $\delta =0$.

In the fine tuned case under consideration,   the constraint
(\ref{constraint2}) yields
\begin{equation}
\zeta =\zeta(\phi,X;M)|_{\delta =0}\equiv b_{g}-\frac{2V_{2}}
{V_{1}+M^{4}e^{-2\alpha\phi/M_{p}}},
\label{zeta-without-ferm-delta=0}
 \end{equation}
 The energy density and pressure take then the canonical form,
\begin{equation}
\rho|_{\delta =0}=\frac{1}{2}\dot{\phi}^{2}+V_{eff}^{(0)}(\phi);
\qquad p|_{\delta
=0}=\frac{1}{2}\dot{\phi}^{2}-V_{eff}^{(0)}(\phi),
\label{rho-delta=0}
\end{equation}
where the effective potential of the scalar field $\phi$ results
from Eq.(\ref{Veff1})
\begin{equation}
V_{eff}^{(0)}(\phi)\equiv V_{eff}(\phi;\zeta ,M)|_{\delta =0}
=\frac{[V_{1}+M^{4}e^{-2\alpha\phi/M_{p}}]^{2}}
{4[b_{g}\left(V_{1}+M^{4}e^{-2\alpha\phi/M_{p}}\right)-V_{2}]}
\label{Veffvac-delta=0}
\end{equation}
and the $\phi$-equation (\ref{phi1}) is reduced to
\begin{equation}
\ddot{\phi}+3H\dot{\phi}+\frac{dV^{(0)}_{eff}}{d\phi}=0.
\label{eq-phief-without-ferm-delta=0}
\end{equation}

Notice that $V_{eff}^{(0)}(\phi)$ is non-negative for any $\phi$
provided
\begin{equation}
b_{g}V_{1}\geq V_{2} \label{bV1>V2},
\end{equation}
that we will assume in this paper.

In the following three subsections  we consider three different
dilaton-gravity cosmological models determined by different choice
of the parameters $V_{1}$ and $V_{2}$: one model with $V_{1}<0$
and two models with $V_{1}>0$. The appropriate three possible
shapes of the effective potential $V_{eff}^{(0)}(\phi)$ are
presented in Fig.1. A special case with the fine tuned condition
$b_{g}V_{1}=V_{2}$ is discussed in Appendix B where we show that
equality of the couplings to measures $\Phi$ and $\sqrt{-g}$ in
the action (equality $b_{g}V_{1}=V_{2}$ is one of the conditions
for this to happen) gives rise to a symmetric form of the
effective potential.

\subsection{Model
with $V_{1}<0$ and $V_{2}<0$}
 \subsubsection{Resolution of the Old Cosmological Constant Problem in TMT}

The most remarkable feature of the effective potential
(\ref{Veffvac-delta=0}) is that it is proportional to the square
of $V_{1}+ M^{4}e^{-2\alpha\phi/M_{p}}$ which is a straightforward
consequence of our basic assumption that $L_1$ and $L_2$ are
independent of the measure fields (see item 2 in Sec.II,
Eq.(\ref{Veff1}) and the discussion after it). Due to this, as
$V_{1}<0$ and $V_{2}<0$, {\it the effective potential has a
minimum where it equals zero automatically}, without any further
tuning of the parameters $V_{1}$ and $V_{2}$ (see also Fig.1c).
This occurs in the process of evolution of the field $\phi$ at the
value of $\phi =\phi_{0}$ where
\begin{equation}
V_{1}+ M^{4}e^{-2\alpha\phi_{0}/M_{p}}=0 \label{Veff=0}.
\end{equation}
This means that the universe evolves into the state with zero
cosmological constant without any additional tuning of the
parameters  and initial conditions.

\begin{figure}[htb]
\begin{center}
\includegraphics[width=15.0cm,height=8.0cm]{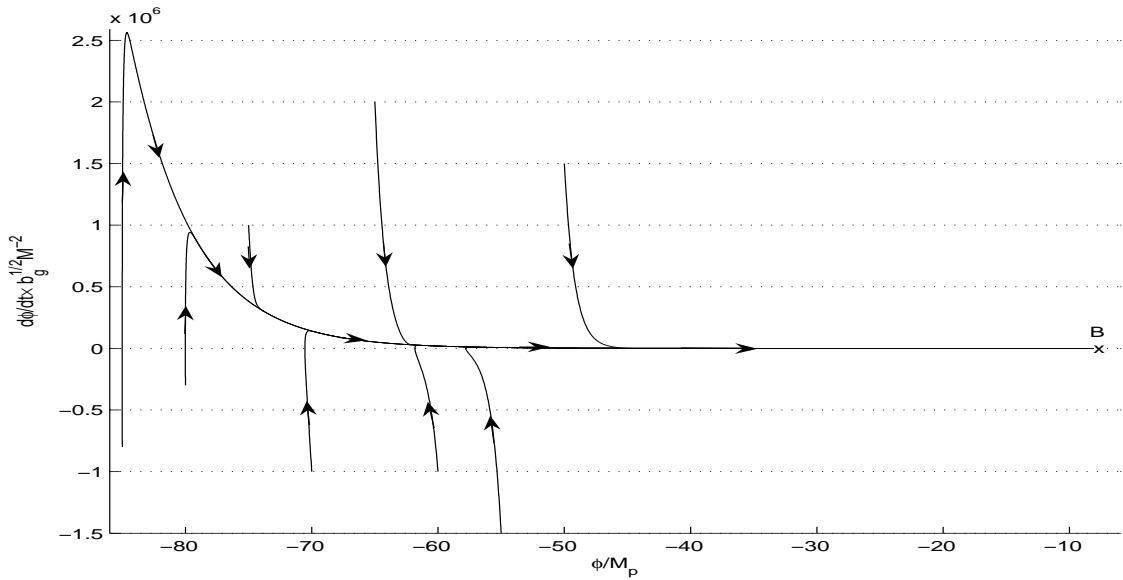}
\end{center}
\caption{Phase portrait (plot of $\frac{d\phi}{dt}$ versus $\phi$)
for the model with $V_{1}<0$ and $V_{2}<0$.   All phase curves
started with $|\phi|\gg M_{p}$ quickly approach the attractor long
before entering the oscillatory regime. The region of the
oscillatory regime is marked by point $B$. The oscillation spiral
is not visible here because of the chosen scales along the
axes.}\label{fig2}
\end{figure}

\begin{figure}[htb]
\begin{center}
\includegraphics[width=16.0cm,height=7.0cm]{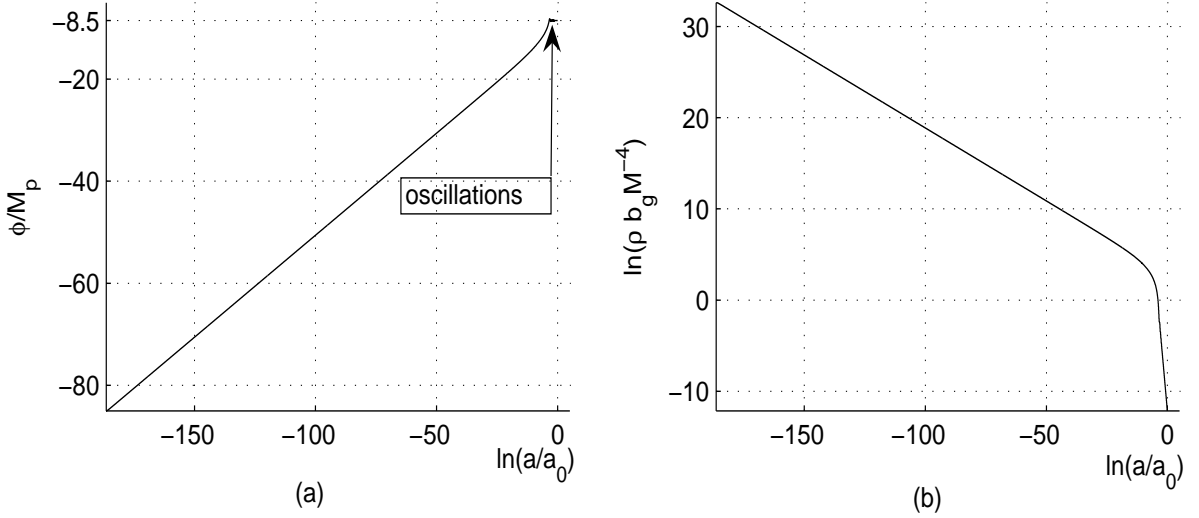}
\end{center}
\caption{(a) Typical dependence of the field $\phi$ (fig. (a)) and
the energy density $\rho$ (fig. (b)) upon $\ln(a/a_{0})$.  Here
and in all the graphs of this paper describing scale factor $a$
dependences, $a(t)$ is normalized such that at the end point of
the described process $a(t_{end})=a_{0}$. In the model with
$V_{1}<0$ the power law inflation ends with damped oscillations of
$\phi$ around $\phi_{0}$ determined by Eq.(\ref{Veff=0}). For the
choice $V_{1}=-30M^{4}$ Eq.(\ref{Veff=0}) gives $\phi_{0}=-8.5M_p$
. (b) The exit from the early inflation is accompanied with
approaching zero of the energy density $\rho$. The graphs
correspond to the evolution which starts from the initial
  values  $\phi_{in} =-85M_{p}$, $\dot{\phi}_{in}=
  -8\cdot 10^5M^2/\sqrt{b_g}$. }\label{fig3}
\end{figure}

\begin{figure}[htb]
\includegraphics[width=8.5cm,height=8.0cm]{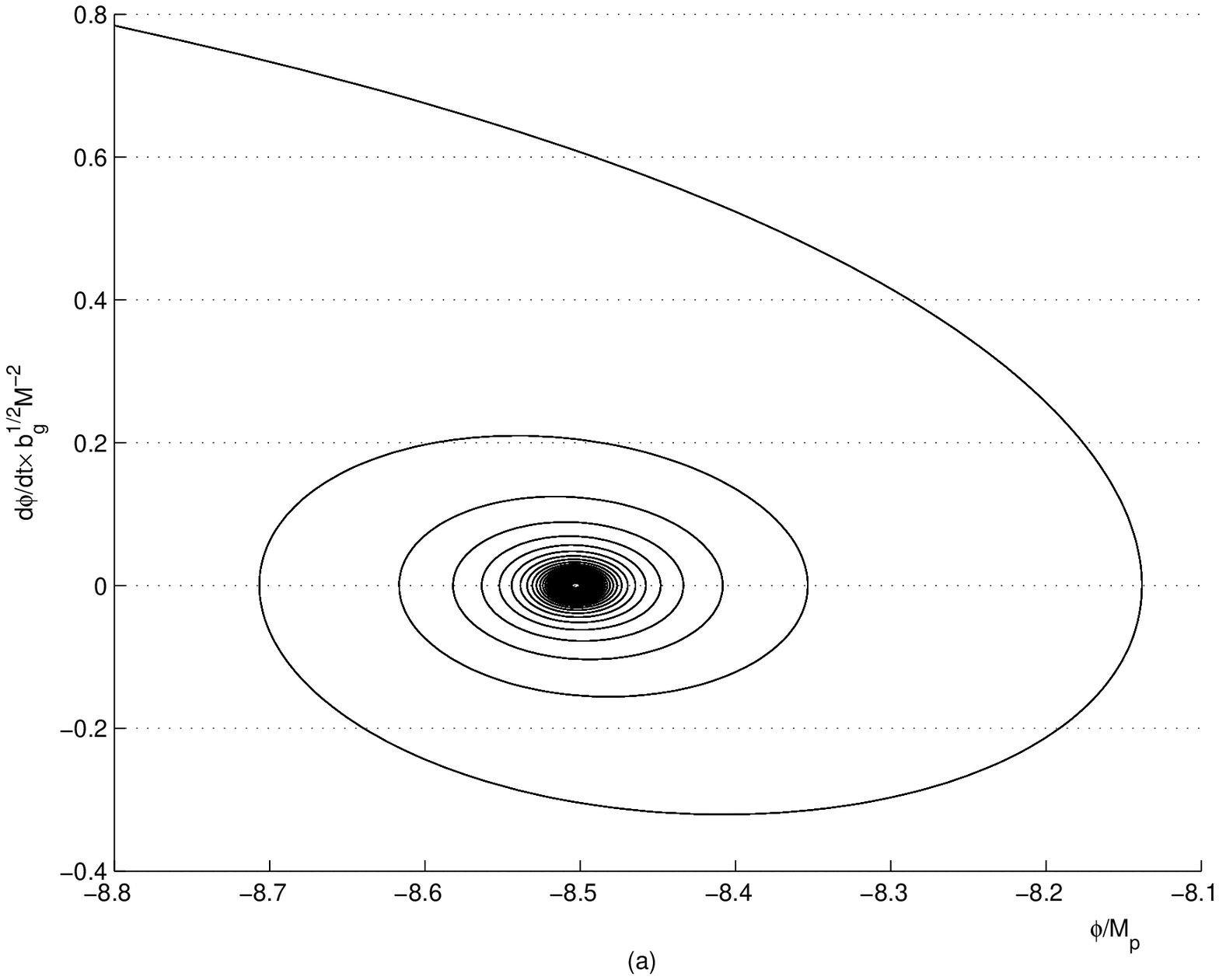}
\includegraphics[width=8.5cm,height=8.0cm]{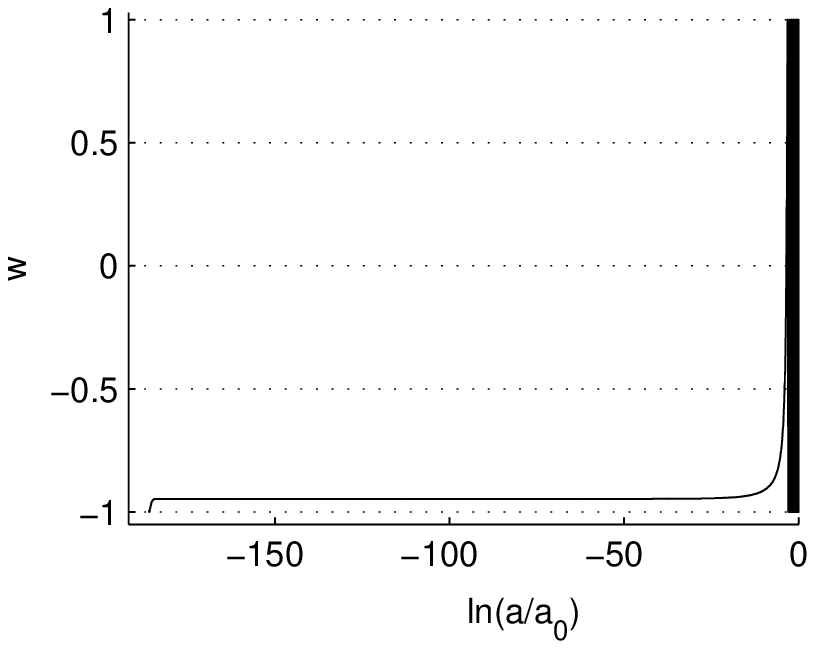}
\caption{Fig.(a) zoom in on the oscillatory regime which marked by
point B in Fig.2. \, (b) Equation-of-state $w=p/\rho$ as function
of the scale factor for the parameters and initial conditions as
in Fig.3. Most of the time the expansion of the universe is a
power law inflation with almost constant $w\approx -0.95$; $w$
oscillates between $-1$ and $1$ at the exit from inflation stage,
i.e. as $\phi\rightarrow\phi_{0}$ and $\rho\rightarrow
0$.}\label{fig4}
\end{figure}

To provide the global scale invariance (\ref{st}), the
prepotentials $V_1$ and $V_2$ enter in the action
(\ref{totaction}) with factor $e^{2\alpha\phi/M_p}$. However, if
quantum effects (considered in the original frame) break the scale
invariance of the action (\ref{totaction}) via modification of
existing prepotentials or by means of generation of other
prepotentials with arbitrary $\phi$ dependence (and in particular
a "normal" cosmological constant term $\int
\tilde{\Lambda}\sqrt{-g}d^4x$), this cannot change the result of
TMT that the effective  potential generated in the Einstein frame
is proportional to a perfect square. Note that the assumption of
scale invariance is not necessary for the effect of appearance of
the perfect square in the effective potential in the Einstein
frame and therefore for the described mechanism of disappearance
of the cosmological constant, see Refs.\cite{GK2}-\cite{G1}.

 If such type of the structure for the scalar field potential in a
conventional (non TMT) model would be chosen "by hand" it would be
a sort of fine tuning. But in our TMT model it is not the starting
point, {\it it is rather a result} obtained in the Einstein frame
of TMT models with spontaneously broken  shift symmetry
(\ref{phiconst}).

At the first glance this effect contradicts the Weinberg's no-go
theorem\cite{Weinberg1} which states that there cannot exist a
field theory model where the cosmological constant is zero without
fine tuning. In Sec.VIIIB we will study in detail the manner our
TMT model avoids this theorem.

\subsubsection{Cosmological Dynamics}

As $M^{4}e^{-2\alpha\phi/M_{p}}\gg Max\left(|V_{1}|,
|V_{2}|/b_{g}\right)$, the effective potential
(\ref{Veffvac-delta=0}) behaves as the exponential potential
$V_{eff}^{(0)}\approx
\frac{1}{4b_{g}}M^{4}e^{-2\alpha\phi/M_{p}}$. So, as $\phi\ll
-M_{p}$  the model describes the well studied power law inflation
of the early universe\cite{power-law}-\cite{Halliwell} if
$0<\alpha < 1/\sqrt{2}$ :
\begin{equation}
a(t)=a_{in}\left(\frac{t}{t_{in}}\right)^{1/2\alpha^{2}}, \qquad
\phi(t)
=\frac{M_p}{\alpha}ln\left(\frac{\alpha^2M^2t}{M_p\sqrt{b_g(3-2\alpha^2)}}\right)
. \label{p-l-solution}
\end{equation}
The only true integration constant in this exact analytic
solution\cite{Liddle} is the initial value of the scale factor
$a_{in}=a(t_{in})$ where $t_{in}>0$. The choice of $t_{in}$
determines both the initial value of $\phi_{in}=\phi(t_{in})$ and
the initial value of $\dot{\phi}_{in}=\dot{\phi}(t_{in})$.
Therefore the initial values
 $\phi_{in}$ and $\dot{\phi}_{in}$ cannot be
chosen independently. This feature of the solutions
(\ref{p-l-solution}) corresponds to the fact that in the phase
plane $(\phi,\dot{\phi})$ there is only one phase curve
representing these solutions and it plays the role of the
attractor\cite{Halliwell} for all other solutions with arbitrary
initial values of $\phi_{in}$ and $\dot{\phi}_{in}$. Excluding
time from $\phi(t)$ and $\dot{\phi}(t)$ we obtain the equation of
the attractor in the phase plane:
\begin{equation}
\frac{\sqrt{b_g}}{M^2}\,\dot{\phi}=\frac{\alpha}{\sqrt{3-2\alpha^2}}e^{-\alpha\phi/M_p}.
\label{p-l-phase-plane}
\end{equation}
In all points of the attractor (\ref{p-l-phase-plane}) the ratio
\begin{equation}
n\equiv\frac{V_{eff}^{(0)}}{\dot{\phi}^2/2}=\frac{3}{2\alpha^2}-1
\label{cons-along-attractor}
\end{equation}
is constant and $n>2$ in the case of the power law inflation, i.e.
if $\alpha <1/\sqrt{2}$. With our choice of $\alpha =0.2$, we have
along the attractor (\ref{p-l-phase-plane}) $n=36.5$ and the
equation-of-state $w\approx -0.95$.

 Note also that for the exact power law inflating solutions
(\ref{p-l-solution}), $\dot{\phi}$ is always positive. Phase
curves corresponding to different independent initial values
$\phi_{in}$ and $\dot{\phi}_{in}$ (including negative
$\dot{\phi}_{in}$) and obtained by means of numerical solutions
are presented in Fig.2. One can see that their shapes are
characterized by much steeper (almost vertical) approaching the
attractor than the exponential shape of the decay of the attractor
itself. Note that in Fig.2 we have presented only phase curves
started from points $(\phi_{in},\dot{\phi}_{in})$ where
$\frac{1}{2}\dot{\phi}_{in}^2\sim V_{eff}^{(0)}(\phi_{in})$. We
have checked that the same very steep approach to the attractor is
peculiar also to the phase curves started from points
$(\phi_{in},\dot{\phi}_{in})$ where
$\frac{1}{2}\dot{\phi}_{in}^2\gg V_{eff}^{(0)}(\phi_{in})$.

Further behavior of the solutions   is qualitatively evident
enough. With the choice $\alpha =0.2$, $V_{1}=-30M^{4}$ and
$V_{2}=-50b_{g}M^{4}$, the results of numerical solutions are
presented in Figs. 2, 3 and 4. Exit from the inflation regime
starts as $\phi$ becomes close to $\phi_0$ determined by
Eq.(\ref{Veff=0}). Then the energy density starts
 to tend to zero very fast, Fig.3b. The numerical solutions show that
for all phase curves corresponding to initial conditions with
$\phi_{in}\ll -M_{p}$ and arbitrary $\dot{\phi}_{in}$,
 the exit from the inflation occurs when these phase curves practically
coincide with the attractor. The process ends with oscillatory
regime, Fig.4a, where $\phi$ performs damped oscillations around
the minimum of the effective potential (see also Fig.1c).

\subsection{Model with $V_1>0$ and $b_{g}V_{1}>
2V_{2}$: Early Power Law Inflation Ending With Small $\Lambda$
Driven Expansion}

\begin{figure}[htb]
\begin{center}
\includegraphics[width=18.0cm,height=6.0cm]{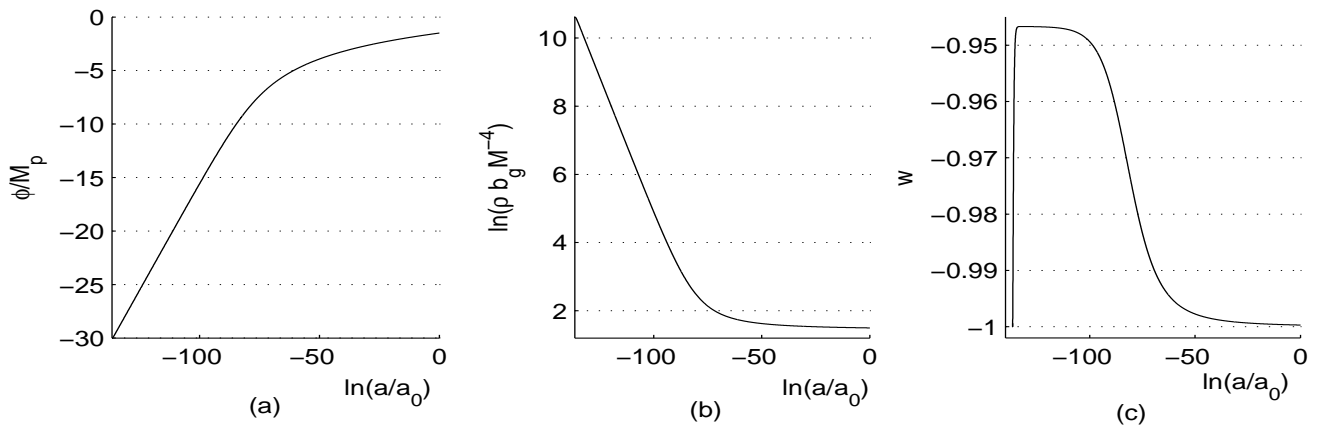}
\end{center}
\caption{ The values of $V_{1}$ and $V_{2}$ are as in Fig.1a. The
graphs correspond to the initial conditions $\phi_{in}=-30M_{p}$,
$\dot{\phi}_{in}=-5M^2b_{g}^{-1/2}$. The early universe evolution
is governed by the almost exponential potential (see Fig.1a)
providing the power low inflation ($w\approx -0.95$ interval in
fig.(c)). After transition to the late time universe the scalar
$\phi$ increases with the rate typical for a quintessence
scenario. Later on  the cosmological constant $\Lambda_{1}$
becomes a dominant component of the dark energy that is displayed
by the infinite region where $w\approx -1$ in
fig.(c).}\label{fig5}
\end{figure}

\begin{figure}[htb]
\begin{center}
\includegraphics[width=14.0cm,height=8.0cm]{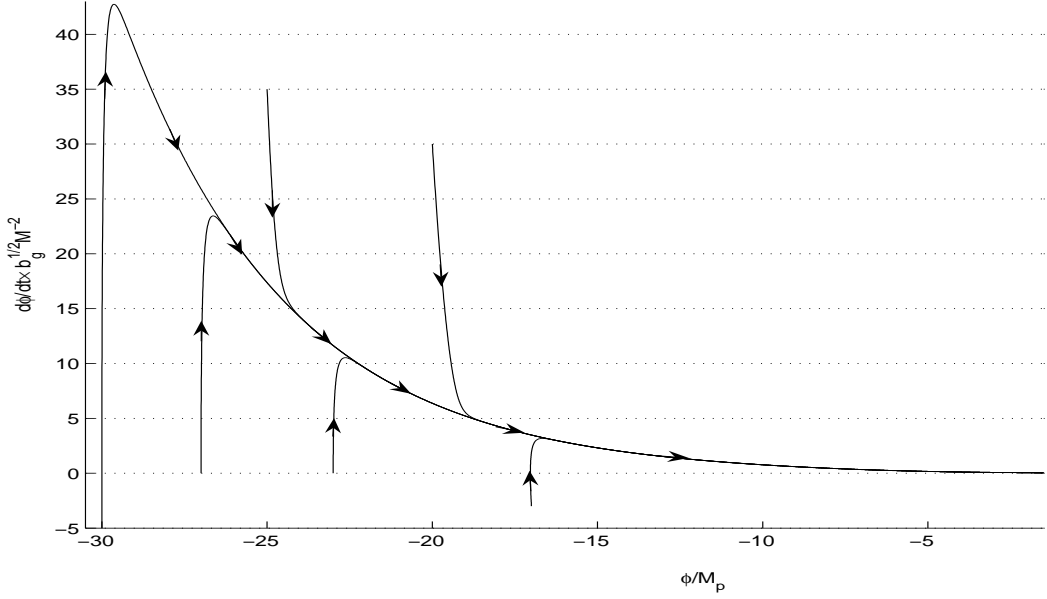}
\end{center}
\caption{Phase portrait (plot of $\frac{d\phi}{dt}$ versus $\phi$)
for the model with $b_{g}V_{1}>2V_{2}$.   All trajectories
 approach the attractor which in its turn
asymptotically (as $\phi\rightarrow\infty$) takes the form of the
straight line $\dot\phi =0$.}\label{fig6}
\end{figure}

 In this model
 the effective potential (\ref{Veffvac-delta=0}) is a
monotonically decreasing function of $\phi$ (see
Fig.{\ref{fig1}}a).  As $\phi\ll -M_{p}$  the model describes the
power law inflation (\ref{p-l-solution}), similar to what we
discussed in the model of previous subsection.

Applying this model to the cosmology of the late time universe and
assuming that the scalar field $\phi\rightarrow\infty$ as
$t\rightarrow\infty$, it is convenient to represent the effective
potential (\ref{Veffvac-delta=0}) in the form
\begin{equation}
V_{eff}^{(0)}(\phi)=\Lambda_1+V_{q-l}(\phi)\qquad
\text{where}\qquad \Lambda_1=\Lambda |_{b_{g}V_1>2V_2},
\label{rho-without-ferm}
\end{equation}
with the definition
\begin{equation}
 \Lambda
=\frac{V_{1}^{2}} {4(b_{g}V_{1}-V_{2})}. \label{lambda}
\end{equation}
Here $\Lambda$ is the positive cosmological constant (see
(\ref{bV1>V2})) and
\begin{equation}
V^{(0)}_{q-l}(\phi)
=\frac{(b_{g}V_{1}-2V_{2})V_{1}M^{4}e^{-2\alpha\phi/M_{p}}+
(b_{g}V_{1}-V_{2})M^{8}e^{-4\alpha\phi/M_{p}}}
{4(b_{g}V_{1}-V_{2})[b_{g}(V_{1}+
M^{4}e^{-2\alpha\phi/M_{p}})-V_{2}]},
\label{V-quint-without-ferm-delta=0}
\end{equation}
that is the evolution of the late time universe  is governed both
by the cosmological constant $\Lambda_1$ and by the
quintessence-like potential $V^{(0)}_{q-l}(\phi)$.

Thus  the effective potential (\ref{Veffvac-delta=0}) provides a
possibility for a cosmological scenario which starts with a power
law inflation and ends with a cosmological constant $\Lambda_1$.
The smallness of $\Lambda_1$ may be achieved without fine tuning
of dimensionfull parameters, that will be discussed in Sec.VIA.
Such scenario may be treated as a generalized quintessential
inflation type of scenario. Recall that the $\phi$-dependence of
the effective potential (\ref{Veffvac-delta=0}) appears here only
as the result  of the  spontaneous breakdown of the global scale
symmetry\footnote{The particular case of this model with $b_{g}=0$
and $V_{2}<0$ was studied in Ref.\cite{G1}. The application of the
TMT model with explicitly broken global scale symmetry to the
quintessential inflation scenario was discussed in Ref\cite{K}.}.

Results of numerical solutions for such type of scenario are
presented in Figs.5 and 6 ($V_{1}=10M^{4}$, $V_{2}=4b_{g}M^{4}$)
The early universe evolution is governed by the almost exponential
potential (see Fig.1a) providing the power low inflation
($w\approx -0.95$ interval in Fig.5c). The choice of $V_{1}$ and
$V_{2}$ such that $b_{g}V_{1}\geq 2V_{2}$ provides a graceful exit
from inflation with transition to the late time universe where the
scalar $\phi$ increases with the rate typical for a quintessence
scenario. Later on  the cosmological constant $\Lambda_{1}$
becomes a dominant component of the dark energy that is displayed
by the infinite region where $w\approx -1$ in Fig.5c. The phase
portrait in Fig.3 shows that all the trajectories started with
$|\phi|\gg M_{p}$ quickly approach the attractor which
asymptotically (as $\phi\rightarrow\infty$) takes the form of the
straight line $\dot\phi =0$. Qualitatively similar results are
obtained also when $V_{1}$ is positive but $V_{2}$ is negative.

\subsection{Model with $V_1>0$ and $V_{2}<b_{g}V_{1}<2V_{2}$}

\begin{figure}[htb]
\begin{center}
\includegraphics[width=16.0cm,height=8.0cm]{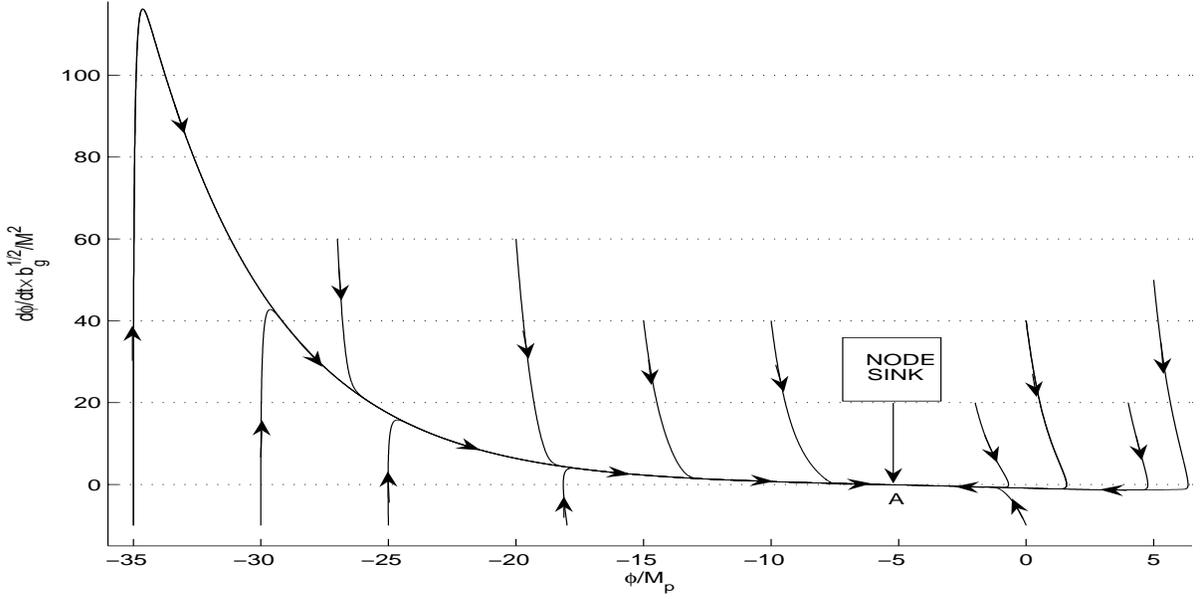}
\end{center}
\caption{Phase portrait (plot of $\frac{d\phi}{dt}$ versus $\phi$)
 for the model with $V_{2}<b_{g}V_{1}<2V_{2}$ and
$V_{1}>0$ (the parameters are chosen here as in Fig.1b ).
Trajectories started anywhere in the phase plane in a finite time
end up at  the same point $A(-5.7,0)$ which is a node sink. There
exist two attractors ending up at $A$, one from the left and other
from the right. All phase curves starting with $|\phi|\gg M_{p}$
quickly approach these attractors.}\label{fig7}
\end{figure}

\begin{figure}[htb]
\begin{center}
\includegraphics[width=18.0cm,height=6.0cm]{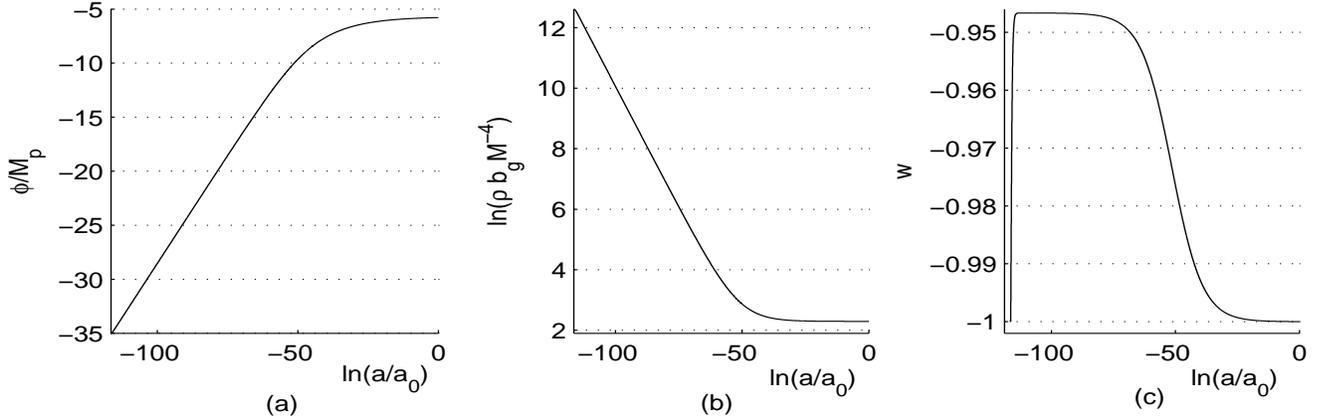}
\end{center}
\caption{Cosmological dynamics in the model with
$V_{2}<b_{g}V_{1}<2V_{2}$ and $V_{1}>0$: typical dependence of
$\phi$ (Fig.(a)), the energy density $\rho$ (Fig.(b)) and the
equation-of-state $w$ (Fig.(c)) upon $\ln(a/a_{0})$ where the
scale factor $a(t)$  normalized as in Fig.3. The graphs correspond
to the initial conditions $\phi_{in}=-35M_{p}$,
$\dot{\phi}_{in}=-10b_{g}^{-1/2}M^{2}$. The early universe
evolution is governed by an almost exponential potential (see
Fig.1b) providing the power low inflation ($w\approx -0.95$
interval in Fig.(c)). After arriving the minimum of the potential
at  $\phi_{min}= -5.7M_p$ (see Fig.1b and
 the point $A(-5.7M_p,0)$ of the phase plane in Fig.4) the scalar $\phi$
remains constant. At this stage the dynamics of the universe is
governed by the  constant energy density $\rho
=V^{(0)}_{eff}(\phi_{min})$  (see the appropriate intervals $\rho
=const$ in Fig.(b) and  $w= -1$ in Fig.(c)).}\label{fig8}
\end{figure}

In this case the effective potential (\ref{Veffvac-delta=0}) has
the minimum (see Fig.1b)
\begin{equation}
V^{(0)}_{eff}(\phi_{min})=\frac{V_{2}}{b_{g}^{2}} \qquad \text{at}
\qquad \phi =\phi_{min}=
-\frac{M_{p}}{2\alpha}\ln\left(\frac{2V_{2}-b_{g}V_{1}}{b_{g}}\right).
\label{minVeff}
\end{equation}

For the choice of the parameters as in Fig.1b, i.e.
$V_{1}=10M^{4}$ and $V_{2}=9.9b_{g}M^{4}$, the minimum is located
at $\phi_{min}=-5.7M_p$. The character of the phase portrait one
can see in Fig.\ref{fig7}.

For the early universe as $\phi\ll -M_{p}$, similar to what we
have seen in the models of the previous two subsections, the model
implies the power law inflation. However, the phase portrait
Fig.\ref{fig7} shows that now all solutions end up without
oscillations at the minimum $\phi_{min}=-5.7M_p$ with
$\frac{d\phi}{dt}=0$. In this final state of the scalar field
$\phi$, the evolution of the universe is governed by the
cosmological constant $V^{(0)}_{eff}(\phi_{min})$ determined by
Eq.(\ref{minVeff}). For some details of the cosmological dynamics
see Fig.\ref{fig8}. The desirable smallness of
$V^{(0)}_{eff}(\phi_{min})$ can be provided again without fine
tuning of the dimensionfull parameters that will be discussed in
Sec.VIA. The absence of appreciable oscillations in the minimum is
explained by the following two reasons: a) the non-zero friction
at the minimum determined by the cosmological constant
$V^{(0)}_{eff}(\phi_{min})$; b) the shape of the potential near to
minimum is too flat.

The described properties of the model are evident enough.
Nevertheless we have presented them here because this model is a
particular (fine tuned) case of an appropriate model with
$\delta\neq 0$ studied in Sec.VII where we will demonstrate a
possibility of states with $w<-1$ without any exotic
contributions, like a phantom term, in the original action.

\section{TMT Cosmology With No Fine Tuning I:
\newline
Absence of the Initial Singularity of the Curvature and
Inflationary Cosmology
 with Graceful Exit to $\Lambda =0$ Vacuum}

 \subsection{General Analysis and Numerical Solutions}

In the following three sections we return to the general case of
our model (see Sec.III) with no fine tuning of the parameters
$b_g$ and $b_{\phi}$, i.e. the parameter $\delta$, defined by
Eq.(\ref{delta}), is non zero. Then the dynamics of the FRW
cosmology is described by Eqs.(\ref{rho1})-(\ref{phi1}) and
(\ref{cosm-phi}). Let us start from the analysis of
Eq.(\ref{phi1}). The interesting feature of this equation is that
 each of the factors $Q_{i}(\phi,X)$ \, ($i=1,2,3$) can get
 to zero and this effect depends on the range of the parameter
 space chosen. This is the
origin of drastic {\it changes of the topology of the phase plane}
comparing with the fine tuned models of Sec.IV.

\begin{figure}[htb]
\begin{center}
\includegraphics[width=17.0cm,height=12.0cm]{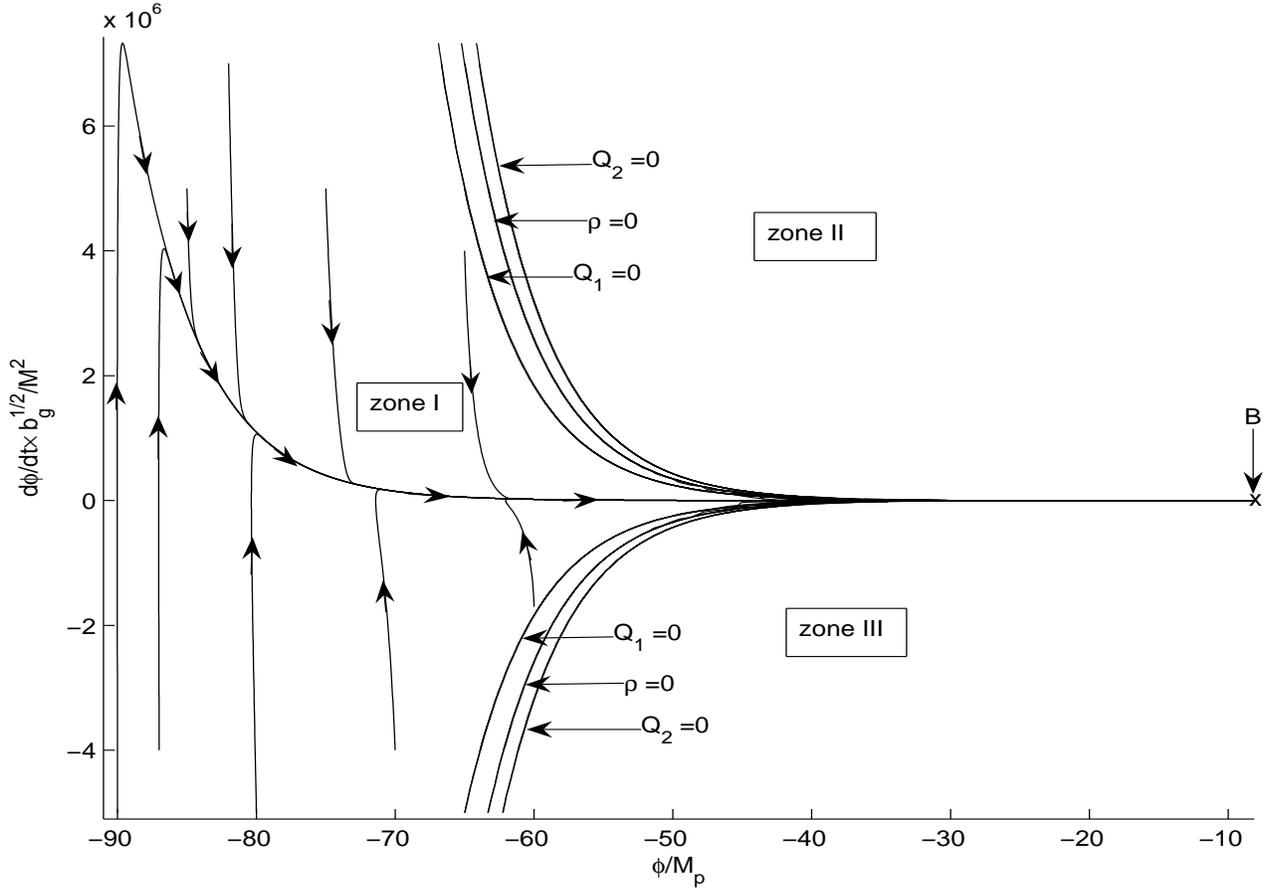}
\end{center}
\caption{The phase portrait for the model with $\delta =0.1$,
$\alpha =0.2$, $V_{1}=-30M^{4}$ and $V_{2}=-50b_{g}M^{4}$.  All
phase curves demonstrate the attractor behavior similar to that in
the fine tuned case  $\delta =0$, Fig.2. One can see that the
attractor does not intersect the line $Q_1=0$, see also Fig.10.
The location of the oscillatory regime marked by point $B$ is
exactly the same as in the model with $\delta =0$. The essential
difference consists in a novel topological structure: in the
neighborhood of the line $Q_1=0$ all the phase curves exhibit a
repulsive behavior from this line and therefore points of the line
$Q_1=0$ are dynamically unachievable. Hence the phase curves
cannot be continued infinitely to the past.}\label{fig9}
\end{figure}

\begin{figure}[htb]
\begin{center}
\includegraphics[width=17.0cm,height=12.0cm]{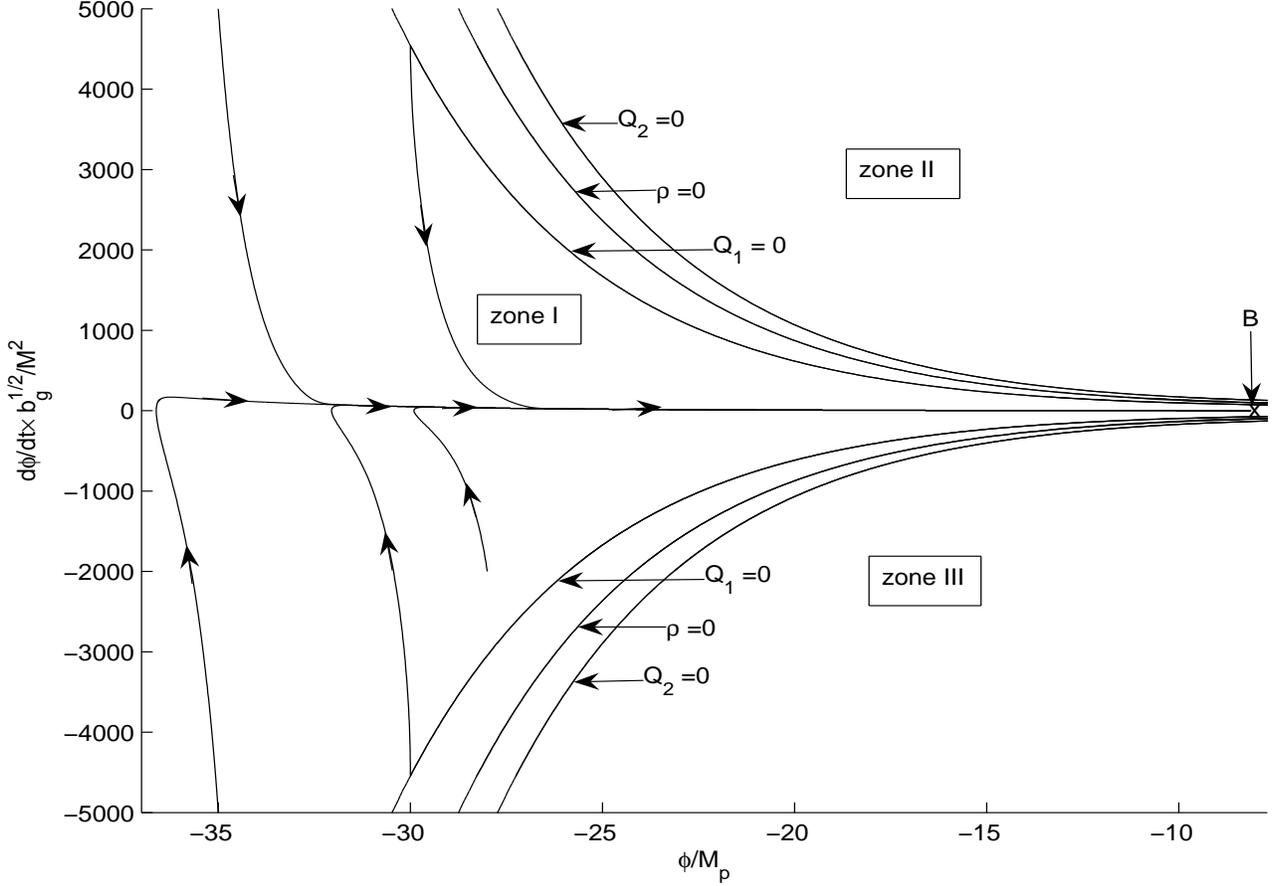}
\end{center}
\caption{The same phase portrait as in Fig.9 but with more clear
exhibition of the structure of the phase plane near to the  point
$B$.}\label{fig10}
\end{figure}

\begin{figure}[htb]
\begin{center}
\includegraphics[width=17.0cm,height=7.0cm]{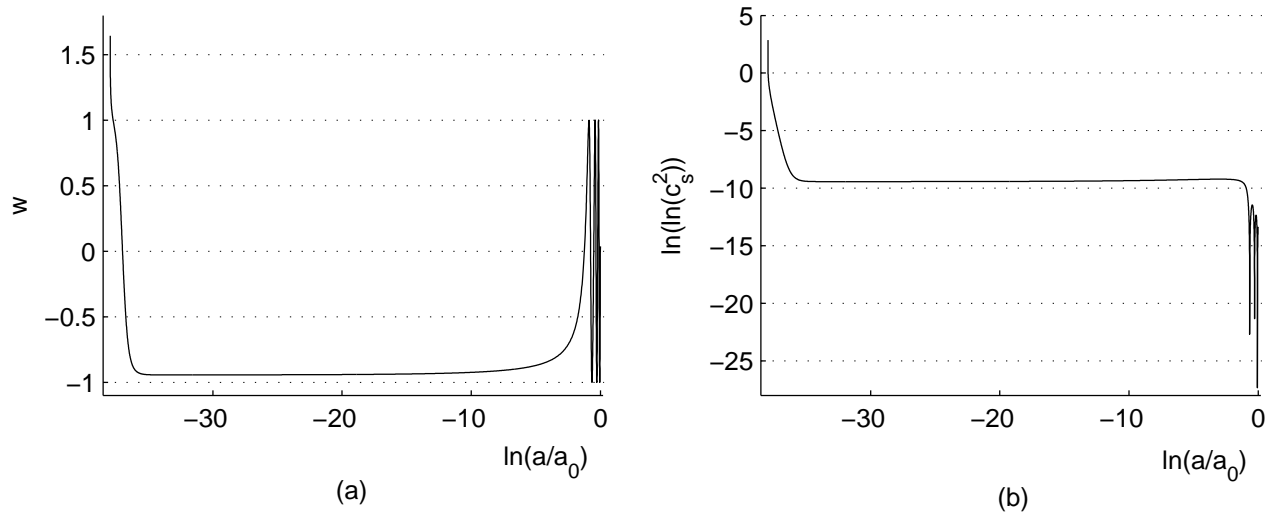}
\end{center}
\caption{Typical scale factor dependence of the equation-of-state
$w$ and the squared sound speed $c^2_s$ for the phase curves
starting from  points very close to the line $Q_1=0$ (in this
figure - for $\phi_{in}=-30M_p$ and $\dot{\phi}_{in}=
4540.7569b_g^{-1/2}M^2$). The squared sound speed in the starting
point is $c^2_s\approx exp(exp (2.86))\approx 3.8\cdot 10^7$. It
is not a problem to obtain more than 75 e-folds during the power
law inflation (the region of $w\approx -0.95$ in Fig.(a)) just by
choosing a larger absolute value $|\phi_{in}|$. We have chosen
$\phi_{in}=-30M_p$ because this allows to show more details in
these and subsequent graphs. }\label{fig11}
\end{figure}

\begin{figure}[htb]
\begin{center}
\includegraphics[width=17.0cm,height=7.0cm]{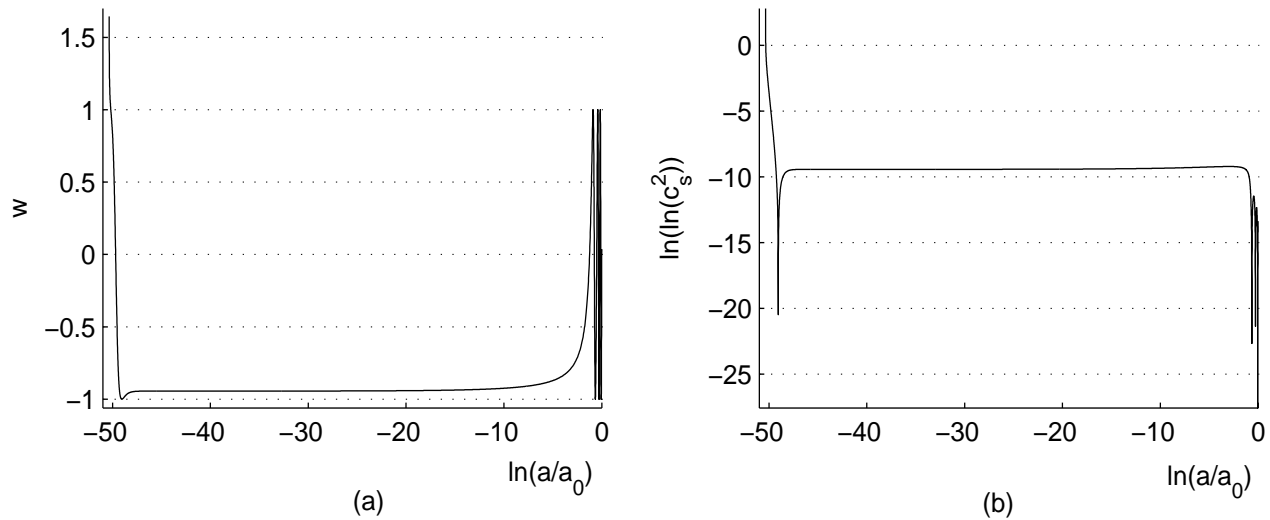}
\end{center}
\caption{The same as in Fig.11 but for $\phi_{in}=-30M_p$ and
$\dot{\phi}_{in}=-4540.7569b_g^{-1/2}M^2$. The squared sound speed
in the starting point is also $c^2_s\approx exp(exp (2.86))\approx
3.8\cdot 10^7$. }\label{fig12}
\end{figure}

\begin{figure}[htb]
\begin{center}
\includegraphics[width=17.0cm,height=7.0cm]{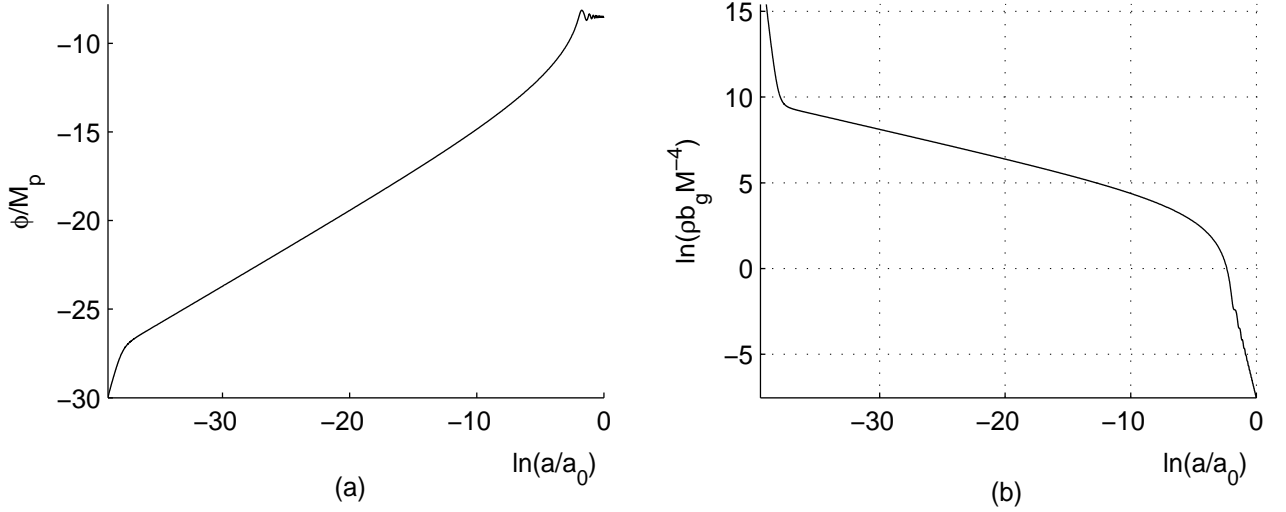}
\end{center}
\caption{Typical scale factor dependence of $\phi$ and $\rho$ for
the phase curves starting from  points very close to the line
$Q_1=0$ (in this figure - for $\phi_{in}=-30M_p$ and
$\dot{\phi}_{in}= 4540.7569b_g^{-1/2}M^2$).}\label{fig13}
\end{figure}

For $Q_{1}\neq 0$, Eqs. (\ref{phi1}), (\ref{cosm-phi}) result in
the well known equation\cite{Vikman}
\begin{equation}
\ddot{\phi}+
\frac{\sqrt{3\rho}}{M_p}c_s^2\dot{\phi}+\frac{\rho_{,\phi}}{\rho_{,X}}
=0, \label{phantom-phi}
\end{equation}
where $c_s$ is the effective sound speed of perturbations\cite{GM}
\begin{equation}
c_s^2=\frac{p_{,X}}{\rho_{,X}}=\frac{Q_2}{Q_1}, \label{c-s-2}
\end{equation}
\begin{equation}
\frac{\rho_{,\phi}}{\rho_{,X}}=- \frac{\alpha}{M_{p}}\cdot
\frac{Q_3}{Q_1}\cdot M^{4}e^{-2\alpha\phi/M_{p}}; \label{Q-rho-p}
\end{equation}
$\rho$ and $p$ are defined by Eqs.(\ref{rho1}), (\ref{p1}) and
$Q_i$ $(i=1,2,3)$ - by Eqs.(\ref{Q1})-(\ref{Q3}).

It follows from the definitions of $Q_1$ and $Q_2$ that $c_s^2>1$
when $Q_1>0$ and $X>0$ (that implies $Q_2>0$). Therefore in the
cosmological FRW background,  the sound speed of perturbations
{\it can be bigger than speed of light}\footnote{We grateful to A.
Vikman for attracting our attention to this point.}, \footnote{The
possibility of a superluminal sound speed of perturbations
(including a possibility of infinite $c_s$) and its possible role
in the tensor-to-scalar perturbation ratio in inflationary models
have been studied in Refs.\cite{GM}, \cite{Mukh-Vik-JCAP} and in
the black hole physics - in
Ref.\cite{Mukh-black-hole-hep-th/0604075 }. Conceptual problems
related to causality have been discussed in
Refs.\cite{Caldwell-Steinhardt-Mukhanov}, \cite{Bruneton}}.
However $c_s^2<1$ when $Q_1<0$ and $Q_2<0$.

In the model with $V_1<0$ and $V_2<0$ (the fine tuned version of
which has been studied in Sec.IVB), the structure of the phase
plane is presented in Figs.9 and 10 for the following set of the
parameters: $\alpha =0.2$, $\delta =0.1$, $V_1=-30$, $V_2=-50$.
With such a choice of the parameters the following condition is
satisfied
\begin{equation}
(b_g+b_{\phi})V_1-2V_2>0 \label{cond-for-phase-structure}
\end{equation}
that determines the structure of the phase plane.
 By the two branches
of the line $Q_1=0$, the phase plane is divided into three large
dynamically disconnected zones. In the most part of two of them
(II and III), the energy density is negative ($\rho <0$). In the
couple of regions between the lines $Q_1=0$ and $\rho =0$ that we
refer for short as ($Q_1\rightarrow\rho$)-regions, $\rho
>0$ but $Q_1<0$. Typical scale factor dependence of the
equation-of-state $w$, the sound speed of perturbations $c_s^2$,
the inflaton $\phi$ and the energy density $\rho$ are presented in
Figs.11, 12  and 13. It is not a problem to obtain more than 75
e-folds during the power law inflation (the region of $w\approx
-0.95$ in Fig.11a) just by choosing a larger absolute value
$|\phi_{in}|$. We have chosen $\phi_{in}=-30M_p$ because this
allows to show more details in these graphs.

The zone I of the phase plane (where $Q_1>0$, $\rho
>0$ and $Q_2>0$) is of a great cosmological interest:

\begin{itemize}

\item

 Similar to what we have seen in the fine tuned model of
 Sec.IVB, all the phase
 curves start with very steep approach to an attractor.  The nonlinearity in $X$
 does not allow to obtain an exact analytic solution in the model under
consideration and therefore we have no here the analytic equation
of the attractor. But taking into account
Eq.(\ref{cons-along-attractor}), it becomes evident that, with our
choice of the parameters $\alpha$ and $\delta$, the emergence of
the additional terms $\propto\delta X$ and $\propto\delta^2X^2$ in
the model under consideration results in small enough corrections
to the equation of the attractor in comparison with
Eq.(\ref{p-l-phase-plane}) of the fined tuned model of Sec.IVB.
For our qualitative analysis below one can use
Eq.(\ref{p-l-phase-plane}) as a good approximation to the true
attractor equation.

\item

Comparing the  equation of the line $Q_1=0$, which for
$\phi\ll-M_p$ can be written in the form
\begin{equation}
\frac{\sqrt{b_g}}{M^2}\,\dot{\phi}=\pm\frac{1}{\delta}
\sqrt{\frac{2}{3}(2-\delta)}\cdot e^{-\alpha\phi/M_p}+{\cal
O}\left(\frac{V_1}{M^4}e^{\alpha\phi/M_p}\right), \label{Q1=0}
\end{equation}
with the equation of the attractor which approximately coincides
with Eq.(\ref{p-l-phase-plane}), we see that the upper brunch of
 the line $Q_1=0$ has actually the same form of a decaying
 exponent as the attractor, but the factor in front of the exponent
  in Eq.(\ref{Q1=0})
 is about $10^2$ times bigger than in Eq.(\ref{p-l-phase-plane}).
 The above analytic estimations are confirmed by the numerical solutions
 as one can see in Fig.9.
 This means that the attractor does not intersect the line
 $Q_1=0$. Therefore all the phase curves starting in zone I arrive
 at
 the attractor (of course asymptotically).

\item

In the neighborhood of the line $Q_1=0$ all the phase curves
exhibit  a {\it repulsive} behavior from this line. In other
words, the shape of two branches of $Q_1=0$ do not allow a
classical dynamical continuation of the phase curves backward in
time without crossing the classical barrier formed by the line
$Q_1=0$. This is true for all finite values of the initial
conditions $\phi_{in}$, $\dot{\phi}_{in}$ in zone I.

\item
Similar to what we have seen already in the model with $\delta =0$
of Sec.IVB, the power law inflation ends with the graceful exit to
a zero CC vacuum state without fine tuning.

\item
As one can see from  Figs.11 and 12, the initial stage of
evolution is very much different from the subsequent one, that is
a power law inflation. This fact may have
 a relation to the results of the study of
completeness of inflationary cosmological models in past
directions\cite{BGV}.

\item
If the phase curves start from points
$(\phi_{in},\dot{\phi}_{in})$ in zone I very close to the line
$Q_1=0$ then the sound speed of perturbations has huge values at
the beginning of the evolution, see Figs.11b and 12b. However in
the power law inflation stage, $c_s$ is too close to the speed of
light and appears to be unable to increase the tensor-to-scalar
perturbation ratio\cite{GM}, \cite{Mukh-Vik-JCAP}.

\end{itemize}

In two regions between the lines $Q_1=0$ and $Q_2=0$ that we refer
for short as ($Q_1\rightarrow Q_2$)-regions,, the squared sound
speed of perturbations is negative, $c_s^2<0$. This means that on
the right hand side of the classical barrier $Q_1=0$, the model is
absolutely unstable. Moreover, this pure imaginary sound speed
becomes infinite in the limit $Q_1\rightarrow 0-$. Thus the
branches of the line $Q_1=0$ divide zone I (of the classical
dynamics) from the ($Q_1\rightarrow Q_2$)-regions where the
physical significance of the model is unclear. Note that the line
$\rho =0$ divides the ($Q_1\rightarrow Q_2$)-regions into two
subregions with opposite signs of the classical energy density.

Thus the structure of the phase plane yields a conclusion that
 the starting point of the classical history in the phase
plane can be only in zone I and {\bf the line $Q_1 =0$ is the
limiting set of points where the classical history might begin}.
For any finite initial values of $\phi_{in}$ and $\dot{\phi}_{in}$
at the initial cosmic time $t_{in}$, the  duration $t_{in}-t_{s}$
of the continuation of the evolution into the past up to the
moment $t_{s}$ when the phase trajectory arrives the line $Q_1
=0$, is finite.

\subsection{Analysis of the Initial Singularity}

Let us analyze what happens  as $t\to t_{s}$ (and
$Q_1(\phi,\dot{\phi})\to 0$) if this continuation to the past
starts from a point in zone I of the phase plane with finite
initial values $\phi_{in}$ and $\dot{\phi}_{in}$ . First note that
the energy density $\rho_{s}=\rho(t_{s})$ and the pressure
$p_{s}=p(t_{s})$ are finite in all the points of the line $Q_1 =0$
with finite coordinates $\phi_s=\phi(t_s)$,
$\dot{\phi_s}=\dot{\phi}(t_s)$, that it is easy to see from
Eqs.(\ref{rho1}), (\ref{p1}) and (\ref{Q1}). The strong energy
condition is satisfied in regions of zone I close to the line $Q_1
=0$ including the line itself. In fact, for any unit time-like
vector $t^{\mu}$ we have on the line $Q_1 =0$
\begin{equation}
\left(T_{\mu\nu}^{eff}-\frac{1}{2}\tilde{g}_{\mu\nu}T^{eff}\right)t^{\mu}t^{\nu}=\frac{1}{2}(\rho_s
+3p_s)\approx
\frac{2b_{\phi}}{(b_g-b_{\phi})^2}M^{4}e^{-2\alpha\phi_s/M_{p}}>0,
\label{energy-codition}
\end{equation}
where $T_{\mu\nu}^{eff}$ is defined in Eq.(\ref{Tmnfluid}),
$T^{eff}\equiv\tilde{g}^{\mu\nu}T_{\mu\nu}^{eff}$ and we have
taken into account our choice of the parameters ($b_g>0$,
$b_{\phi}>0$ and Eq.(\ref{bV1>V2})) and assumed that
$M^{4}e^{-2\alpha\phi_s/M_{p}}\gg |V_1|$. This result is in the
total agreement with the numerical solutions of the previous
subsection. In particular, the described analytic approximation on
the line $Q_1 =0$ yields $w_s=p_s/\rho_s\approx 5/3$ which is in a
very good agreement with the numerical results obtained in regions
of zone I close enough to the line $Q_1 =0$, see Figs.11a and 12a.

It follows from the Einstein equations (\ref{cosm-phi}) and
\begin{equation}
\frac{\ddot{a}}{a}=-\frac{1}{6M_p^2}(\rho
+3p)\label{ddotEinsteinEq}
\end{equation}
that the first and second time derivatives of the scale factor,
$\dot{a}_{s}=\dot{a}(t_{s})$ and $\ddot{a}_{s}=\ddot{a}(t_{s})$,
and therefore the curvature, are finite on the line $Q_1 =0$. The
time derivative of the energy density also approaches a finite
value
\begin{equation}
\dot{\rho}_s=3\frac{\dot{a}_{s}}{a_s}(\rho_s +p_s)<\infty
\label{dot-rho-finite}
\end{equation}
 It is interesting to
see how this result follows from the scalar field dynamics.
  Using Eqs.(\ref{rho1}),
(\ref{Q1}) and (\ref{Q3-rho-phi}) we obtain
\begin{equation}
\dot{\rho}=\rho,_{\phi}\cdot\dot{\phi}+\rho,_X\cdot\dot{X}=-
\frac{\alpha}{2M_{p}}\cdot
\frac{M^{4}e^{-2\alpha\phi/M_{p}}}{b_{g}(M^{4}e^{-2\alpha\phi/M_{p}}+V_{1})-V_{2}}\cdot
Q_{3}+\frac{Q_1}{2[b_{g}(M^{4}e^{-2\alpha\phi/M_{p}}+V_{1})-V_{2}]}\cdot\dot{\phi}\ddot{\phi}
\label{dot-rho}
\end{equation}
The first term in the r.h.s. of Eq.(\ref{dot-rho}) is evidently
finite on the line $Q_1 =0$. To analyze the behavior of the second
term in the r.h.s. of Eq.(\ref{dot-rho}) one should note that
 the last two terms of Eq.(\ref{phi1}) remain finite as $t\to
t_{s}$ and $Q_1(\phi(t),\dot{\phi}(t))\to 0$. We infer from this
that
\begin{equation}
|\ddot{\phi}|\to\infty \label{ddot-phi-infty}
\end{equation}
but in such a way that
\begin{equation}
Q_1\ddot{\phi}\to \text{finite  value depending on}\, \phi_s \,
\text{and}\,  \dot{\phi}_s .\label{Q1ddot-phi-finite}
\end{equation}
 Therefore the second term in the
r.h.s. of Eq.(\ref{dot-rho}) has a finite limit too.

Similar manipulations for $\dot{p}$ give
\begin{equation}
\dot{p}=p,_{\phi}\cdot\dot{\phi}+p,_{X}\cdot\dot{X}\label{dot-p}
\end{equation}
where again the first term is evidently finite on the line $Q_1
=0$ but for the second term we have using Eq.(\ref{Q2} )
\begin{equation}
p,_{X}\cdot\dot{X}=\frac{Q_2}{2[b_{g}(M^{4}e^{-2\alpha\phi/M_{p}}+V_{1})-V_{2}]}
\cdot\dot{\phi}\ddot{\phi} \label{p-X-dotX}
\end{equation}
Recall that $Q_2>0$ in zone I including the line $Q_1 =0$.
Moreover, the  analysis of the phase curves in Figs.9 and 10 shows
that if $\dot{\phi}_{in}>0$ than $\ddot{\phi}_{in}<0$ and vice
versa. Thus we conclude that
\begin{equation}
\dot{p}\to -\infty \qquad as \qquad t\to t_s\label{dot-p-infty}
\end{equation}
It is now clear from Eq.(\ref{ddotEinsteinEq}) that the third time
derivative of the scale factor is singular at $t=t_s$:
\begin{equation}
\frac{\dddot{a}}{a}\approx -\frac{\dot{p}}{2M_p^2}\to\infty \qquad
as \qquad t\to t_s \label{dddot-a-infty}
\end{equation}
Therefore although the scalar curvature
\begin{equation}
R=-6(\frac{\ddot{a}}{a}+\frac{\dot{a}^2}{a^2})\label{ScalarCurvature}
\end{equation}
is finite as $t\to t_s$ but its time derivative is singular:
\begin{equation}
\dot{R}\approx -6\frac{\dddot{a}}{a}\to -\infty \qquad \text{as}
\qquad t\to t_s\label{ScalarCurvature}
\end{equation}

The regular behavior of $\dot{a}$ and $\ddot{a}$ together with
singularity of $\dddot{a}$ implies that
\begin{equation}
a(t)\approx a_s +A(t-t_s)^n \qquad \text{as} \qquad t\to t_s,
\qquad where \qquad 2<n<3 \label{a(t)-n}
\end{equation}
and $A>0$ is a constant.
 This type of singularity we discover here in the framework of
the dynamical model is present in the classification of "sudden"
singularities given by Barrow\cite{Barrow1} on purely kinematic
grounds\footnote{We are grateful to the referee of the present
paper for attracting our attention to the papers of
Barrow\cite{Barrow1}. Note that Barrow\cite{Barrow1} was
interested in a classification of future singularities while we
study initial singularities. However this difference is not
essential.}.

We are going now to solve the equations of motion in order to find
out the value of $n$ in Eq.(\ref{a(t)-n}). For this, as $t$ is
very close to $t_s$ one can represent $\phi(t)$ and
$\dot{\phi}(t)$ in the form $\phi(t)=\phi_s
+\frac{M_p}{2\alpha}\chi$,
$\dot{\phi}(t)=\dot{\phi}_s+\frac{M_p}{2\alpha}\dot{\chi}$ where
$|\chi|\ll \frac{2\alpha}{M_p}|\phi_s|$ and $|\dot{\chi}|\ll
\frac{2\alpha}{M_p}|\dot{\phi}_s|$. Then keeping in the
$\phi$-equation (\ref{phi1}) only leading terms as $t\to t_s$,  we
obtain
\begin{equation}
\dot{\chi}\cdot\ddot{\chi}\approx B,\label{chi-B}
\end{equation}
where the constant $B$ is defined by
\begin{equation}
B=\frac{4\alpha^2}{\delta^2b_g^2M_p^2}\left[H_sQ_2^{(s)}-\frac{\alpha
Q_3^{(s)}}{3M_p\dot{\phi}_s}M^4e^{-2\alpha\phi_s/M_p}\right],\label{B}
\end{equation}
$H_s$ is the Hubble parameter at $t=t_s$ and
$Q_i^{(s)}=Q(\phi_s,\dot{\phi}_s)>0$, \, $i=2,3$. Eq.(\ref{chi-B})
results in
\begin{equation}
\dot{\phi}=\dot{\phi}_s\pm
\frac{M_p}{2\alpha}\sqrt{2B(t-t_s)}\label{chi-solution}
\end{equation}
provided $B>0$. If we study the evolution starting from
$\dot{\phi}_{in}<0$ then the nearest $\dot{\phi}_s<0$ and $B>0$
without any additional restrictions on the parameters. However if
the evolution starts from $\dot{\phi}_{in}>0$ then the nearest
$\dot{\phi}_s>0$ and  one should test whether the condition $B>0$
implies additional restrictions on the parameters. Using
Eqs.(\ref{rho1}), (\ref{cosm-phi}), (\ref{Q2}) and (\ref{Q3}) one
can show that for $\dot{\phi}_s>0$
\begin{equation}
B=\frac{8\alpha^2}{9\sqrt{6}\delta^3b_g^4\sqrt{b_g+b_{\phi}}}\cdot
[\sqrt{6}(b_g+b_{\phi})(b_g^2+b_{\phi}^2)- 2\alpha
b_g(b_g^2+b_{\phi}^2-b_gb_{\phi})]
\frac{M^6}{M_p^3}e^{-3\alpha\phi_s/M_p} \label{B1}
\end{equation}
Therefore $B>0$ if
\begin{equation}
\alpha<\sqrt{\frac{3}{2}}\cdot\frac{(b_g+b_{\phi})(b_g^2+b_{\phi}^2)}
{b_g(b_g^2+b_{\phi}^2-b_gb_{\phi})}\label{alpha-B}
\end{equation}
The inequality (\ref{alpha-B})
 provides that the condition $\alpha<1/\sqrt{2}<1$, we
have assumed earlier for a power law inflation, confidently holds.
Recall that our choice from the beginning (see Sec. III.A) was
$b_g>0$ and $b_{\phi}>0$.

Solution (\ref{chi-solution}) enable now to find out $n$ in
Eq.(\ref{a(t)-n}). Using Eqs.(\ref{dddot-a-infty}), (\ref{dot-p})
and (\ref{p-X-dotX}) we obtain the following singular behavior of
$\dddot{a}$ as $t\to t_s$:
\begin{equation}
\frac{\dddot{a}}{a}=\frac{Q_2^{(s)}\sqrt{B}|\dot{\phi}_s|}
{8\alpha\sqrt{2}M_p[b_g(M^4e^{-2\alpha\phi/M_p}+V_1)-V_2]}
\cdot\frac{1}{\sqrt{t-t_s}}\label{dddotasingularity}
\end{equation}
Therefore $n$ in Eq.(\ref{a(t)-n}) is $n=5/2$.

Finally we want to discuss possible scales of the energy when the
initial conditions are close to the line $Q_1 =0$. If
$\phi_{in}\approx\phi_s\ll -M_p$ then approximately
\begin{equation}
\rho_{in}\approx\rho_s\approx\frac{(b_g+b_{\phi})^2}{12b_g(b_g-b_{\phi})^2}M^4e^{-2\alpha\phi_s/M_p}
\label{rho-very-big}
\end{equation}
Therefore depending on the parameters and initial conditions, the
described mild singular initial behavior is possible for the
energy densities close to the Planck scale as well as for the
energy densities much lower the quantum Planck era. It is
interesting that the singularity of the time derivative of the
curvature on the line $Q_1 =0$ is accompanied with singularity of
the sound speed of perturbations $c_s^2$ which results directly
from Eqs.(\ref{c-s-2}), (\ref{Q1ddot-phi-finite}) and
(\ref{chi-solution}):
\begin{equation}
c_s^2=\frac{Q_2}{Q_1}\sim\ddot{\phi}\sim\frac{1}{\sqrt{t-t_s}}.
\label{singularity-c-s}
\end{equation}

\subsection{Describing the Initial Stage of the Classical Evolution in
 "Natural" Time Coordinate}

Recall that the reason of the repulsive behavior of the phase
curves in two sides of the line $Q_1=0$ is that the line $Q_1=0$
is a continuous set of singular points of the scalar field $\phi$
equation (\ref{phi1}) while all other equations are regular as
$Q_1=0$.  It is interesting that by a change of the time
coordinate one can achieve a picture where all equations of motion
remain regular in the limit when $Q_1(\phi,\dot{\phi})\rightarrow
0$ . To define such a {\it natural time coordinate} let us
consider a solution with a certain initial conditions $a_{in}$,
$\phi_{in}$, $\dot{\phi}_{in}$. Let the appropriate phase curve
$C$ have equations $\phi =\phi(t), \,\dot{\phi}=\dot{\phi}(t)$. We
want that when using the new time coordinate $\tau$, the first
term in the scalar field equation (\ref{phi1}) takes the form
$d^2\phi/d\tau^2$. This may be achieved (up to a common constant
factor $2|V_2|$ in front of all terms of the $\phi$-equation) by
defining the new time coordinate $\tau$ as follows:
\begin{equation}
d\tau
=\left(\frac{2|V_2|}{Q_1(\phi(t),\dot{\phi}(t))}\right)^{1/2}dt
\label{time-tau}
\end{equation}

\begin{figure}[htb]
\begin{center}
\includegraphics[width=18.0cm,height=6.0cm]{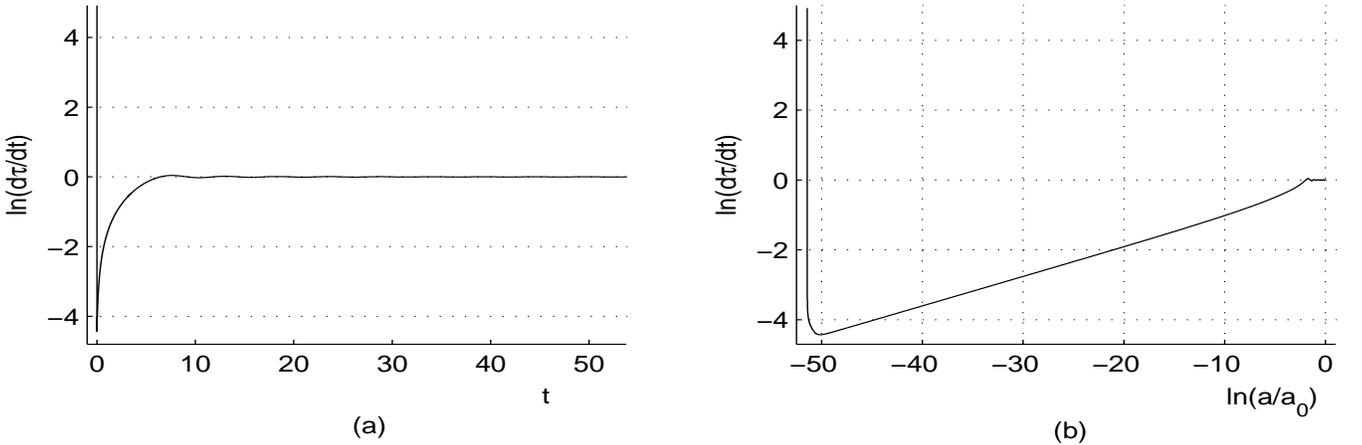}
\end{center}
\caption{Typical cosmic time dependence, Fig.(a), and scale factor
dependence, Fig.(b), of $\frac{d\tau}{dt}$ defined by
Eq.(\ref{time-tau}) as the phase curves start from points very
close to the line $Q_1=0$. In this figure, the initial conditions
are as in Fig.12.}\label{fig14}
\end{figure}

 The $t$ and scale factor dependence of $d\tau/dt$ are presented in Fig.14.
  Note that the origin
 $t_{in}=0$ in Fig.14a is chosen just because this simplifies the numerical
  computations and it has no a real physical sense. After the
exit from inflation, when $\phi(t)\rightarrow \phi_0$ (see
Eq.(\ref{Veff=0})) and $\rho$ starts to fall to zero, the time
coordinate $\tau$ tends to coincide with the cosmic time $t$:
$d\tau/dt\rightarrow 1$. Fig.14a shows that the most of the time
the coordinate $\tau$ practically coincides with the cosmic time
$t$ and this happens after the exit from inflation, as one can
conclude from Figs. 12 and 14b. But at the very beginning of the
scenario, if the phase curve $C$ starts from a point very close to
the line $Q_1=0$, $d\tau/dt$ is very big.  In the limit
$t_{in}\rightarrow t_s$ (where $t_s$ is defined in the beginning
of the previous subsection), $d\tau/dt$ approaches infinity very
fast such that for any $t_1>t_{in}$
\begin{equation}
\tau(t_1)=\lim_{t_{in}\rightarrow
t_s}\int_{t_{in}}^{t_1}\frac{d\tau}{dt}dt=\infty
\label{infinite-evol}
\end{equation}
By continuation of the classical solution to the past we
inevitably arrive a regime where the appropriate phase curve
becomes infinitely close to the line $Q_1=0$. If the classical
evolution had started from this regime then the age of the
universe will be infinite when it is measured in the natural time
$\tau$. It is also interesting to see what kind of information one
can obtain analytically about the behavior of the universe in this
starting regime. For this let us rewrite the $\phi$-equation
(\ref{phi1}) in terms of the $\tau$ dependent $\phi$,
$\dot{\phi}=Q_1^{-1/2}d\phi/d\tau$ and
$H=\dot{a}/a=Q_1^{-1/2}d(\ln a)/d\tau$:
\begin{equation}
\frac{d^2\phi}{d\tau^2}-\left(\frac{1}{2}\frac{dQ_1}{d\tau}-3Q_2\frac{d\ln
a}{d\tau}\right)\frac{1}{Q_1}\frac{d\phi}{d\tau}-
\frac{\alpha}{2M_{p}|V_2|}Q_{3} M^{4}e^{-2\alpha\phi/M_{p}}=0.
\label{phi-in-tau}
\end{equation}
As we know, the $\phi$-equation is regular in all points of zone I
(excluding  the line $Q_1=0$) both in terms of the cosmic time $t$
and the natural time $\tau$. Besides, the transformation
(\ref{time-tau}) is also regular in zone I and it has a
singularity only in points of the line $Q_1=0$. Since the first
and the third terms in Eq.(\ref{phi-in-tau}) are regular
everywhere in zone I, the singular behavior of the factor
$Q_1^{-1}$ in the second term as $Q_1\rightarrow 0$, has to be
compensated by the expression in the brackets:
\begin{equation}
\frac{1}{2}\frac{dQ_1}{d\tau}-3Q_2\frac{d\ln a}{d\tau}\rightarrow
0 \qquad \text{as} \qquad Q_1\rightarrow 0. \label{compensation}
\end{equation}
We are interested in the solution $a(\tau)$ corresponding to the
interval of the phase curve very close to the line $Q_1=0$, i.e.
for very small $t_1$ in Eq.(\ref{infinite-evol}). This allows to
represent $Q_2$ in the form $Q_2=Q_2^{(0)}+{\cal O}(t_1)$, where
$Q_2^{(0)}$ is the value of $Q_2$ in the point of the line $Q_1=0$
nearest to the phase curve. It follows from this that in the
starting regime
\begin{equation}
a\approx
a_{in}\exp\left(\frac{Q_1}{6Q_2^{(0)}}\right),\label{starting-regime}
\end{equation}
where $a_{in}$ is the value of the scale factor in the beginning
of the starting regime. Since according to
Eq.(\ref{infinite-evol}) $Q_1$ remains extremely close to zero
during indefinitely long time $\tau$, the scale factor also
remains very close to $a_{in}$ for a very long time $\tau$. This
may be compared with the picture addressed by the emergent
universe models\cite{emergent} where the scale factor also remains
constant during indefinitely long time in the beginning. Another
similar feature is that in Ref.\cite{emergent} the starting regime
can also be realized at an energy scale much lower the quantum
Planck era. However, in contrast to the models\cite{emergent} in
our case there are no needs  neither of a spatial curvature, nor
of the fine tuning of the initial conditions.

\section{TMT Cosmology With No Fine Tuning II:
\newline
 Absence of the Initial Singularity of the Curvature and Inflationary
Cosmology with Graceful Exit to a Small Cosmological Constant
State}

\begin{figure}[htb]
\begin{center}
\includegraphics[width=18.0cm,height=6.0cm]{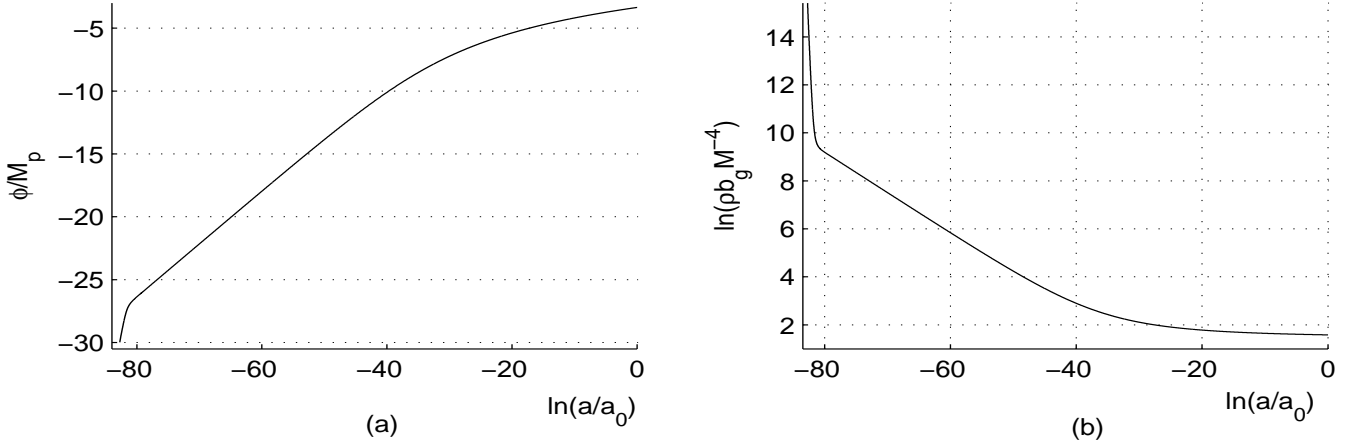}
\end{center}
\caption{In the model with $\delta =0.1$, $\alpha =0.2$,
$V_{1}=10M^{4}$ and $V_{2}=4b_{g}M^{4}$: (a) Typical scale factor
dependence of the inflaton $\phi$ and (b) the energy density
$\rho$ for the phase curves starting from points very close to the
line $Q_1=0$ (in this figure - for $\phi_{in}=-30$,
$\dot{\phi}_{in}= 4540.52202 M^2$).}\label{fig15}
\end{figure}

\begin{figure}[htb]
\begin{center}
\includegraphics[width=18.0cm,height=6.0cm]{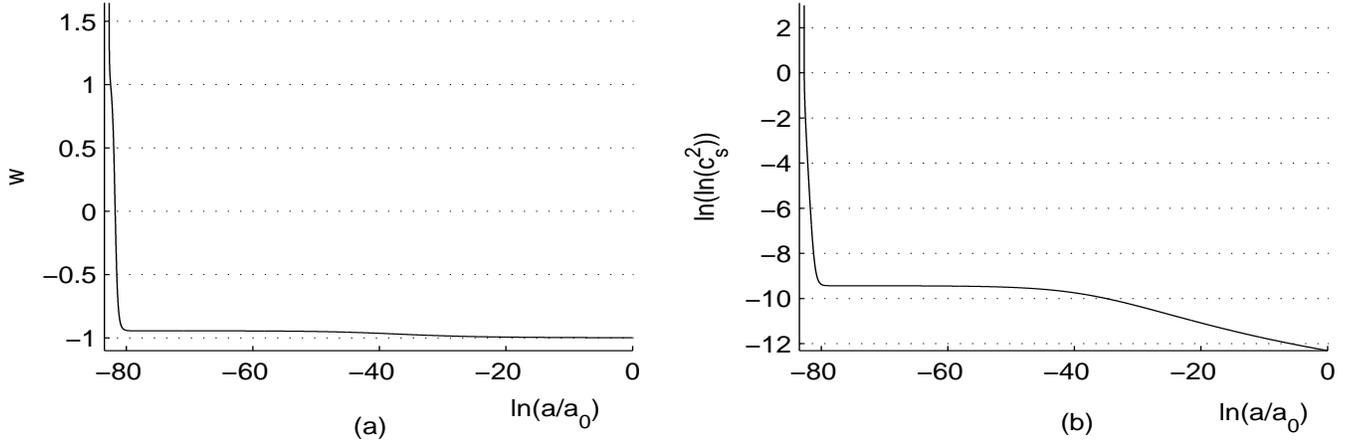}
\end{center}
\caption{The same model and the same initial conditions as in
Fig.15: (a) Typical scale factor dependence of $w$ and (b) the
squared sound speed $c^2_s$ for the phase curves starting from
points very close to the line $Q_1=0$.}\label{fig16}
\end{figure}

With other choice of the parameters $V_1$ and $V_2$, but keeping
the condition (\ref{cond-for-phase-structure}), one can realize a
scenario where the pre-inflationary epoch and inflation
practically coincide with those of the previous section but the
exit from inflation and subsequent evolution are similar to those
studied in Sec.IVC. We present here the results of numerical
solutions for the model with $\delta =0.1$, $\alpha =0.2$,
$V_{1}=10M^{4}$ and $V_{2}=4b_{g}M^{4}$. The structure of the
phase plane, the behavior of the phase curves and the attractor
are very similar to what we have seen in Figs.9 and 10, i.e zones
I, II and III are present in the same manner, and for this reason
we do not repeat the plane phase picture here. Therefore the same
effects, namely the sudden singularity at $S_1=0$ and power law
inflation present in this model as well. The only essential
difference consists in the absence of the oscillatory regime
labeled as the point $B$ in Figs.9 and 10. Now instead of this,
all trajectories
 approach the attractor which in its turn
asymptotically (as $\phi\rightarrow\infty$) takes the form of the
straight line $\dot\phi =0$.

Typical scale factor dependence of the inflaton $\phi$, the energy
density $\rho$, the equation-of-state $w$ and the sound speed of
perturbations $c_s^2$ are presented in Figs.15 and 16. Their
behavior in the very early universe is actually identical to that
in the model of the previous section. Again, it is not a problem
to obtain more than 75 e-folds during the power law inflation (the
region of $w\approx -0.95$ in Fig.16a) just by choosing a larger
absolute value $|\phi_{in}|$. We have chosen $\phi_{in}=-30M_p$
because this allows to show more details in these graphs. During
the power low inflation and in the late universe the behavior of
$\phi$, $\rho$ and $w$ is very similar to what we have seen in the
model of Sec.IVC, see Fig.5. $c_s^2$ starting from a huge value
decreases to a value slightly bigger than 1 and remains
practically constant during the power low inflation; afterward it
asymptotically approaches the value $c_s=1$.

\section{TMT Cosmology With No Fine Tuning III: Superaccelerated Universe}

\begin{figure}[htb]
\begin{center}
\includegraphics[width=17.0cm,height=12.0cm]{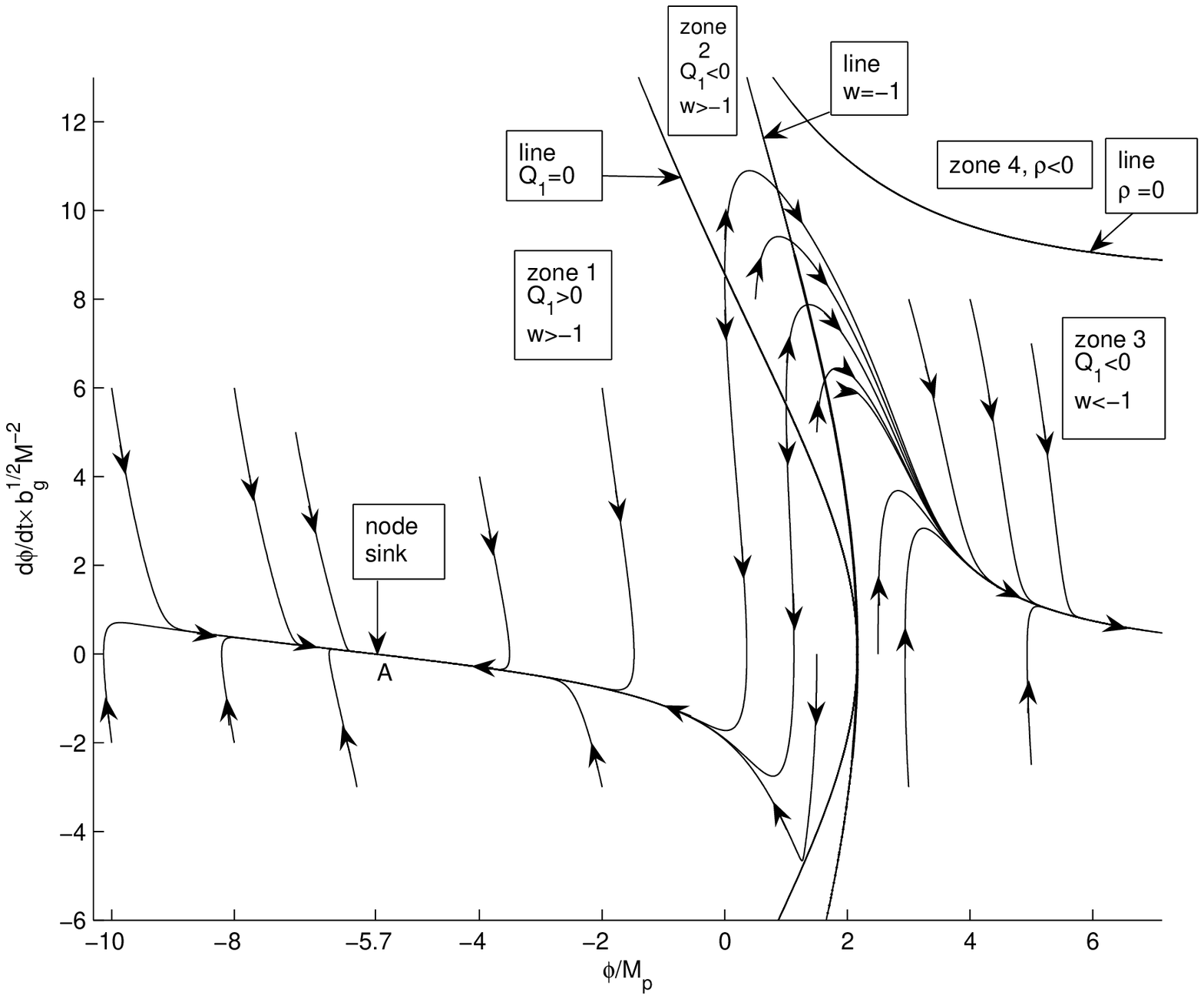}
\end{center}
\caption{The phase portrait for the model with $\alpha =0.2$,
$\delta =0.1$, $V_{1}=10M^{4}$ and $V_{2}=9.9b_{g}M^{4}$. The
region with $\rho >0$ is divided into two dynamically disconnected
regions by the line $Q_{1}(\phi,\dot{\phi})=0$. To the left of
this line $Q_{1}>0$ (the appropriate zone we call zone 1) and to
the right \, $Q_{1}<0$. In zone 1, the phase portrait in the
neighborhood of the node sink is similar to that of Sec.IVD.
However the dynamical protection from initial singularities takes
place here for the same reasons as those considered in Secs.V and
VI. The $\rho
>0$ region to the right of the line $Q_{1}(\phi,\dot{\phi})=0$ is
divided into two zones (zone 2 and zone 3) by the line $Q_2=0$
(the latter coincides with the line where $w=-1$). In zone 2 \,
$w>-1$ but $c_s^2<0$. In zone 3 \,$w<-1$ and $c_s^2>0$. Phase
curves started in zone 2 cross the line $w=-1$. All phase curves
in zone 3 exhibit processes with super-accelerating expansion of
the universe. Besides all the phase curves in zone 3 demonstrate
dynamical attractor behavior to the line which asymptotically, as
$\phi\to\infty$, approaches  the straight line
$\dot{\phi}=0$.}\label{fig17}
\end{figure}

\begin{figure}[htb]
\begin{center}
\includegraphics[width=16.0cm,height=5.0cm]{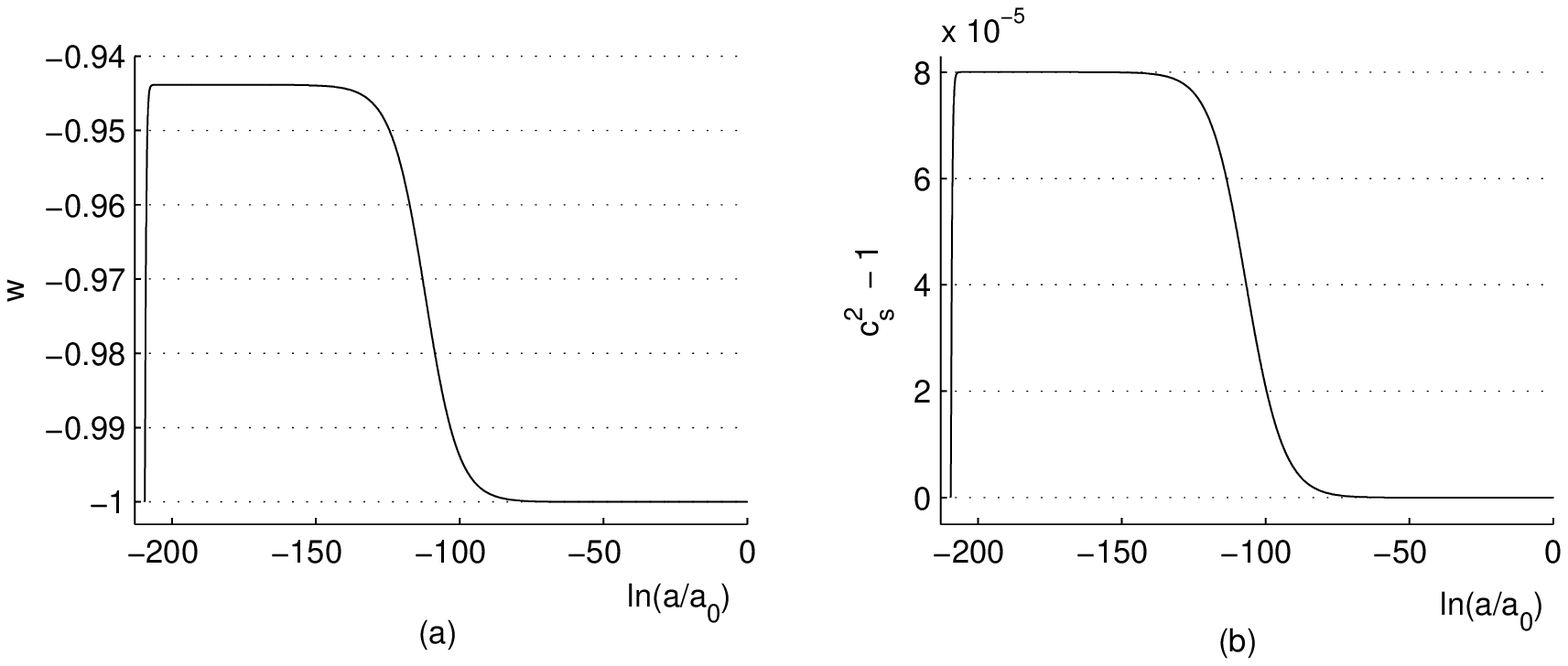}
\end{center}
\caption{For the model with $\alpha =0.2$, $\delta =0.1$,
$V_{1}=10M^{4}$ and $V_{2}=9.9b_{g}M^{4}$: typical scale factor
dependence of the equation-of-state $w$ (Fig.(a)) and effective
speed of sound for perturbations $c_s^2$ (Fig.(b)) for phase
curves in zone 1 (in the figure the initial conditions are
$\phi_{in}=-50M_{p}$, $\dot\phi_{in} =0$).} \label{fig18}
\end{figure}
\begin{figure}[htb]
\begin{center}
\includegraphics[width=16.0cm,height=5.0cm]{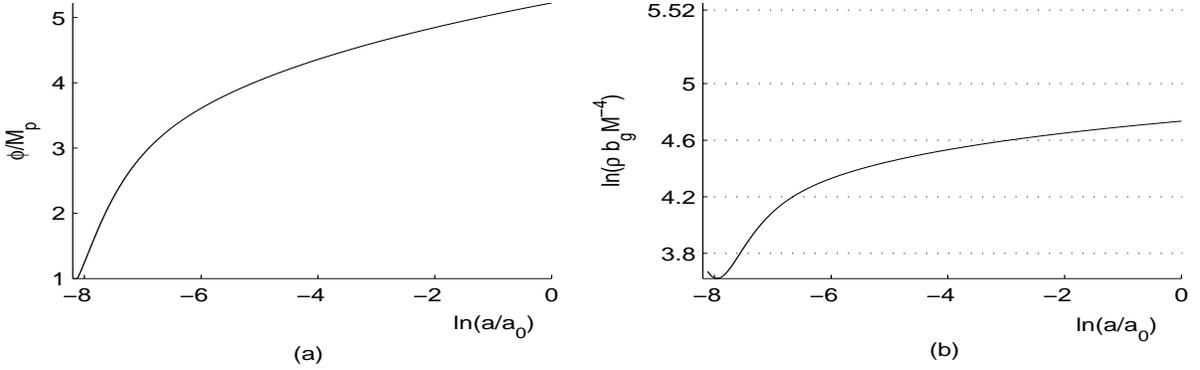}
\end{center}
\caption{For the model with $\alpha =0.2$, $\delta =0.1$,
$V_{1}=10M^{4}$ and $V_{2}=9.9b_{g}M^{4}$: typical scalar factor
dependence of $\phi$  (Fig.(a)) and
 of the energy density $\rho$, defined by Eq.(\ref{rho1}),
(Fig(b)) in the regime corresponding to the phase curves started
in zone 2. Both graphs correspond to the initial conditions
$\phi_{in}=M_{p}$, $\dot\phi_{in} =5.7M^2/\sqrt{b_g}$; $\rho$
increases approaching asymptotically $\Lambda_{2}=
\frac{M^{4}}{b_{g}}e^{5.52}$, where $\Lambda_{2}$ is the value of
$\Lambda$ determined by Eq.(\ref{lambda}); see also
Fig.1b.}\label{fig19}
\end{figure}

\begin{figure}[htb]
\begin{center}
\includegraphics[width=18.0cm,height=7.0cm]{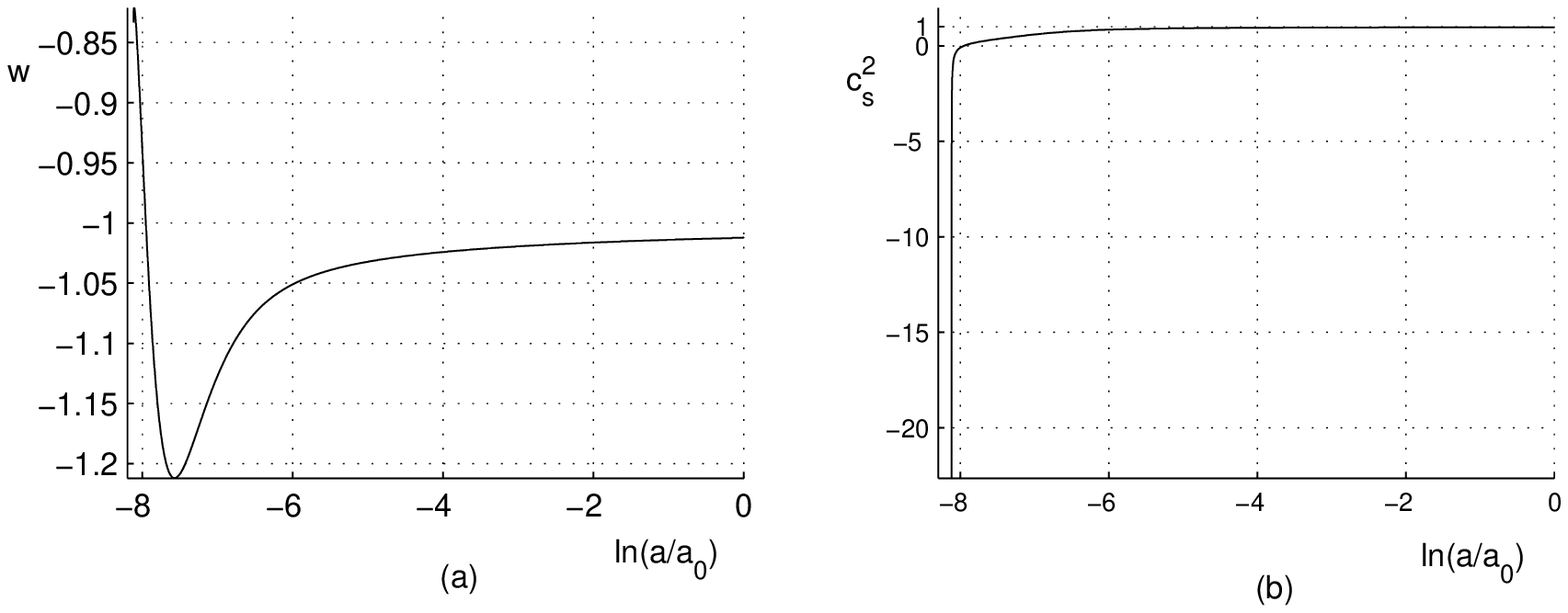}
\end{center}
\caption{For the model with $\alpha =0.2$, $\delta =0.1$,
$V_{1}=10M^{4}$ and $V_{2}=9.9b_{g}M^{4}$: typical scale factor
dependence of the equation-of-state $w$ (Fig.(a)) and effective
speed of sound for perturbations $c_s^2$ (Fig.(b)) for phase
curves in zones 2 and 3 (in the figure the initial conditions are
$\phi_{in}=M_{p}$, $\dot\phi_{in} =5.7M^2/\sqrt{b_g}$.}
\label{fig20}
\end{figure}

  Equations
$Q_{i}(\phi,\dot\phi)=0$ \, ($i=1,2,3$) \, determine lines in the
phase plane $(\phi,\dot\phi)$. In terms of a mechanical
interpretation of Eq.(\ref{phi1}), the change of the sign of
$Q_{1}$ can be treated as the change of the mass of "the
particle". Therefore one can think of situation where "the
particle" climbs up in the potential with acceleration. It turns
out that when the scalar field is behaving in this way, the flat
FRW universe may undergo a super-acceleration.

With simple algebra one can see that the following "sign rule"
holds for the equation-of-state $w=p/\rho$:
\begin{equation}
sign(w+1)=sign (Q_2) \label{sgnw+1}
\end{equation}
Therefore if in the one side from the line $Q_2(\phi,\dot\phi)=0$
\, $w>-1$ then in the other side $w<-1$. To incorporate the region
of the phase plane where $w<-1$ into the cosmological dynamics one
should provide that the line $Q_2(\phi,\dot\phi)=0$ lies in a
dynamically permitted zone. As we have seen in Secs.V and VI, the
condition (\ref{cond-for-phase-structure}) disposes the line
$Q_2(\phi,\dot\phi)=0$ in the physically unacceptable region where
$\rho<0$. It turns out that if the opposite condition holds
\begin{equation}
(b_g+b_{\phi})V_1-2V_2<0 \label{cond-for-phase-structure-1}
\end{equation}
then the line $Q_2(\phi,\dot\phi)=0$ may be in a dynamically
permitted zone.

There are a lot of  sets of parameters providing the $w<-1$ phase
in the late universe. For example we are demonstrating here this
effect with the following set of the parameters of the original
action(\ref{totaction}): $\alpha =0.2$, $V_{1}=10M^{4}$ and
$V_{2}=9.9b_{g}M^{4}$ used in Sec.IVD but now we choose $\delta
=0.1$ instead of $\delta =0$. The results of the numerical
solution are presented in Figs.17-20.

The phase plane, Fig.17, is divided into two regions by the line
$\rho =0$. The region $\rho >0$ is divided into two dynamically
disconnected regions by the line $Q_{1}(\phi,\dot\phi)=0$.

To the left of the line $Q_{1}(\phi,\dot\phi)=0$ - zone 1 where
$Q_{1}>0$. Comparing carefully the phase portrait in the zone 1
with that in Fig.7 of Sec.IVD, one can see an effect of
$\delta\neq 0$ on the shape of phase trajectories. However the
general structure of these two phase portraits is very similar. In
particular, they have the same node sink $A(-5.7M_{p},0)$. At this
point "the force" equals zero since $Q_{3}|_A=0$. The value $\phi
=-5.7M_{p}$ coincides with the position of the minimum of
$V_{eff}^{(0)}(\phi)$ because in the limit $\dot{\phi}\rightarrow
0$ the role of the terms proportional to $\delta$ is negligible.
Among trajectories converging to node $A$ there are also
trajectories corresponding to a power low inflation of the early
universe, which is just a generalization to the case $\delta\neq
0$ of the similar result discussed in Sec.IVD. For illustration we
present some features of one of the solutions in Fig.18. Note that
the same mechanism of the dynamical protection from the initial
singularity we have discussed in Secs.V and VI, holds also here
for solutions whose phase curves are located in zone 1.

In the region to the right of the line $Q_{1}(\phi,\dot\phi)=0$,
all phase curves approach the attractor which in its turn
asymptotically (as $\phi\rightarrow\infty$) takes the form of the
straight line $\dot\phi =0$. This region is divided into two zones
by the line $Q_2(\phi,\dot\phi)=0$. In all points of this line
$w=-1$. In zone 2, i.e. between the lines $Q_1=0$ and $w=-1$, the
equation-of-state $w>-1$ and the sound speed $c_s^2<0$. Therefore
in zone 2 the model is absolutely unstable and has no any physical
meaning. In zone 3, i.e. between the line $w=-1$ and the line
$\rho =0$, the equation-of-state $w<-1$ and the sound speed
$c_s^2>0$.

For a particular choice of the initial data $\phi_{in}=M_{p}$,
$\dot\phi_{in} =9M^4/\sqrt{b_g}$, the features of the solution of
the equations of motion are presented in Figs.19 and 20. The main
features of the solution as we observe from the figures are the
following: 1) $\phi$ slowly increases in time; 2) the energy
density $\rho$ slowly increases approaching the constant $\Lambda
=\Lambda_{2}$  defined by the same formula as in
Eq.(\ref{lambda}), see also Fig.1b; for the chosen parameters
$\Lambda_{2}\approx\frac{M^{4}}{b_{g}}e^{5.52}$; \quad
 3) in zone 3 $w$ becomes
less than $-1$ and after achieving a minimum $w\approx -1.2$ it
then increases asymptotically approaching $-1$ from below.

Using the classification of Ref.\cite{Vikman} of conditions for
 the dark energy to evolve from the state with $w>-1$ to the
phantom state, we see that transition of the phase curves from
zone 2 (where $w>-1$) to the phantom zone 3 (where $w<-1$) occurs
under the conditions  $p_{,X}=0$, $\rho_{,X}\neq 0$, $X\neq 0$.
Qualitatively the same behavior one observes for all initial
conditions $(\phi_{in},\dot{\phi}_{in})$ disposed in the zone 2.
The question of constructing a realistic scenario where the dark
energy can evolve from the power low inflation state disposed in
zone 1 to the phantom zone 3 is beyond of the goal of this paper.

\section{Resolutions of
the  Cosmological Constant Problems and Connection Between TMT and
Conventional Field Theories with Integration Measure $\sqrt{-g}$}

\subsection{The new cosmological constant problem}
The smallness of the observable cosmological constant $\Lambda$ is
known as the new cosmological constant problem\cite{Weinberg2}. In
TMT, there are two ways to provide the observable order of
magnitude of $\Lambda\sim (10^{-3}eV)^{4}$ by an appropriate
choice of the parameters of the theory (see Eqs.(\ref{lambda}) and
(\ref{bV1>V2})) but {\it without fine tuning of the dimensionfull
parameters}.

\subsubsection{Seesaw mechanism}

 If $V_{2}<0$ then there is no need for $V_{1}$ and $V_{2}$ to be
small: it is enough that $b_{g}V_{1}<|V_{2}|$ and $V_{1}/
|V_{2}|\ll 1 $.  This possibility is a kind of  {\it seesaw}
 mechanism\cite{G1},\cite{seesaw}). For instance, if $V_{1}$ is
determined by the
 energy scale of electroweak symmetry breaking $V_{1}\sim
(10^{3}GeV)^{4}$ and $V_{2}$ is determined by the Planck scale
$V_{2} \sim (10^{18}GeV)^{4}$ then $\Lambda_{1}\sim
(10^{-3}eV)^{4}$. The range of the possible scale of the
dimensionless parameter $b_{g}$ remains very broad.

\subsubsection{The TMT correspondence principle and the smallness of
$\Lambda$}

 Let us start from the notion that if $V_2>0$ or alternatively
 $V_2<0$ and $b_gV_{1}>|V_{2}|$
then $\Lambda\sim \frac{V_1}{b_g}$. Hence the second possibility
to ensure the needed smallness of $\Lambda$ is to choose the {\it
dimensionless} parameter $b_{g}>0$ to be a huge number.   In this
case the order of magnitudes of $V_{1}$ and $V_{2}$ could be
either as in the above case of the seesaw mechanism or to be not
too much different from each other (or even of the same order).
For example, if $V_{1}\sim (10^{3}GeV)^{4}$ then for getting
$\Lambda_1\sim (10^{-3}eV)^{4}$ one should assume that $b_{g}\sim
10^{60}$. It is important to stress that as it was explained in
footnote 69, {\it the huge value of $b_{g}$ can be equivalently
regarded as an extremely small ($\sim 10^{-60}$) value of the
coupling constant of the scalar curvature to the measure} $\Phi$.
Below we will use this value just for illustrative purposes.

Note that $b_{g}$ is the ratio of the coupling constants of the
scalar curvature to the measures $\sqrt{-g}$ and $\Phi$
respectively in the fundamental action (\ref{totaction}). The
Lagrangians $L_1$ and $L_2$ have the same structure: both of them
contain the scalar curvature, kinetic and pre-potential terms. It
is natural to assume that the ratio of couplings of all the
corresponding terms in $L_1$ and $L_2$ to the measures $\sqrt{-g}$
and $\Phi$ have the same or close orders of magnitude. This is why
in Sec.III we have made an assumption that the dimensionless
parameters $b_g$ and $b_{\phi}$ have close orders of magnitude.
For the same reason we will also assume that $V_2/V_1\sim b_g\sim
10^{60}$. If this is the case\footnote{Note that if $V_{2}<0$ then
the choice $|V_2|/V_1\sim b_g$ means that in this case the second
way of resolution of the new CC problem is a particular case of
the seesaw mechanism. However the second way is applicable also if
$V_2>0.$} then the huge value of $b_g$ can be treated as an
indication that TMT implies a certain sort of {\bf the
correspondence principle between TMT and conventional field
theories} (i.e theories with only the measure of integration
$\sqrt{-g}$  in the action). In fact, using the notations of the
general form of the TMT action (\ref{S}) in the case of the action
(\ref{totaction}), one can conclude that the relation between the
"usual" (i.e. entering in the action with the usual measure
$\sqrt{-g}$) Lagrangian density $L_2$ and the new one $L_1$
(entering in the action with the new measure $\Phi$) is roughly
speaking $L_2\sim 10^{60}L_1$. In the case $\int L_1\Phi d^4x$
becomes negligible, the remaining term of the action $\int
L_2\sqrt{-g} d^4x$ would describe GR instead of TMT. It seems to
be very interesting that such {\it a correspondence principle for
the TMT action (\ref{S}) may have a certain relation to the
extreme smallness of the cosmological constant}.

Appearance of a large dimensionless constant in particle field
theory is usually associated with hierarchy of masses and/or
interactions describing by {\it different} terms in the
Lagrangian. The way large numbers can appear in the TMT action is
absolutely different. It is easier to see this difference in the
case of a fine tuned model, where $b_g=b_{\phi}$ and $b_gV_1=V_2$,
see Appendix B. In such a case the Lagrangians $L_1$ and $L_2$ not
only have the same type of terms but they are just proportional:
$L_2=b_gL_1$. Therefore {\it the nature of the huge value of $b_g$
differs here very much from the conventional hierarchy issue}.

If such the ratio between  $L_1$ and $L_2$ is actually realized,
then taking into account the fact that $L_1$ and $L_2$ describe
the same matter and gravity degrees of freedom in a very similar
manner, the question arises why $L_1$ is not dynamically
negligible in comparison with $L_2$. To answer this question we
have to turn to the fundamental action (\ref{totaction}) that it
is convenient to rewrite in the following form
\begin{eqnarray}
&S=&\int \sqrt{-g}d^{4}x e^{\alpha\phi /M_{p}}
\left[-\frac{b_{g}}{\kappa}R(\Gamma
,g)\left(\frac{\zeta}{b_{g}}+1\right)+\left(\frac{\zeta}{b_{g}}
+\frac{b_{\phi}}{b_{g}}\right)\frac{b_{g}}{2}g^{\mu\nu}\phi_{,\mu}\phi_{,\nu}-e^{\alpha\phi
/M_{p}}\left(\zeta\frac{V_{1}}{V_2} +1\right)V_2\right]
 \label{totaction1}
\end{eqnarray}
where one can see that the ratio $\zeta/b_g$ has an important
dynamical role. Analyzing  the constraint (\ref{constraint2}) and
cosmological dynamics studied in Secs.IV-VI it is easy to see that
the order of magnitude of the scalar field
$\zeta\equiv\Phi/\sqrt{-g}$ is generically close to that of $b_g$
(recall that $b_{\phi}/b_{g}\sim 1$). In other words, it turns out
that {\it on the mass shell the ratio of the measures}
$\Phi/\sqrt{-g}$ {\it generically compensates the smallness of}
$L_1/L_2$. Thus, similar terms ($R$-terms, kinetic terms and
pre-potential terms) appearing in the action (\ref{totaction})
with measures $\Phi$ and $\sqrt{-g}$ respectively, are both
dynamically important in general.

In the light of this understanding of the general picture it is
interesting to check the TMT dynamics in situations where
$\zeta/b_g$ becomes very small. Let us start from the fine tuned
model where $b_gV_1=2V_2$ and $b_g=b_{\phi}$ (recall that  $b_g>0$
and we consider the case $V_1>0$). In this case it follows from
the constraint (\ref{zeta-without-ferm-delta=0}) that
\begin{equation}
\frac{\zeta}{ b_{g}}=\frac{M^{4}e^{-2\alpha\phi/M_{p}}}
{V_{1}+M^{4}e^{-2\alpha\phi/M_{p}}}, \label{zeta-special}
 \end{equation}
Then the effective potential (\ref{Veffvac-delta=0}) (see also
Eqs.(\ref{rho-without-ferm})-(\ref{V-quint-without-ferm-delta=0}))
reads
\begin{equation}
V_{eff}^{(0)}(\phi)=\frac{V_2}{b_g^2}+\frac{M^{8}e^{-4\alpha\phi/M_{p}}}
{2b_g(V_1+ 2M^{4}e^{-2\alpha\phi/M_{p}})}. \label{Veff-special}
\end{equation}
In such a fine tuned model, $\zeta/b_g$ approaches zero
asymptotically as $\phi\rightarrow\infty$ where the effective
potential becomes flat. However when looking into the TMT action
written in the form (\ref{totaction1}) we see that the asymptotic
disappearance of $\zeta/b_g$ means that we deal with an asymptotic
transition from TMT to a conventional field theory model with only
measure of integration $\sqrt{-g}$ and only one Lagrangian
density. In the limit $\zeta/b_g\rightarrow 0$, the
transformation to the Einstein frame (\ref{ct}) takes the form
$\tilde{g}_{\mu\nu}=b_{g}e^{\alpha\phi/M_{p}}g_{\mu\nu}$.
Therefore in the Einstein frame, the limit of the action
(\ref{totaction1}) as $\zeta/b_g\rightarrow 0$ is reduced to the
following  model:
\begin{equation}
S|_{\zeta=0,\, Eistein\, frame}=\int \sqrt{-\tilde{g}}d^{4}x\left[
-\frac{1}{\kappa}R(\tilde{g})+\frac{1}{2}\tilde{g}^{\mu\nu}\phi_{,\mu}\phi_{,\nu}-\frac{1}{b_g^2}V_2
\right].
 \label{totaction-zeta0}
\end{equation}
It is easy to see that for example in the  FRW universe, the
asymptotic (as the scale factor $a(t)\rightarrow\infty$) behavior
of the universe in the model (\ref{totaction-zeta0}) coincides
with the appropriate asymptotic result of TMT model under
consideration: both of them asymptotically describe the universe
governed by the cosmological constant $V_2/b_g^2=V_1/2b_g$.

Similar conclusion is obtained in a model where $V_2<b_gV_1<2V_2$,
i.e. with no fine tuning of the prepotentials, which have been
studied in Sec.IVD. The only difference is that now
$\zeta/b_g\rightarrow 0$ and $V_{eff}\rightarrow V_2/b_g^2\sim
V_1/b_g$ as $\phi\rightarrow\phi_{min}$, see Eq.(\ref{minVeff});
in a small neighborhood of $\phi_{min}$ the TMT action presented
in the Einstein frame looks like (\ref{totaction-zeta0}).

Note however that in the context of the k-essence model studied in
Sec.VII where $\delta\neq 0$ and  $V_2<b_gV_1<2V_2$, the
asymptotic value of $\zeta$ in the late time universe ( as
$\phi\rightarrow\infty$ and $X\rightarrow 0$ which is an
asymptotic regime of the phantom behavior) is $|\zeta|\sim b_g$
but nevertheless the energy density tends to $\Lambda =\Lambda_2
=\frac{V_{1}^{2}} {4(b_{g}V_{1}-V_{2})}\sim V_1/b_g$ as well (see
Figs.1b and 19b).

 Similar situation takes place in the model where $b_gV_1>2V_2$, studied
in Secs.IVC and VI. The asymptotic value of $\zeta$ in the late
time universe is again $\zeta\sim b_g$  and the energy density
tends to $\Lambda =\Lambda_1 =\frac{V_{1}^{2}}
{4(b_{g}V_{1}-V_{2})}\sim V_1/b_g$ as well (see Figs.1a, 5b and
15b).

Thus in all the models, the huge value of $b_g$ can ensure the
needed smallness of the dark energy density in the late time
universe but it is not always realized due to the limit
$\zeta/b_g\rightarrow 0$.

\subsection{The Old Cosmological Constant Problem
Is Solved in the Dynamical Regime where the  Fundamental TMT
Action Tends to a Limit Opposite to Conventional Field Theory
(with only measure $\sqrt{-g}$).}

As we have seen in Sec.IVB, in the model with $V_{1}<0$ and
$V_{2}<0$, the old cosmological constant problem is resolved
without fine tuning: the effective potential
(\ref{Veffvac-delta=0}) is proportional to the square of
$V_{1}+M^{4}e^{-2\alpha\phi/M_{p}}$, and  $\phi =\phi_{0}$ where
$V_{1}+ M^{4}e^{-2\alpha\phi_{0}/M_{p}}=0$, is the minimum of the
effective potential without any further tuning of the parameters
and initial conditions. Now we want to analyze some of the
essential differences we have in TMT as compared with the
conditions of the Weinberg's no-go theorem and show what are the
reasons providing solution of the old CC problem in TMT. This has
to be done when TMT is considered in the original frame since in
the Einstein frame we observe only the results in the effective
picture after some of the symmetries are broken.

\begin{itemize}

\item
The basic assumption of the Weinberg's  theorem is that in the
vacuum all the fields (metric tensor $g_{\mu\nu}$ and matter
fields $\psi_n$) are constant. As it was pointed out by S.Weinberg
in the review\cite{Weinberg1}, the Euler-Lagrange equations  for
such constant fields (with the action $\int{\cal L}\left(
g_{\mu\nu},\psi_n \right)d^4x$) have the form
\begin{equation}
\frac{\partial{\cal L}}{\partial g_{\mu\nu}}=0, \label{W1}
\end{equation}
\begin{equation}
\frac{\partial{\cal L}}{\partial\psi_n}=0
 \label{W2}
\end{equation}
and these equations constitute the basis for further Weinberg's
arguments. In particular, if $GL(4)$ symmetry
\begin{equation}
g_{\mu\nu}\rightarrow
A^{\alpha}_{\mu}A^{\beta}_{\nu}g_{\alpha\beta}, \qquad
\psi_i\rightarrow D_{ij}(A)\psi_j \label{W3}
\end{equation}
survives as a vestige of general covariance when all the fields
are constrained to be constant, the Lagrangian ${\cal L}$
transforms as a density:
\begin{equation}
{\cal L}\rightarrow detA\cdot{\cal L}. \label{W4}
\end{equation}
 Weinberg concludes that when
Eq.(\ref{W1}) is satisfied then the unique form of ${\cal L}$ is
\begin{equation}
{\cal L}=c\sqrt{-g}, \label{W5}
\end{equation}
where $c$ is independent of $g_{\mu\nu}$. As a matter of fact this
means that for example in the case of a scalar matter field $\phi$
model considered by Weinberg in Sec.VI of the review
\cite{Weinberg1}, $c$ is determined by the value of the scalar
field $\phi$ potential as $\phi$ is a constant determined by
Eq.(\ref{W2}).

However, if  for example one of the fields $\psi_n$ appears in
${\cal L}$ only via  a term linear in space-time derivatives of
this field then Eq.(\ref{W2}) turns out to be an identity, but
instead  the Euler-Lagrange equations take another form. This is
what happens in TMT where the first term in the action (\ref{S})
is linear in space-time derivatives of $A_{\alpha\beta\gamma}$
(when using the definition (\ref{Aabg}))(see also \footnote{A
possibility of a vacuum with non constant 3-form gauge field has
been discussed in Footnote 8 of the Weiberg's
review\cite{Weinberg1}}). Then instead of Eq.(\ref{W2}) which
appears to be an identity in this case, the Euler-Lagrange
equations for $A_{\alpha\beta\gamma}$ look
\begin{equation}
\partial_{\mu}\frac{\partial({\Phi L_1})}{\partial
A_{\alpha\beta\gamma,\mu}}=0, \label{W6}
\end{equation}
which are nontrivial even for constant $A_{\alpha\beta\gamma}$,
and resulting in Eq.(\ref{AdL1}). Note that $\Phi L_1$ is a scalar
density and transforms exactly according to Eq.(\ref{W4}).
Therefore generically (i.e. if $A_{\alpha\beta\gamma}$ are not
constant while other fields are constant), the Lagrangian ${\cal
L}$ satisfying (\ref{W4}) can have the following form
\begin{equation}
{\cal L}=c_1\Phi +c_2\sqrt{-g}\label{W7}
\end{equation}
where $c_1$ and $c_2$ are independent of $g_{\mu\nu}$ and $\Phi$.
This is why the equation
\begin{equation}
\partial{\cal L}/\partial\phi=T^{\mu}_{\mu}\sqrt{-g},\label{W8}
\end{equation}
where $T^{\mu}_{\mu}$ is the trace of the energy-momentum tensor,
used by Weinberg\cite{Weinberg1} for all constant $g_{\mu\nu}$ and
matter fields, is generically no longer valid.

\item
Let us now note that $\zeta =\zeta_{0}(\phi)$,
Eq.(\ref{zeta-without-ferm-delta=0}), becomes singular
\begin{equation}
|\zeta|\approx\frac{2|V_2|}{|V_{1}+
M^{4}e^{-2\alpha\phi/M_{p}}|}\rightarrow\infty \qquad \text{as}
\qquad \phi\rightarrow\phi_{0}. \label{zeta-singular}
\end{equation}
  In this limit the effective potential
 (\ref{Veff1}) (see also
 Eq.(\ref{Veffvac-delta=0})) behaves as
\begin{equation}
V_{eff}\approx\frac{|V_2|}{\zeta^2}. \label{V-singular}
\end{equation}

Thus, disappearance of the cosmological constant occurs in the
regime where $|\zeta|\rightarrow\infty$. In this limit, the
dynamical role of the terms of the Lagrangian $L_2$ (coupled with
the measure $\sqrt{-g}$) in the action (\ref{totaction}) becomes
negligible in comparison with the terms of the Lagrangian $L_1$
(see also the general form of the action (\ref{S})). A particular
realization of this we observe in the behavior of $V_{eff}$,
Eq.(\ref{V-singular}). It is evident that the limit of the TMT
action (\ref{S}) as $|\zeta|\rightarrow\infty$ is opposite to the
conventional field theory (with only measure $\sqrt{-g}$) limit of
the TMT action discussed in subsection VIIIA. {\bf From the point
of view of TMT, this is the answer to the question why the old
cosmological constant problem cannot be solved (without fine
tuning) in theories with only the measure of integration}
$\sqrt{-g}$ {\bf in the action}.

\item
Recall that one of the basic assumptions of the Weinberg's no-go
theorem is that all fields in the vacuum must be constant. This is
also assumed for the metric tensor, components of which  in the
vacuum must be {\bf nonzero} constants. However, this is not the
case in the fundamental TMT action (\ref{totaction}) defined in
the original (non Einstein) frame if we ask what is the metric
tensor $g_{\mu\nu}$ in the $\Lambda =0$ vacuum . To see this let
us note that in the Einstein frame all the terms in the
cosmological equations are regular. This means that the metric
tensor in the Einstein frame $\tilde{g}_{\mu\nu}$ is always well
defined, including the $\Lambda =0$ vacuum state $\phi =\phi_{0}$
where $\zeta$ is infinite. Taking this into account and using the
transformation to the Einstein frame (\ref{ct}) we see that {\bf
all components of the metric in the original frame $g_{\mu\nu}$ go
to zero overall in space-time as $\phi$ approaches the $\Lambda
=0$ vacuum state}:
\begin{equation}
g_{\mu\nu}\sim \frac{1}{\zeta}\sim V_{1}+
M^{4}e^{-2\alpha\phi/M_{p}}\rightarrow 0 \qquad (\mu,\nu =0,1,2,3)
\qquad \text{as} \quad \phi\rightarrow\phi_{0}. \label{g-degen}
\end{equation}
This result shows that the Weinberg's analysis based on the study
of the trace of the energy-momentum tensor misses any sense in the
case $g_{\mu\nu}=0$.

The metric is an attribute of the space-time term. Hence
disappearance of the metric $g_{\mu\nu}$ in the limit
$\phi\rightarrow\phi_{0}$ means that the strict formulation of the
TMT model (\ref{totaction}) with $V_1<0$ and $V_2<0$ may require a
new mathematical basis. A manifold which is not equipped with the
metric (corresponding to the $\Lambda =0$ vacuum state) emerges as
a certain limit of a sequence of space-times. Thus the model under
consideration might be formulated not in a space-time manifold but
rather by means of a set of space-time manifolds. A limiting point
of a sequence of space-times is a "vacuum space-time manifold"
 one of the differences of which from a regular space-time
consists in the absence of the metric $g_{\mu\nu}$.

It follows immediately from (\ref{g-degen}) that $\sqrt{-g}$ tends
to zero like
\begin{equation}
\sqrt{-g}\sim \frac{1}{\zeta^2}\sim \left(V_{1}+
M^{4}e^{-2\alpha\phi/M_{p}}\right)^2\rightarrow 0 \quad \text{as}
\quad \phi\rightarrow\phi_{0}. \label{non-degen-tends}
\end{equation}
Then the definition $\zeta\equiv\Phi/\sqrt{-g}$ implies that the
integration measure $\Phi$ also tends to zero but rather like
\begin{equation}
\Phi\sim \frac{1}{\zeta}\sim V_{1}+
M^{4}e^{-2\alpha\phi/M_{p}}\rightarrow 0 \quad \text{as} \quad
\phi\rightarrow\phi_{0}. \label{Phi-tends}
\end{equation}
Thus both the measure $\Phi$ and the measure $\sqrt{-g}$ become
degenerate  in the $\Lambda =0$ vacuum state $\phi =\phi_{0}$.
However $\sqrt{-g}$ tends to zero more rapidly than $\Phi$.

\item
As we have discussed in detail (see Secs.II, IIIB and
Refs.\cite{GK2},\cite{GK3}), with the original set of variables
used in the fundamental TMT action it is very hard or may be even
impossible to display the physical meaning of TMT models. One of
the reasons is that in the framework of the postulated need to use
the Palatini formalism, the original metric $g_{\mu\nu}$ and
connection $\Gamma^{\alpha}_{\mu\nu}$ appearing in the fundamental
TMT action describe a non-Riemannian space-time. The
transformation to the Einstein frame (\ref{ct}) enables to see the
physical meaning of TMT because the space-time becomes Riemannian
in the Einstein frame. Now we see that the transformation to the
Einstein frame (\ref{ct}) plays also the role of a {\bf
regularization of the space-time metric}: the singular behavior of
the transformation (\ref{ct}) as $\phi\approx\phi_{0}$ compensates
the disappearance of the original metric $g_{\mu\nu}$ in the
vacuum $\phi =\phi_{0}$. As a result of this the metric in the
Einstein frame $\tilde{g}_{\mu\nu}$ turns out to be well defined
in all physical states including the $\Lambda =0$ vacuum state.

\end{itemize}

\section{Discussion and Conclusion}

\subsection{Differences of TMT from the standard field theory in
curved space-time} The main  idea of TMT is that the general form
of the action $\int L\sqrt{-g}d^{4}x$ is not enough in order to
account for some of the fundamental problems of particle physics
and cosmology. The key difference of TMT from the conventional
field theory in curved space-time consists in the
hypothesis\cite{GK2}-\cite{GKatz} that in addition to the term in
the action with the volume element $\sqrt{-g}d^{4}x$ there should
be one more term where the volume element is metric independent
but rather it is determined either by four (in the 4-dimensional
space-time) scalar fields $\varphi_{a}$ or by a three index
potential $A_{\alpha\beta\gamma}$,  see
Eqs.(\ref{S})-(\ref{Aabg}). We would like to emphasize that
including in the action of TMT the coupling of the Lagrangian
density $L_{1}$ with the measure $\Phi$, we modify in general both
the gravitational and matter sectors as compared with the standard
field theory in curved space-time. Besides we made two more
assumptions: the measure fields ($\varphi_{a}$ or
$A_{\alpha\beta\gamma}$) appear only in the volume element; one
should proceed in the first order formalism. {\it These
assumptions constitute all the modifications of the general
structure of the theory we have made as compared with the
conventional  field theory where only the measure of integration
$\sqrt{-g}$ is used in the action principle}. In fact, the
Lagrangian densities $L_{1}$ and $L_{2}$ studied in the present
paper, contain  only such
 terms which should be present in a conventional model with
 minimally coupled to gravity scalar field. In particular there is no need
for the non-linear kinetic term as well as for the phantom type
term in the fundamental Lagrangian densities $L_{1}$ and $L_{2}$
in order to obtain a super-acceleration phase at the late time
universe.

After making use of the variational principle and formulating the
resulting equations in the Einstein frame, we have seen that the
effective action (\ref{k-eff}) represents a concrete realization
of the $k$-essence\cite{k-essence} obtained from first principles
of TMT without any exotic terms in the fundamental Lagrangian
densities.

\subsection{Short summary of results}

\subsubsection{The classical pre-inflation epoch of very early universe:
absence of initial singularity of the curvature.}

As $\delta =0$, i.e. in the fine-tuned model, the dynamics of
$\phi$ can be analyzed by means of its effective potential
(\ref{Veffvac-delta=0}). As $\phi\ll -M_{p}$ the effective $\phi$
potential has the exponential form and it is proportional to the
integration constant $M^4$. In other words, the effective
potential governing the dynamics of the early universe results
from the spontaneous breakdown of the global scale invariance
(\ref{st}) caused by the intrinsic feature of TMT (see Eqs.
(\ref{varphi}) and (\ref{app1})). We have seen that independently
of the values of the parameters $V_{1}$, $V_{2}$ and under very
general initial conditions, solutions rapidly achieve a regime of
the well explored power law inflation. In this fine-tuned case the
initial singularity is present in the usual
sense\cite{Borde-Vilenkin-PRL}.

If $\delta\neq 0$, we deal with the intrinsically $k$-essence
dynamics. Here solutions also rapidly achieve a regime of   power
law inflation. However the solutions describing the
pre-inflationary stage cannot be continued into the past till the
singularity of the curvature. The reason is that the specific
structure of the phase plane (when $\delta\neq 0$)
 does not allow classical evolution for which the phase curve
 crosses
the line $Q_1(\phi,\dot{\phi})=0$ (on this line $\ddot{\phi}$
becomes infinite). Independently of how close is the starting
point to the border line $Q_1=0$ in zone I (where $Q_1>0$, see
Figs. 9 and 10), duration of the cosmic time evolution from the
start up to the transition to the regime of the power law
inflation is finite. This result takes place for all finite
initial conditions $\phi_{in}$, $\dot{\phi}_{in}$ in zone I.
However the energy density $\rho$, pressure $p$, the first two
derivatives of the scale factor $\dot{a}$ and $\ddot{a}$ (and
therefore curvature) remain finite on the line
$Q_1(\phi,\dot{\phi})=0$ while only $\dot{p}$ and $\dddot{a}$ (and
therefore time derivative of the curvature) become singular. It is
clear that this type of "sudden" singularity studied by
Barrow\cite{Barrow1} on purely kinematic grounds  has no relation
to the initial singularity theorems\cite{Borde-Vilenkin-PRL}. Note
also that the strong energy condition holds on the line
$Q_1(\phi,\dot{\phi})=0$.

It is worthwhile to pay attention to the fact that the sound speed
$c_s^2$ tends to infinity as the starting point
$(\phi_{in},\dot{\phi}_{in})$ approaches the line
$Q_1(\phi,\dot{\phi})=0$. Therefore generation of any mode of
scalar fluctuations in states extremely close to the line $Q_1=0$
requires extremely large energy. Thus the initial state formed in
the close neighborhood of the line $Q_1=0$ must be practically the
ground state. This allows to hope that the effect $c_s^2\to\infty$
could help to solve the problem of the initial conditions in
inflationary cosmology.

\subsubsection{Graceful exit from inflation}

 In our toy model there
are three regions of the parameters $V_{1}$ and $V_{2}$ and
corresponding three shapes of the effective potentials, Fig.1.
Consequently three different types of scenarios for exit from
inflation can be realized (note that when the kinetic term $X$
becomes small, the terms where $\delta$ appears are negligible):

a)  $V_{1}<0$ and $V_{2}<0$, \,  Sec.IVB. In this case the power
law inflation ends with damped oscillations of $\phi$ approaching
the point of the phase plane ($\phi =\phi_0, \dot\phi =0$) where
the vacuum energy $V_{eff}^{(0)}(\phi_0)=0$. This occurs without
fine tuning  of the parameters  and the initial conditions.

b) $V_{1}>0$ and $b_{g}V_{1}>2V_{2}$, \, Sec.IVC. In this case the
power law inflation monotonically transforms to the late time
inflation asymptotically governed by the cosmological constant
$\Lambda_1$.

c) $V_{2}<b_{g}V_{1}<2V_{2}$, \, Sec.IVD. In this case the power
law inflation ends without oscillations at the final value
$\phi_{min}$,
 corresponding to the (non zero) minimum of the effective
potential.

The model we have studied in this paper may be extended by
including the Higgs field, as well as gauge fields and fermions.
It turns out that the scalar sector of such an extended model
enables a scenario\cite{hep-th/0603150} which resembles a hybrid
inflation\cite{hybrid}. These results will be presented in a
future publication.

\subsubsection{Cosmological constant problems}

1) {\it The old cosmological constant problem}.
 In Sec.IVB
we have seen in details that if $V_1<0$ then, for a broad range of
other parameters, the vacuum energy turns out to be zero without
fine tuning. This effect is a direct consequence of the TMT
structure which yields the following results: a) the effective
scalar field potential generated in the Einstein frame is
proportional to {\it a perfect square} of a $\phi$-depending
expression, which gets zero at some value of $\phi$; b) one of the
terms in this expression is proportional to the integration
constant $\pm M^4$ the appearance of which is also the intrinsic
feature of TMT. If such type of the structure for the scalar field
potential in a conventional (non TMT) model would be chosen "by
hand" it would be a sort of fine tuning. Note that the
spontaneously broken global scale invariance is not necessary to
achieve this effect\cite{GK3}.

In Sec.VIIIB we have explained in details how this result avoids
the well known no-go theorem by Weinberg\cite{Weinberg1} stating
that generically in field theory one cannot achieve zero value of
the potential in the minimum without fine tuning. It is
interesting that the resolution of the old CC problem in the
context of TMT happens in the regime where
$\zeta\rightarrow\infty$. From the point of view of TMT, the
latter is the answer to the question why the old cosmological
constant problem cannot be solved (without fine tuning) in
theories with only the measure of integration $\sqrt{-g}$  in the
action.

2) {\it The new cosmological constant problem}. Interesting result
following from the general structure of the scale invariant TMT
model with $V_1>0$ is that the cosmological constant $\Lambda$,
Eq.(\ref{lambda}), is a ratio of quantities constructed from
pre-potentials $V_{1}$, $V_{2}$ and the dimensionless parameter
$b_{g}$. Such structure of $\Lambda$ allows to propose two  ways
(see Sec.VIIIA) for the resolution of the problem of the smallness
of $\Lambda$ that should be $\Lambda\sim (10^{-3}eV)^{4}$:

\quad a) The first way is a kind of a {\it seesaw}
mechanism\cite{seesaw}. For instance, if  $V_{1}\sim
(10^{3}GeV)^{4}$ and $V_{2} \sim (10^{18}GeV)^{4}$ then
$\Lambda_1\sim (10^{-3}eV)^{4}$.

\quad b) The second way is realized if the dimensionless
parameters $b_{g}$, $b_\phi$ and $V_2/V_1$ of the action
(\ref{totaction}) are huge numbers of the close orders of
magnitude.  For example, if $V_{1}\sim (10^{3}GeV)^{4}$ then for
getting $\Lambda\sim (10^{-3}eV)^{4}$ one should assume that
$b_{g}\sim 10^{60}$. The latter means that the ratio of the
Lagrangian density $L_1$ coming into the underlying action with
the 'new' measure $\Phi$, to the Lagrangian density $L_2$ coming
with the 'usual' measure $\sqrt{-g}$, is extremely small:
$L_1/L_2\sim 10^{-60}$. Possibility of this idea means that the
resolution of the new cosmological constant problem may have a
certain relation to
 {\it the correspondence principle} between TMT and
conventional field theories (see details in Sec.VIII.A.2).

\subsubsection{Super-acceleration phase of the Universe.
 }
If no fine tuning of the parameters is made in the fundamental
action, namely if $b_g\neq b_{\phi}$, then our TMT model has big
enough regions in the parameter space where the super-acceleration
phase in the late time universe becomes possible. The appropriate
phantom dark energy asymptotically approaches a cosmological
constant.
 However, in the framework of our model it is impossible to obtain {\it a pure classical
solution} which connects the early universe power law inflation
with the late time super-acceleration. This problem  is apparently
related with the toy character of our TMT model. First, possible
dependence of $V_1$ and $V_2$ on the Higgs field has not been
taken into account. Second, the role of the matter creation has
been ignored: in TMT the fermionic matter generically contributes
to the constraint equation for the scalar field $\zeta$ and so can
effect the field $\phi$ dynamics as well. One can speculate that
matter creation at the end of inflation may cause a 'phase
transition' from the structure of the phase plane of the type
shown in Figs.9 and 10  to the structure as in Fig.17. Then
without any fine tuning, a scenario involving a start from finite
curvature, a power law inflation in the early universe and
super-accelerated late universe with small asymptotic CC would be
possible.

\subsection{What can we expect from quantization}

In this paper we have studied only classical TMT and its possible
effects in the context of cosmology. However quantization of TMT
as well as influence of quantum effects on the processes explored
in this paper may have a crucial role. We summarize here some
ideas and speculations which gives us a hope that quantum effects
can keep the main results of this paper.

 Recall first two fundamental facts of TMT as a classical field
theory: (a) The measure degrees of freedom appear in the equations
of motion only via the scalar $\zeta$, Eq.(\ref{zeta}); (b) The
scalar $\zeta$  is determined (as a function of matter fields, in
our toy model - as a function of $\phi$) by the constraint which
is nothing but a consistency condition of the equations of motion
(see Eqs.(\ref{app1})-(\ref{app3}) in Appendix A and
Eq.(\ref{constraint2})). Therefore the constraint plays a key role
in TMT. Note however that if we were ignore the gravity from the
very beginning in the action (\ref{totaction}) then instead of the
constraint (\ref{app3}) we would obtain Eq.(\ref{app1}) (where one
has to put zero the scalar curvature). In such a case we would
deal with a different theory. This notion shows that the gravity
and matter intertwined in TMT in a much more complicated manner
than in GR. Hence introducing the new measure of integration
$\Phi$ we have to expect that the quantization of TMT may be a
complicated enough problem. Nevertheless we would like here to
point out that in the light of the recently proposed idea of
Ref.\cite{Giddings}, the incorporation of four scalar fields
$\varphi_a$  together with the scalar density $\Phi$,
Eq.(\ref{Phi}), (which in our case are the measure fields and the
new measure of integration respectively), is  a possible way to
define local observables in the local quantum field theory
approach to quantum gravity. We regard this result as an
indication that the effective gravity $+$ matter field theory has
to contain the new measure of integration $\Phi$ as it is in TMT.

 The assumption  that the
measure fields $\varphi_a$ (or $A_{\alpha\beta\gamma}$) appear in
the action (\ref{S}) {\it only} via the measure of integration
$\Phi$, has a key role in the TMT results and in particular for
the resolution of the old cosmological constant problem. In
principle one can think of breakdown of such a structure by
quantum corrections. However, TMT possesses an infinite
dimensional symmetry mentioned in item 2 of Sec.II which, as we
hope, is able to protect  this feature of the structure of the
action from a deformation caused by quantum corrections. Another
effect of quantum corrections is the possible appearance of a
nonminimal coupling of the scalar field $\phi$ to gravity in the
form like for example $\xi R\phi^2$. Proceeding in the first order
formalism of TMT one can show that the nonminimal coupling can
affect the k-essence dynamics but the mechanism for resolution of
the old CC problem exhibited in this paper remains unchanged. This
conclusion together with expected effect of quantum corrections on
the scale invariance, including appearance of a "normal"
cosmological constant term $\int \tilde{\Lambda}\sqrt{-g}d^4x$
(see our discussion in the paragraph after Eq.(\ref{Veff=0})),
allows us to hope that the quantization of the underlying TMT
model (\ref{totaction}) will not spoil the exhibited resolution of
the old CC problem.

Quantization of TMT being a constrained system requires developing
the Hamiltonian formulation of TMT. Preliminary consideration
shows that the Einstein frame appears in the canonical formalism
in a very natural manner. A systematic exploration of TMT in the
canonical formalism will be a subject of forthcoming research.

\subsection{Possible quantum  effects on the initial singularity related
to  the line $Q_1=0$}

Possibility of the classical evolution of the universe to start
from finite  values of $\rho_{in}$, $p_{in}$ such that
$\dot{a}_{in}$ and $\ddot{a}_{in}$ are finite and
$Q_1(\phi_{in},\dot{\phi}_{in})\approx 0$, requires certain
arguments addressed to explain how such an initial state could be
formed.

One can think of the action (\ref{totaction}) of the underlying
model as a tree approximation of the entire TMT effective action
at energies  below the Planck scale. Then one can expect that the
next terms in the expansion of the entire effective action  will
be like $R^2(\Gamma,g)$, $(g^{\mu\nu}\phi_{,\mu}\phi_{,\nu})^2$
etc. coupled both to the measure $\sqrt{-g}$ and to the measure
$\Phi$. It is evident that in such more general TMT model
formulated in the Einstein frame, the $X$ dependence of the matter
Lagrangian density $p(\phi,X;M)$ in Eq.(\ref{k-eff}) may be very
much different from that in Eq.(\ref{p1}). As a result of this the
structure of the phase plane may change quantitatively in
comparison with that of the model studied in the present paper.
For example, it may be found that the new factor
$Q_1(\phi,\dot{\phi};M)$ has no zeros. In such a case the effect
of the initial singularity related to the line $Q_1=0$ vanishes.
However this is only one of the possibilities. In the absence of a
reliable knowledge about the entire effective action one can also
expect, with  not less grounds, that the changed structure of the
phase plane still involves the line $Q_1=0$. So we return the
problem formulated in the previous paragraph. At the qualitative
level we can suggest the following two ideas.

1. As we have already mentioned after Eq.(\ref{rho-very-big}) the
described initial state can be realized at the energy density
close to the Planck scale. In this case, similar to the
cosmological models in the framework of the standard Einstein
gravity, there is a need to use the quantum gravity and therefore
the  singularity effects related to the line $Q_1=0$ are expected
to be smoothed out.

2. If $\rho_{in}<\rho_{Planck}$, then quantum effects from the
scalar field dynamics itself can play a crucial role in smoothing
the singularity related to  the line $Q_1=0$. Recall that
$c_s^2<0$ in the regions between $Q_1=0$ and $\rho =0$ (where
$Q_1<0$, see Figs.9 and 10), which means that from this side of
the line $Q_1=0$ arbitrarily small scalar field fluctuation is
amplified exponentially. Besides
 $c_s^2\to -\infty$ when approaching the line $Q_1=0$ from
this side. Therefore we expect that the classical or quantum chaos
in the regions with $Q_1<0$ should affect the classical dynamics
on the other side of the border, i.e. in the close neighborhood of
the line $Q_1=0$ in zone I. Moreover, this influence can work as a
mechanism responsible for creation of our universe starting from
an excitation in zone with $Q_1<0$.

\section{Acknowledgements}

  We  acknowledge L. Amendola, S.
Ansoldi, R. Barbieri, J. Bekenstein, A. Buchel, A. Dolgov, A.
Feinstein, S.B. Giddings, V. Goldstein, P. Gondolo, A. Guth, B-L.
Hu, P.Q. Hung, D. Kazanas, Y. Lyubarski, D. Marolf, D.G. McKeon,
J. Morris, V. Miransky,
 H. Nielsen, Y.Jack Ng, H. Nishino, E. Nissimov,  S. Pacheva, L.
Parker, R. Peccei, M. Pietroni, S. Rajpoot, R. Rosenfeld,  V.
Rubakov, J. Senovilla, E. Spallucci, A. Vikman, A. Vilenkin, M.
Visser, S. Wetterich, F. Wilczek and A. Zee for helpful
conversations and communications on different stages of this
research. Our special gratitude to L. Prigozhin for his help in
the implementation of numerical solutions.

\appendix

\section{Equations of motion  in
the original frame}

Variation of the measure fields $\varphi_{a}$ with the condition
$\Phi\neq 0$ leads, as we have already seen in Sec.II, to the
equation $ L_{1}=sM^{4}$ where $L_{1}$ is now defined, according
to  Eq. (\ref{S}), as the part of the integrand of the action
(\ref{totaction}) coupled to the measure $\Phi$.  Equation
(\ref{varphi}) in the context of the model (\ref{totaction}) reads
(with the choice $s=+1$):
\begin{equation}
\left[-\frac{1}{\kappa}R(\Gamma, g)+
\frac{1}{2}g^{\mu\nu}\phi_{,\mu}\phi_{,\nu}\right]e^{\alpha\phi
/M_{p}} -V_{1}e^{2\alpha\phi /M_{p}} =M^{4}, \label{app1}
\end{equation}
It can be noticed that the appearance of a nonzero integration
constant $M^{4}$ spontaneously breaks the scale invariance
(\ref{st}).

Variation of the action (\ref{totaction}) with respect to
$g^{\mu\nu}$ yields
\begin{equation}
-\frac{1}{\kappa}(\zeta +b_{g})R_{\mu\nu}(\Gamma)  +(\zeta
+b_{\phi})\frac{1}{2}\phi_{,\mu}\phi_{,\nu}
+\frac{1}{2}g_{\mu\nu}\left[\frac{b_{g}}{\kappa}R(\Gamma ,g)
-\frac{b_{\phi}}{2}g^{\alpha\beta}\phi_{,\alpha}\phi_{,\beta}
+V_{2}e^{\alpha\phi /M_{p}}\right]
 =0.\label{app2}
\end{equation}

We see that in contrast to field theory models with only the
measure $\sqrt{-g}$, in TMT there are two independent equations
containing curvature. Contracting Eq.(\ref{app2}) with
$g^{\mu\nu}$ and solving Eq.(\ref{app1}) for $R(\Gamma, g)$  we
obtain the following {\it consistency condition} of these two
equations:
\begin{equation}
(\zeta -b_{g})\left(M^{4}e^{-\alpha\phi /M_{p}}+V_{1}e^{\alpha\phi
/M_{p}}\right)+2V_{2}e^{\alpha\phi
/M_{p}}+(b_{g}-b_{\phi})\frac{1}{2}g^{\mu\nu}\phi_{,\mu}\phi_{,\nu}=0,
 \label{app3}
\end{equation}
that we will call {\it the constraint in the original frame}.

It follows from Eqs.(\ref{app1}) and (\ref{app2}) that
\begin{eqnarray}
 \frac{1}{\kappa}R_{\mu\nu}(\Gamma)&=&
 \frac{\zeta +b_{\phi}}{\zeta +b_g}\cdot\frac{1}{2}\phi_{,\mu}\phi_{,\nu}
 \nonumber\\
 &-&\frac{g_{\mu\nu}}{2(\zeta
+b_g)}\left[b_gM^{4}e^{-\alpha\phi /M_{p}}
      +(b_gV_{1}-V_{2})e^{\alpha\phi /M_{p}}
      -(b_g-b_{\phi})\frac{1}{2}g^{\alpha\beta}\phi_{,\alpha}\phi_{,\beta}\right]
\label{app4}
\end{eqnarray}

The scalar field $\phi$ equation of motion in the original frame
can be written in the form
\begin{eqnarray}
&& \frac{1}{\sqrt{-g}}\partial_{\mu}\left[e^{\alpha\phi /M_{p}}
(\zeta +b_{\phi})\sqrt{-g}g^{\mu\nu}\partial_{\nu}\phi\right]
\nonumber\\
&-&\frac{\alpha}{M_{p}}e^{\alpha\phi /M_{p}}\left[(\zeta
+b_g)M^{4}e^{-\alpha\phi /M_{p}}+
[(b_g-\zeta)V_{1}-2V_{2}]e^{\alpha\phi /M_{p}}
-(b_g-b_{\phi})\frac{1}{2}g^{\alpha\beta}\phi_{,\alpha}\phi_{,\beta}\right]
=0 \label{phi-orig}
\end{eqnarray}
where Eq.(\ref{app1})  has been used.

Variation of the action (\ref{totaction}) with respect to the
connection degrees of freedom leads to the equations we have
solved earlier\cite{GK3}. The result is
\begin{equation}
\Gamma^{\lambda}_{\mu\nu}=\{
^{\lambda}_{\mu\nu}\}+\frac{1}{2}(\delta^{\alpha}_{\mu}\sigma,_{\nu}
+\delta^{\alpha}_{\nu}\sigma,_{\mu}-
\sigma,_{\beta}g_{\mu\nu}g^{\alpha\beta}) \label{GAM2}
\end{equation}
where $\{ ^{\lambda}_{\mu\nu}\}$  are the Christoffel's connection
coefficients of the metric $g_{\mu\nu}$ and
\begin{equation}
\sigma,_{\mu}\equiv \frac{\alpha}{M_p}\phi,_{\mu}+\frac{1}{\zeta
+b_g}\zeta,_{\mu}\label{sigma-mu}
\end{equation}

\section{Asymmetry between early and late time dynamics of the
universe as result of asymmetry in the couplings to measures
$\Phi$ and $\sqrt{-g}$ in the action. }

The results obtained in Secs.IV and V depend very much on the
choice of the parameters $V_{1}$, $V_{2}$ and $\delta$ in the
action (\ref{totaction}). Let us recall  that the curvature term
in the action (\ref{totaction}) couples to the measure $\Phi
+b_{g}\sqrt{-g}$ while the $\phi$ kinetic term couples to the
measure $\Phi +b_{\phi}\sqrt{-g}$. This is the reason of
$\delta\neq 0$. If we were choose the fine tuned condition $\delta
=0$ then both the curvature term and the $\phi$ kinetic term would
be coupled to the same measure $\Phi +b_{g}\sqrt{-g}$. One can
also pay attention that depending on the choice of one of the
alternative  conditions $b_{g}V_{1}>2V_{2}$ or $b_{g}V_{1}<2V_{2}$
we realize different shapes of the effective potential if
$b_{g}V_{1}>V_{2}$ (see Fig.1). And again, if instead we were
choose the fine tuned condition $b_{g}V_{1}=V_{2}$ then the action
would contain only one prepotential coupled to the measure $\Phi
+b_{g}\sqrt{-g}$.

So, in order to avoid fine tunings we have  introduced asymmetries
in the couplings of the different terms in the Lagrangian
densities $L_{1}$ and $L_{2}$ to measures $\Phi$ and $\sqrt{-g}$.
In order to display the role of these asymmetries it is useful to
consider what happens if such asymmetries are absent in the action
at all. In other words we want to explore here the gravity+dilaton
model where both $\delta =0$ and $b_{g}V_{1}=V_{2}$. In such a
case the action contains only one Lagrangian density coupled to
the measure $\Phi +b_{g}\sqrt{-g}$:
\begin{equation}
S=\int (\Phi +b_{g}\sqrt{-g}) d^{4}x e^{\alpha\phi /M_{p}}
\left(-\frac{1}{\kappa}R +
\frac{1}{2}g^{\mu\nu}\phi_{,\mu}\phi_{,\nu}-Ve^{\alpha\phi /M_{p}}
\right), \label{S-symm}
\end{equation}
where $V=V_{1}=V_{2}/b_{g}$. An equivalent statement is that
$L_{1}=b_{g}L_{2}$; it is an example of the very special class of
the TMT models where $L_{1}$ is proportional to $L_{2}$.

To see the cosmological dynamics in this model one can use the
results of Sec.IIIB. If we assume in addition $b_{g}>0$ and
$V_{1}>0$, then after the shift $\phi\rightarrow \phi +\Delta\phi$
where $\Delta\phi =-{M_{p}}{2\alpha}\ln(V/M^{4})$ (which is not a
shift symmetry in this case), the effective potential
(\ref{Veffvac-delta=0}) takes the form
\begin{equation}
V_{eff}^{(symm)}(\phi)=\frac{V^{2}}{b_{g}M^{4}}\cosh^{2}(\alpha\phi/M_{p}).
\label{Veff-symm}
\end{equation}
In contrast to general cases ($b_{g}V_{1}\neq V_{2}$)  this
potential has no flat regions and it is symmetric around a certain
point in the $\phi$-axis. This form of the potential (with an
additional  constant) has been used in a model of the early
inflation\cite{Barrow}.

\section{Some remarks on the measure fields independence of
$L_{1}$ and $L_{2}$}

Although we have assumed in the main text that $L_{1}$ and $L_{2}$
are $\varphi_{a}$ independent, a contribution equivalent to the
term $\int f(\Phi/\sqrt{-g})\Phi d^{4}x$ can be effectively
reproduced in the action (\ref{S})  if a nondynamical field
(Lagrange multipliers) is allowed in the action. For this purpose
let us consider the contribution to the action of the form
\begin{equation}
S_{auxiliary}=\int[\sigma\Phi+l(\sigma)\sqrt{-g}]d^{4}x
\label{L-m-action}
\end{equation}
where $\sigma$ is an auxiliary nondynamical field and $l(\sigma)$
is an analytic function. Varying $\sigma$ we obtain
$dl/d\sigma\equiv l^{\prime}(\sigma)=-\Phi/\sqrt{-g}$ that can be
solved for $\sigma$: $\sigma =l^{\prime -1}(-\Phi/\sqrt{-g})$
where $l^{\prime -1}$ is the inverse function of $l^{\prime}$.
Inserting this solution for $\sigma$ back into the action
(\ref{L-m-action}) we obtain
\begin{equation}
S_{aux.integrated}=\int f(\Phi/\sqrt{-g})\Phi d^{4}x
\label{L-m-action-int}
\end{equation}
where the auxiliary field has disappeared and
\begin{equation}
f(\Phi/\sqrt{-g})\equiv l^{\prime -1}(-\Phi/\sqrt{-g})+l(l^{\prime
-1}(-\Phi/\sqrt{-g}))\frac{\sqrt{-g}}{\Phi}.\label{f-aux}
\end{equation}

To see the difference between effect of this type of auxiliary
fields as compared with a model where the $\sigma$ field is
equipped with a kinetic term, let us consider two toy models
including gravity and $\sigma$ field: one - without kinetic term
\begin{equation}
S_{toy}=\int\left[\left(-\frac{1}{\kappa}R+\sigma\right)\Phi
+b\sigma^{2}\sqrt{-g}\right]d^{4}x \label{toy-1}
\end{equation}
and the other - with a kinetic term
\begin{equation}
S_{toy,k}=\int\left[\left(-\frac{1}{\kappa}R+\sigma +
\frac{1}{2}g^{\alpha\beta}\frac{\partial_{\alpha}\sigma
\partial_{\beta}\sigma}{\sigma^{2}}\right)\Phi
+b\sigma^{2}\sqrt{-g}\right]d^{4}x \label{toy-2}
\end{equation}
where $b$ is a real constant. For both of them it is assumed the
use of the first order formalism. The first model is invariant
under local transformations $\Phi\rightarrow J\Phi$, \,
$g_{\mu\nu}\rightarrow Jg_{\mu\nu}$, \, $\sigma\rightarrow
J^{-1}\sigma$ where $J$ is an arbitrary space-time function while
in the second model the same symmetry transformations hold only if
 $J$ is  constant.

Variation of the measure fields $\varphi_{a}$ in the model
(\ref{toy-2}) leads (if $\Phi\neq 0$) to
\begin{equation}
-\frac{1}{\kappa}R+\sigma +
\frac{1}{2}g^{\alpha\beta}\frac{\partial_{\alpha}\sigma
\partial_{\beta}\sigma}{\sigma^{2}}=M^{4},\label{toy-1-phi}
\end{equation}
where $M^{4}$ is the integration constant. On the other hand
varying the action (\ref{toy-2}) with respect to $g^{\mu\nu}$
gives
\begin{equation}
\chi\left(-\frac{1}{\kappa}R_{\mu\nu} +
\frac{1}{2}\frac{\partial_{\mu}\sigma
\partial_{\nu}\sigma}{\sigma^{2}}\right)-
\frac{1}{2}b\sigma^{2}g_{\mu\nu}=0,\label{toy-1-gmunu}
\end{equation}
where $\chi\equiv\frac{\Phi}{\sqrt{-g}}$. The corresponding
equations in the model (\ref{toy-1}) are obtained from
(\ref{toy-1-phi}) and (\ref{toy-1-gmunu}) by omitting the terms
with gradients of $\sigma$. It follows from Eqs.(\ref{toy-1-phi})
and (\ref{toy-1-gmunu}) that
\begin{equation}
\frac{1}{\chi}=\frac{M^{4}-\sigma}{2b\sigma^{2}}
\label{toy-constr}
\end{equation}
This result holds  in both models.

 In the model (\ref{toy-1}), variation of $\sigma$ results in
 $\frac{1}{\chi}=-\frac{1}{2b\sigma}$ which is consistent with
 Eq.(\ref{toy-constr}) only if the integration constant $M=0$.
 This means that through the classical mechanism displayed in TMT,
  it is impossible to achieve
 spontaneous breakdown of the {\it local} scale invariance  the first model
 possesses. This appears consistent with arguments by
 Elitzur\cite{Elitzur} concerning impossibility of a spontaneous
 breaking of a local symmetry without gauge fixing.

 Transition to the Einstein frame where the space-time becomes
 Riemannian is implemented by means of the
 transformation $\tilde{g}_{\mu\nu}=\chi g_{\mu\nu}$. For the
 model (\ref{toy-1}) the gravitational equations in the Einstein frame
 read
\begin{equation}
\frac{1}{\kappa}G_{\mu\nu}(\tilde{g}_{\alpha\beta})=\frac{1}{8b}\tilde{g}_{\mu\nu}.
\label{toy-1-grav}
\end{equation}
This means that the model (\ref{toy-1}) with auxiliary
(nondynamical) field $\sigma$ intrinsically contains a constant
vacuum energy.

In the model (\ref{toy-2}), where $\sigma$ appears as a dynamical
field,  the gravitational equations in the Einstein frame results
from Eq.(\ref{toy-1-gmunu})
\begin{equation}
\frac{1}{\kappa}G_{\mu\nu}(\tilde{g}_{\alpha\beta})=
\frac{(M^{4}-\sigma)^{2}}{8b\sigma^{2}}\tilde{g}_{\mu\nu}
+\frac{1}{2}\left(\frac{\partial_{\mu}\sigma
\partial_{\nu}\sigma}{\sigma^{2}}-
\frac{1}{2}\tilde{g}_{\mu\nu}\tilde{g}^{\alpha\beta}\frac{\partial_{\alpha}\sigma
\partial_{\beta}\sigma}{\sigma^{2}}\right). \label{toy-2-grav}
\end{equation}
It is convenient to rewrite this equation in terms of the scalar
field $\ln\sigma\equiv\phi$:
\begin{equation}
\frac{2}{\kappa}G_{\mu\nu}(\tilde{g}_{\alpha\beta})=V_{eff}(\phi)
\tilde{g}_{\mu\nu} +\frac{1}{2}\left(\partial_{\mu}\phi
\partial_{\nu}\phi-
\frac{1}{2}\tilde{g}_{\mu\nu}\tilde{g}^{\alpha\beta}\partial_{\alpha}\phi
\partial_{\beta}\phi\right), \qquad \text{where} \qquad
V_{eff}(\phi)=\frac{1}{4b\sigma^{2}}(M^{4}e^{-\phi}-1)^{2}.
\label{toy-2-grav-phi}
\end{equation}
 The
$\phi$-equation reads $\Box\phi +V_{eff}^{\prime}(\phi)=0$.
Similar to the general discussion in the main text we see that if
the $\sigma$ field is dynamical then TMT provides the vacuum with
zero energy without fine tuning.

Hence the main difference between the TMT models with auxiliary
and dynamical scalar fields consists in radically different
results concerning the cosmological constant problem.

However it is very unlikely that a nondynamical scalar field will
not acquire a kinetic term after quantum
corrections\cite{Gross-Neveau}. Then it  becomes dynamical which
restores the above results for the model (\ref{toy-2}). This is
why we have ignored the rather formal possibility of introducing
the nondynamical scalars into the fundamental action of the models
studied in this paper.

\section{The model with $V_1=V_2=0$ and its relation to
models with scaling solutions\cite{Tsujikawa}}

With the choice of the parameters $V_1=V_2=0$, the scalar field
$\phi$ Lagrangian density (\ref{p1}) takes the form $p(\phi,X;M)
=Xg(Y)$ where
\begin{equation}
Y=Xe^{2\alpha\phi/M_{p}}
 \label{XY}
\end{equation}
\begin{equation}
g(Y) = \left[1-\frac{M^4}{4b_gY}\left(1+\frac{\delta\cdot
b_g}{M^4}Y\right)^2\right]. \label{g(Y)}
\end{equation}
Such a model is a particular realization of the models studied in
Ref.\cite{Tsujikawa}, but recall that here it follows from first
principles. It is very important that the factor $Q_1(\phi,X)$ in
front of $\ddot{\phi}$ in the $\phi$-equation (\ref{phi1}) is now
\begin{equation}
Q_1
=(b_{g}+b_{\phi})M^{4}e^{-2\alpha\phi/M_{p}}-3\delta^{2}b_{g}^{2}X,
\label{Q100}
\end{equation}
and the line $Q_1=0$ in the plane $(\phi,\dot{\phi})$ provides the
same mechanism for the dynamical avoidance of the initial
singularity we have studied in detail in Secs.V and VI. In the
models considered in those sections,  the role of nonzero $V_1$
and $V_2$ becomes negligible as $\phi\ll M_p$. Therefore the only
essential difference of the model with $V_1=V_2=0$ from the models
of Secs.V and VI is that the power law inflation ends with a
graceful exit to a quintessence phase (with zero CC in contrast
with the model of Sec.VI).

\end{document}